\documentclass[a4paper]{article}
\usepackage[margin=0.75in]{geometry}

\usepackage{tocloft}
\usepackage{graphicx}
\usepackage{amsmath}
\usepackage{amsfonts}
\usepackage{mathrsfs}
\usepackage[all,cmtip]{xy}
\usepackage{soul}
\usepackage{epigraph}
\usepackage{hyperref}

\newtheorem{theorem}{Theorem}

\title{Elements of Consciousness and Cognition. Biology, Mathematic, Physics and Panpsychism: an Information Topology Perspective}
\author{Pierre Baudot \\ Inserm UNIS UMR1072 - Universit\'{e} Aix-Marseille AMU, \\ Facult\'{e} de M\'{e}decine - Secteur Nord, 51, Boulevard Pierre Dramard,\\ 13015 Marseille, France\\
	pierre.baudot@gmail.com}

\date{16th July 2018}

\begin{document}
\maketitle

\begin{abstract}
	This review presents recent and older results on elementary quantitative and qualitative aspects of consciousness and cognition and tackles the question "What is consciousness?" conjointly from biological, neuroscience-cognitive, physical and mathematical points of view. It proposes to unify various results and theories by means of algebraic topology and puts forward the suggestion that information topology is a particularly appropriate formalism to achieve such an aim. The resulting discrete probabilistic and group theoretic principles and structures governing the theory of consciousness underline its Galoisian nature.  \\
	The first chapter presents the postulates and results on elementary perception in psychophysics and neuroscience at various organizational scales of the nervous system and proposes the hypothesis of an electrodynamic intrinsic nature of consciousness which is sustained by an analogical code. It underlines the diversity of the learning mechanisms that sustain the dynamics of perception and consciousness, including adaptive and homeostatic processes on multiple scales, and details their current generic expression within probability and information theory. \\
    The second chapter investigates the logical aspects of cognition and consciousness and proposes an axiomatization based on measure and probability theory. Topos and constructive logic are presented as providing an intrinsic non-deterministic-probabilistic logic, with the long-term aim of avoiding the paradoxical decomposition induced by the Axiom of Choice. Using such a basis, we sketch an elementary procedure allowing an expression of the information of a mathematical formula a la G\"{o}del. We then present the formalism of information topology and propose that it provides a preliminary basis for synthesizing the main models of cognition and consciousness within a formal Gestalt theory. Information topology establishes a characterization of information theory functions, allowing for a precise expression of information structures and patterns. It provides a quantification of the structure of statistical interactions and their expression in terms of statistical physics and machine learning. Notably, those topological methods allow conciliation of some of the main theories of consciousness, namely integrated information theory, the global neuronal workspace model, the free energy principle and logical dynamics. The topological approach points out that consciousness is a structural phenomenon arising from collective interactions. Underlining the central role of invariance to transformation in neuroscience and perception, we further propose a possible correspondence of information topology with dynamical system theory and the related quantification of arousal states. 
\end{abstract}

\tableofcontents

\section{Introduction}

A theory of consciousness concerns anyone and should be a theory of anyone: a theory of everybody and everybody's theory. It should be consensual and hence should acknowledge and account for the diversity of all beings (bodies). It should account for and respect the consciousness of anybody, and encompass without contradiction all the hardly countable investigations that have treated consciousness in its different forms and aspects: biological, physical, psychological, mathematical, computational, etc. Consciousness and qualitative perception is also one of the main topics of theology and art; hence, a theory of consciousness should also be theological and artistic, at least minimally, such that it does not contradict the diversity of theologies and art that human minds have formalized and which are some of the central forms of human consciousness and cognition. To avoid the usual dualist oppositions, it is necessary to precise that seen from the world of probability explored here, atheism is also a system or a form of human belief, which also enriches the complex landscape of diverse consciousness and thoughts. 
As a consequence, the road towards such a theory appears difficult, and, while we do not achieve it here, we instead propose some ideas towards what the aims of a theory of consciousness that respects and harmoniously verifies its own axioms (which we consider firstly and in a literal sense to be unity and diversity, as proposed by Tononi and Edelman \cite{Tononi1998}), would be. In the mathematical section of the paper following \cite{Baudot2015a}, we present the formalization of the probability theory within topos theory and constructive logic, a logic with multi-valuations in which the excluded third is not a theorem (independent). Such constructive logic could underline the idea that those beliefs classically considered as complementary opposite statements - dualism - may indeed refer to a diversity of beliefs - pluralism. It provides a preliminary  soft non-deterministic rationality that further provides a legitimate rational status to free will. 
This should not be understood as a novel, personal or established theory of consciousness, and all the more a closed and definitive framework. Information topology is simply a name proposed because two existing, partially established theories, information theory and a central branch of algebraic topology  appear indistinguishable, and should ultimately be, just one. Such unification  is currently only partially understood. 
As emphasized in the ecological mind conclusion, we simply present, recycle and combine, in a consistent fashion, well-established results (which is a cognitive task of associative memory), such that the resulting theory is the least partial possible. In short, there is no claim of originality or novelty, just as in the case of consciousness itself: "Novelty is as old as the world" (Prevert); hence this paper's  status as a review and perspective. 
An important part of the ideas presented here are inherited from Bennequin and are the result of a long-lasting collaboration. Notably, the formalization of visual processing as regards invariance is developed at length in \cite{Bennequin2014}. In the world of ideas, nothing is lost, nothing is created, everything transforms. We will focus on an old intuitionist idea of a mathematical and physical nature of our subjective being, and even of our most elementary perceptions. \\
The main ideas of the review (without analytical requirements) are expressed quite synthetically in the following citations of Riemann and Poincar\'{e} that introduce both consciousness and a topological view on it:
\label{riemanncitation} \textit{"When we think a given thought, then the meaning of this thought is expressed in the shape of the corresponding neurophysiological process."} Riemann \cite{Riemann1876}
\textit{"Now what is science? ...it is before all a classification, a manner of bringing together facts which appear separate, though they are bound together by some natural and hidden kinship. Science, in other words, is a system of relations. ...it is in relations alone that objectivity must be sought. ...it is relations alone which can be regarded as objective. External objects... are really objects and not fleeting and fugitive appearances, because they are not only groups of sensations, but groups cemented by a constant bond. It is this bond, and this bond alone, which is the object in itself, and this bond is a relation."} \cite{Poincare1905}\\
\textit{"Mathematicians do not study objects, but the relations between objects; to them it is a matter of indifference if these objects are replaced by others, provided that the relations do not change. Matter does not engage their attention, they are interested in form alone."  Poincar\'{e} \cite{Poincare1902}.}
\textit{When you use the word information, you should rather use the word form} Thom \cite{Thom1983}.

\section{Neurobiology and psychophysics, electrophysiology of elementary perception}

\subsection{"Unity and Diversity" \cite{Tononi1998}}
This section investigates the question of the limit from which a particular cognitive process or a particular living species can be considered as conscious or not. 
It is not relevant or possible to review all the results concerning consciousness that neuroscience imaging, electrophysiological studies, psychophysic and psychology studies have already presented. All of those studies concern consciousness more or less directly, and most researchers we have encountered or worked with are quite aware that their work more or less directly concerns consciousness, although they may not refer to such a generic concept and usually prefer much more precise, specific, and less grandiose ones. In what follows, we cite only a few examples of such works, not because they are the most pertinent but because we are already familiar with them; the rest can be found in research libraries. The results of such studies, as advocated and centrally underlined by the Integrated Information Theory of Tononi and Edelmann, tend to be that forms of consciousness are very diverse \cite{Tononi1998}. Neuroscience and cognitive sciences have developed specialized concepts and taxonomy for these different forms, such as attention, low-level vision, audition, multi-modal integration, decision, motor planning, short-term memory, etc. In a sense, there exists a given, particular name of consciousness for each function and associated structure in nervous systems. Moreover, there exist a wide diversity of nervous systems: human, macaque, cat, rat, mouse, zebra finch, bat, turtle, elephantfish, cricket, fly, squid, aplysia, worms (caenorhabditis elegans), to cite just a few generic experimental models. Such a diversity reveals the richness of cognitive forms \cite{Yartsev2017}. Each of them have remarkably different structures and functions; hence, a satisfying theory of consciousness would have to be very basic and generic such that all those fields of research can converge. The point of view adopted here, now more accepted in neuroscience (thanks most notably to Koch, Tononi and Edelmann \cite{Paulson2017}), is that if one accepts that there exists a qualitative low-level perception in humans, and admits it provides a quite elementary form of consciousness, one should accept from the purely empirical criterion of observability that the echolocation of a bat, for example, is also associated with elementary forms of consciousness, albeit likely to be different from the one we experience, as can be inferred from electro-physiological studies and discussed by Nagel \cite{Nagel1974}. The boundaries of consciousness have been the subject of numerous social debates with important social ramifications and consequences; notably, in justifying slavery, part of humanity was considered as not being conscious \cite{Montesquieu1748}. From the philosophical perspective, it has been clear since Hegel's "phenomenology of spirit", which is built on the dialectic of the slave and the master, that the question of consciousness and the problem of its multiplicity, of alterity, can be investigated in terms of competitive or dominating purposes \cite{Hegel1807}. The quite recent emergence of biological sciences, introducing the multiplicity and diversity of natural structures and functions and underlining their constitutive inter-dependencies has started to promote a more co-operative, symbiotic or synergistic view of alterity. More generally, the problem of consciousness poses the question of humanity's place with respect to nature and physics. 

\subsection{The neuronal postulate - neural assemblies - neural coding - shapes of memory and qualia}

\subsubsection{The neuronal/biological postulate - physical reductionism} \label{neuronal_postulate}

Neuroscience and cognitive research play a particular role in science, in that they aim to objectively study, by empirical and physical means and with mathematical models or data analysis, the subjectivity of perceptions, actions and decisions. The main postulate of those investigations was clearly stated by Changeux \cite{Changeux1983} and can be summarised by the hypothesis that for any mental subjective state there exists an empirical observable phenomenon, most commonly an electrical activity, that corresponds to or generates it. One debate regarding consciousness theory is whether such correspondence is one to one, what we call the Tononi-Edelmann model (for reasons that will become clear in the course of the paper), or if it is injective-only, implying the existence of some unconscious states - what we call the Dehaenne-Changeux model. The usual neuronal-biological dualist hypothesis, however, forbids metaphysical subjective states (subjective states without any physical observable correlate). This has meant that a major part of neuroscience and cognitive research has adopted physics' reductionist approach and respects the observability axiom of physical theories. Hence, they deserve the name of physical investigations into qualitative experience. What is presented here reconciles the Tononi-Edelmann \cite{Tononi1998,Edelman2000} and Dehaenne-Changeux models \cite{Dehaene2000,Dehaene2011,Dehaene2006} by proposing  that what one may consider as an unconscious state is "someone else's" consciousness. The general proposition that an unconscious state is "someone else's" consciousness can be simply illustrated by the development in patients, specifically called "split-brain" patients, of two quite independent consciousness streams following a callosotomy, as studied notably in the celebrated work of Sperry and Gazzaniga  \cite{Sperry1961,Gazzaniga1967}. Here, we start from the postulate that the objects of our subjective experiences or perceptions exist. We also postulate the existence of the subject that perceives (the "I think therefore I am" of Descartes) and complete it with a statement along the lines of "It moves therefore it is", a phenomenological definition of anima based on animation.  \\
\textbf{Reflexive and qualitative consciousness: feedback and the hard problem. From mind-body dualism to a synthetic monadic view.} The question investigated in this section is whether there necessarily exists an ontological difference between the mind and the body.
An important part of studies into consciousness, following classical, at least partially Platonic dualism and standard mind-body problems, assumes a fundamental distinction between reflexive and qualitative consciousness called qualia, as investigated  by Chalmers \cite{Chalmers1995}. According to this view, reflexive consciousness, the fact of a consciousness being conscious of its own "states", is an easy problem which it has been possible to solve with cybernetic and control theory, which formalize the concept of feedback and gain controls further pursued in neural networks studies. Qualitative consciousness, on the other hand, the elementary qualia, provides what is known as the "hard problem" of consciousness. The "knowledge argument" is a typical thought experiment given to illustrate what a qualia is \cite{Jackson1982}: a scientist, Mary, is living in a black and white room with books providing her all the "reflexive" knowledge about color, including its physical, artistic and neuroscientific aspects. Jackson argues that a qualia is what Mary experiences when she first sees colors that she could not know from the reflexive knowledge contained in her books. Such a thought experiment appears to be more like a linguistic human problem, equating very high level cognitive linguistic abstraction ("reflexive knowledge") with very elementary color perception. Even color perceptions result from a learning or adaptive process, and as in any learning task, we can only assume that she would gain a new qualitative experience by seeing color, which would be in full agreement with her highly abstract qualitative, linguistic and scientific experience of color - probably what she had expected or even spontaneously experienced by synesthetic completion as proposed by Ramachandran \cite{Ramachandran2003}. In other words, we propose here a more monist point of view that reconciles the reflexive and qualitative aspects of consciousness. In this sense, there is a reflexive mechanism, that is further developed here in terms of a self-interaction or internal energy (cf. \ref{infotopo_synthesis}), to any qualitative experience and respectively there is a qualitative mechanism associated with any reflexive experience. Such a view was notably developed at length by Leibniz in his explorations of the nature of what he called 'monads' \cite{Leibniz1686,Leibniz1714}, a view that was further pursued in works that will be partially reviewed later in the paper. 
In this review, we will focus on elementary, "low-level" qualia and highlight the fact that electrophysiology and neuroscience results have demonstrated that they rely on feedback and gain controls on virtually all scales of  nervous system organization and recordings.

\subsubsection{Consciousness: the electromagnetic view - "Where is my mind?"}

\begin{figure} [!h]
	\centering
	\includegraphics[height=14cm]{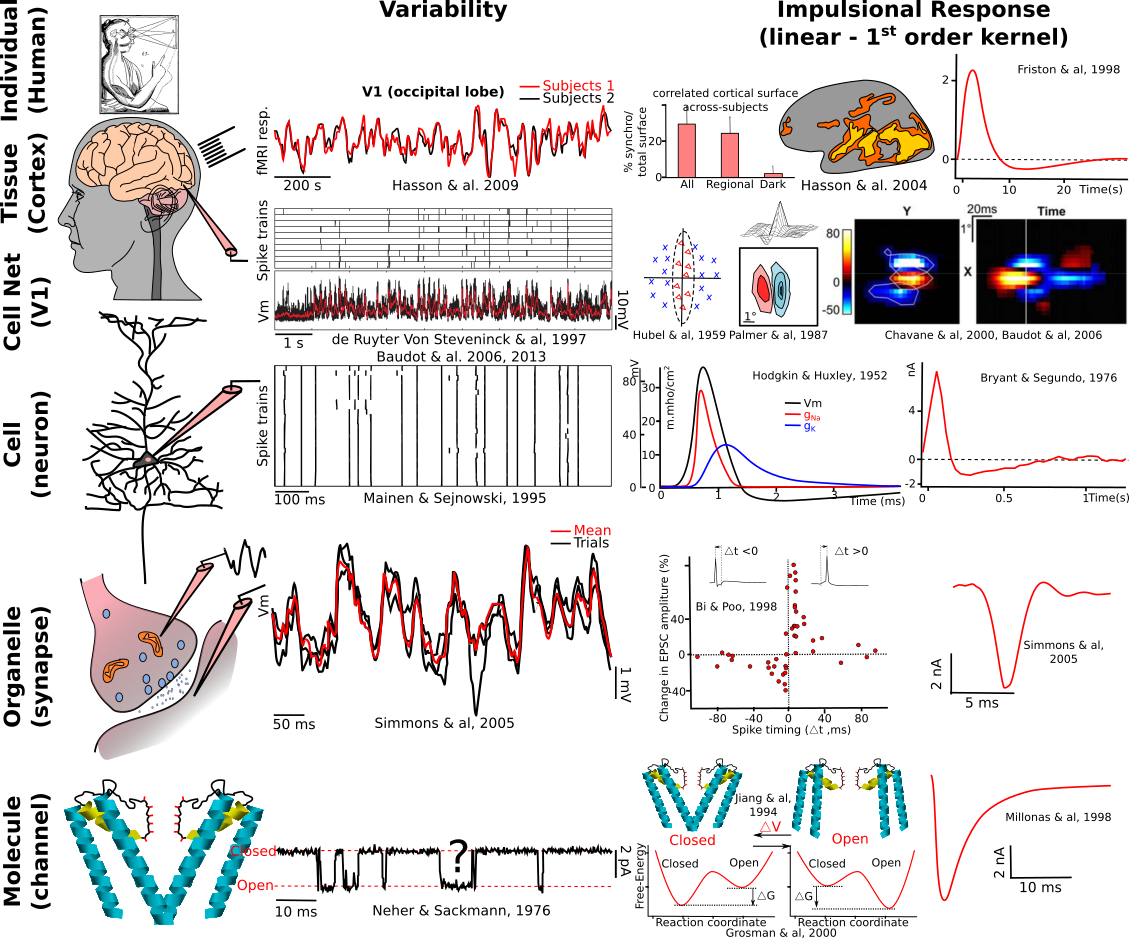}
	\caption{\textbf{Impulsional responses and variability at different organization scales of the nervous system}. See legend \ref{legend scale_function}. }
	\label{scale_function}
\end{figure}

This section asks at what biological organizational scale consciousness arises and the nature of its physical support.
Since the work of Galvani in 1771 on "animal electricity" \cite{Galvani1791}, electric and more generally electromagnetic signals have provided the main sources of observable phenomena for neuroscience and cognitive studies, and yet provide the basis of consciousness theory, at least in this review. It is indeed a posteriori justified from the physical point of view not to take into account other forces such as gravity, except in some particular cases such as the study of the vestibular system, in the dynamic of nervous system activity. However, neglecting gravity is only an occasionally justified and possibly imprecise course of action. Since Galvani, experiments have become more precise and provide empirical measurements of electromagnetic signals at many different scales, that is, with varying space-time resolutions, ranging from single molecule channels to the whole brain, as is the case in fMRI or EEG recordings. An out of date and non-exhaustive comparative figure of the space-time resolutions of some measures of the electromagnetic field given off by activity in the central nervous system is given in \cite{Gazzaniga1998,Sarraf2016}. Figure \ref{scale_function} shows recordings of electromagnetic activity in response to "noisy" or naturally fluctuating stimulation at some of the different organizational scales of the nervous system. Studies into impulsional response and variability is reviewed in the following sections. 

\textbf{Legend of Figure \ref{scale_function} (from bottom to top and from left to right).} \label{legend scale_function} \textbf{Molecule (channel):} A representation of the structure of a potassium ion channel (KcsA, adapted and redrawn from MacKinnon \cite{MacKinnon2003} and \cite{Salari2017}). A single-channel (AcetylCholine, ACh) current recording (redrawn and modified with permission from Neher and Sakmann \cite{Neher1976,Neher1991,Sakmann1991}). To our knowledge, a variability study of a single channel response has never been made. The "gating" conformation change from open to close of a potassium channel (redrawn and modified with permission from Jiang and colleagues \cite{Jiang2002}) and the free-energy landscape transition (redrawn and modified with permission from Grosman and colleagues \cite{Grosman2000} for Ach receptor channel). The linear response of a single Sodium channel (cardiac isoform, hH1a) to colored (100Hz) dichotomous noise (redrawn and modified with permission from Millonas and Hanck \cite{Millonas1998}). \textbf{Organelle (synapse):} a simplified drawing of a synapse. Recordings of several trials of postsynaptic voltage (Vm) in response to presynaptic white noise (black) and the mean response (red) in the graded synapse of the locust (redrawn and modified with permission from Simmons and de Ruyter van Steveninck \cite{Simmons2005}). A Spike Timing Dependent Plasticity profile representing the synaptic potentiation and depression of a synapse in the rat hippocampal neurons as a function of the time interval ($\Delta t$) between the onset of Excitatory Post-Synaptic Potential (EPSP) and the peak of the postsynaptic spike (redrawn and modified with permission from Bi and Poo \cite{Bi1998}). It should be possible and of interest to express such plasticity rule by means of the impulsonal response of a synapse (within the nonlinear higher order kernels). The postsynaptic current evoked by a presynaptic spike (approximated as impulsional) in the study of Simmons and de Ruyter van Steveninck \cite{Simmons2005}. \textbf{Cell (neuron):} 25 trials of spike trains recorded in patch from neocortical slices, responding to the same white noise stimulation (redrawn and adapted from Mainen and Sejnowski \cite{Mainen1995}). The Vm, Sodium and Potassium conductance responses of the Hodgkin-Huxley model of a giant squid axon (redrawn and modified with permission from Hodgkin and Huxley \cite{Hodgkin1952}). The impulsional response of an Aplysia neuron to white noise (redrawn and modified with permission from Bryant and Segundo \cite{Bryant1976}, see also \cite{Mainen1995}). \textbf{Sub-tissular structures - cell networks (V1 area cortical network):}  10 trials of spike trains and Vm responses (mean Vm in red), recorded intracellularly in vivo, to natural image animated by eye movements (redrawn and modified with permission from Baudot and colleagues \cite{Baudot2006,Baudot2013}). A similar study was conducted extracellularly in the H1 neuron of the fly by de Ruyter van Steveninck and colleagues \cite{RuytervanSteveninck1997}. Spatial profile of a spiking receptive field of a simple cell  in V1 (A17), recorded extracellularly; $\times$ and $\bigtriangleup$ denote the visual areas giving excitation and inhibition, respectively, to bright light spot stimulation ("ON response", redrawn and modified with permission from Hubel and Wiesel \cite{Hubel1959}). The more quantitative spatial profile of the linear response of a simple cell spiking receptive field obtained by sparse noise reverse correlation;  blue and red color-scales denote the visual areas giving excitatory response to bright (ON) and dark (OFF) stimulus, respectively response (redrawn and modified with permission from Jones and Palmer \cite{Jones1987}). Above this is presented the Gabor wavelet-like spatial profile of the receptive field. A spatial and space-time profile of the linear response of a simple cell Vm receptive field obtained with dense noise stimuli (redrawn and modified with permission from Baudot and colleagues \cite{Baudot2006,Baudot2013}). \textbf{Tissue (cortical area):} the fMRI responses of the V1 area averaged over two groups of subjects (red and black) while watching a popular movie is illustrated together with a diagram representing the percentage of intersubject correlated cortical areas during viewing and the cortical localization of intersubject synchronized areas (redrawn and modified with permission from Hasson and colleagues \cite{Hasson2004,Hasson2009}). The impulsional linear fMRI response of a voxel in the left superior temporal gyrus to a random sequence of words (redrawn and modified with permission from Friston and colleagues \cite{Friston1998}).\\

The basic proposition of this review from a physical point of view is that the theory of consciousness is the theory of electromagnetic fields (leaving aside the effects of gravity). The electromagnetic theory of consciousness has been developed on the basis of the theory of oscillating neural assemblies (cf. section on neural coding \ref{Neural coding}) most notably by John \cite{John2001}, Pockett  \cite{Pockett2000} and McFadden \cite{McFadden2002}, and basically considers the idea that the spectrum of electrical activity observable in Electroencephalograms (EEGs), typically ranging from 0 to 100 Hz, sustains consciousness. The proposition here is to broaden the spectrum to any frequency and to take into account the temporal and phase dynamics of the activity in question. The beta (12-40Hz) and gamma (40-100Hz) frequencies are simply particular activity ranges evoked by conscious states in humans and in primates more generally, and are mainly generated by primates' peculiar cortical (or cortical-like, e.g. olfactory bulb) excitatory-inhibitory microcircuits. They do not account for the activity related to consciousness observed using other methods at different scales and in other species. This proposition is in fact simply an up-to-date reconsideration of the statement attributed to Pythagoras: \textbf{"All things feel!"} and developed in a poem by de Nerval in his "golden verse reproduced in annex \ref{The objective poetry}. 
By no means should such a proposition be understood as either a simplification or a definitive theory: electromagnetism is neither a simple nor a closed theory (all the more if one considers its relation to gravity). It simply proposes, taking a scientific interpretation of Blake's statement "to see a world in a grain of sand", that there are no more fundamental mysteries in the black box of a human brain, nor any fewer, than in the black box of a particle collider or bubble chamber. 

\begin{figure} [!h]
	\centering
	\includegraphics[height=6cm]{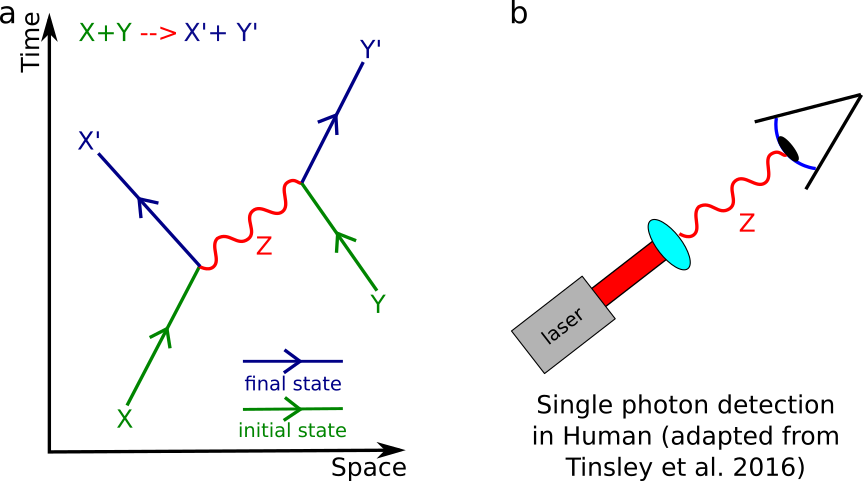}
	\caption{\textbf{a, Feynman diagram of an interaction between an electron $X$ and a proton $Y$ via the photon $Z$, b} A simplified illustration of the experimental set up for single photon detection in a human, constructed by Tinsley and colleagues \cite{Tinsley2016}.}
	\label{feynman_diag}
\end{figure}

Such a proposition includes non-spiking activity, for example graded potential neural activity as reviewed by Juusola \cite{Juusola2007}, and also the activity of non-neural cells such as Glial cells, which display sensory responses although very slowly (due to their large capacitance) and even tuning, as shown by Sur et al \cite{Schummers2008a}.  Such Glial activity can be conceived of as a component of consciousness, albeit a slowly-operating one. 
This proposition of the electromagnetic nature of consciousness does not exclude chemical reactions. Bio-cellular signaling or even metabolic chains are, from the physical point of view, biochemical instances of electromagnetism. For example, Preat and colleagues showed the involvement of intracellular signaling in Drosophila behavior and long-term memory formation \cite{Preat2008}. Genetic expressions and regulations are also electromagnetic processes, and their impact on macroscospic electrical activity is further underlined by the fact that they are involved in the electrical phenotypes, such as phasic or tonic, of neurons, as shown by Soden and colleagues \cite{Soden2013}. 
As cited by Monod, Wyman, Changeux in their work on allostery, "It is certain that all bodies whatsoever, though they have no sense, yet they have perception … and whether a body be alterant or altered, evermore a perception precedeth operation; for else all bodies would be alike to one another" (Francis Bacon, 1967, \cite{Monod1965}). To give an example of an information-theoretic treatment of such a cellular perception, chemotaxis, the chemically guided movement of cells, can be looked at in terms of considering the mutual information between the input gradient and the spatial distribution \cite{Fuller2010}. Such a view includes plants, as action potentials occur in most if not all plants \cite{Davies1987}.
How far in the elementary organizations of matter is it possible to pursue the consideration of some elementary perception, action and consciousness? What is an electrodynamic theory of consciousness at the elementary level? Consider the simple Feynman diagram of elementary particle interaction included in Figure \ref{feynman_diag}, representing the scattering process $X+Y\rightarrow X'+Y'$. As long as one only considers the empirical and observable considerations, that is, if one takes a phenomenological point of view, it is legitimate to consider that the proton $Y$ perceived the electron $X$ via the photon $Z$, with a "reaction" of the proton leading it to $Y'$. Any signal received or propagated by our nervous system is at the elementary level in this way and mediated by boson-like particles.   

Psychophysical experiments can partially illustrate the establishing of such an elementary percept. Holmes showed that humans can sense light flashes containing as few as three photons \cite{Holmes2015}, and Tinsley and colleagues showed that humans can detect a single-photon incident on the cornea with a probability significantly above chance \cite{Tinsley2016}. Elementary auditory perception was also studied by Bialek and Schweitzer, who established that the sensitivity of ears can reach the limit dictated by the quantum uncertainty principle \cite{Bialek1985}. The conclusion of this study is that the measurement apparatus, i.e. the receptor cell, operates in a condition analogous to a 0 Kelvin ground state which maintains quantum coherence. From a more biophysical perspective, single action quanta and quantum formalism have been shown to be relevant to the model of the potassic ion channel selectivity filter that generates important macroscopic patterns of electrophysiological activity in neurons \cite{Roy2009}. From an experimental point of view, it is clear that quantum effects are relevant to nervous system models and that attempts to model with precision should take quantum formalism into account. Bohr originally gave a cognitive and biologic view of quantum physics in his book "Atomic Physics and Human Knowledge" \cite{Bohr1958}, further highlighting that quantum physics is not just a theory of physics, but also a theory of what one can objectively know about physics. Since Bohr's work, many works have proposed to examine consciousness and the nervous system  on the basis of quantum  entanglement and decoherence, or even quantum gravity principles, as in the celebrated works of Hameroff and Penrose \cite{Hameroff1996}, which proposed a specific involvement of cytoskeletal microtubules. Penrose's propositions \cite{Penrose1989} fall within the bounds of the present framework from a physical point of view, while his biological proposition involving microtubules, over-restrictive with respect to the current corpus of knowledge on the dynamics of the nervous system, is extended here to the whole nervous system's dynamic. Recent reviews of some results of the application of quantum formalism to cognition can be found in the book of Busemeyer and Bruza \cite{Busemeyer2014} and in the work of Khrennikov \cite{Khrennikov2015}. \\

With regard to the question, "Where is my mind?", we conclude that biological studies have reported that it can be found at all organizational scales and locations of the nervous system. To lead into the next section on plasticity, computational models such as that of Fusi and Abbott \cite{Fusi2005} have proposed that the nervous system adapts to its environment with a cascading of adaptive processes operating at different time scales, allowing it to fill the gap between traditional short- and long-term memory formation. This multiplicity of scales has an important functional role in controlling the interplay between plasticity and stability-homeostasis (or metaplasticity) as adapting processes operating at different scales.
As a result, the nervous system can be seen as a constantly adapting system with a range of plasticity and homeostatic processes operating at different scales of time and space. Such a view explains why biological studies aiming to localize plasticity and memory in certain biological structures (for example the synapse) or tissues (for example the hippocampus, often called "the site of long-term memory") have found relevantly memory-forming process characterization in virtually all scales and all structures of the nervous system. Open a randomly-chosen journal to a randomly-chosen page in a neuroscience library, and you are likely to come upon a memory-plasticity-learning related paper. By this, we mean that the substrate of memory in the nervous system can be and has been found virtually everywhere, from genetic expression, kinase and/or calcium intracellular signaling cascades, the synaptic NMDA mechanism, to neuronal morphology including synaptic formation, cortical maps of areas remodeling etc. In electrodynamics, the formalism accounting for such multi-scale dynamics is still accepted and is one of its core tenets: the renormalization theory, as reviewed by Shirkov \cite{Shirkov1999} and Huang \cite{Huang2013}. The expression of renormalization in condensed statistical physics based on Ising systems was achieved by Kadanov \cite{Kadanoff1966} and  Wilson \cite{Wilson1974b}, who iteratively constructed Hamiltonians for each scale by aggregating spins within "short" scale distances into blocks. There exist classical versions of the renormalization group, already extensively used in complex system studies, and Dyson developed renormalization in perturbation theory \cite{Dyson1949}.

\subsubsection{Neural coding - neural assemblies - synchrony - noise}\label{Neural coding}

This section investigates consciousness from a coding and engineering point of view and asks the question: what is the code of consciousness? 
A quantitative, formal and typical approach to consciousness relies on investigating how information is processed, stored and retrieved within the nervous system, a field generically known as neural and sensory coding. In such a context, consciousness and information can be considered synonymous (more precisely mutual-information as we will see). 
The word coding comes from the original engineering context of information theory, and may not be appropriate since it suggests that there exists an external structure to decode and gain access to the meaning of the information, which is equivalent to the homonculus problem. Barlow has previously explained how to solve the homonculus problem using biological and probabilistic learning arguments \cite{Barlow1995}. However, Bayesian statistical inference can be interpreted in terms of an ideal homonculus reading the neural code (see Foldiack \cite{Foldiack1993}), and we here consider the phenomenological principle that considers that what reads the neural code are the structures that effectively receive the signal-responses of the system (the "physical observer" rather than an ideal one). In this sense, there is no need to consider such an external 'homonculus' structure, or equivalently, one can consider that there are chains of homonculi. It is sufficient to consider that the structure of the "code" is its meaning and conveys its semantics, and we give an algebraic definitions of structures in the mathematical section of the paper. In terms of information theory, there is no need to consider another coding scheme than the random variables themselves, and we consider here a bijective coding function from the codomain of the random variable to the alphabet. Put simply, the electrical message and its structure are the code itself.   \\

\textbf{Cell assemblies, neural ensembles, synchrony and polyphony, cortical songs and beta-gamma oscillations}. The mainstream historical development of neuroscience has come to consider the nervous system as an \textbf{associative dynamic memory}. This central role of associativity is probably further sustained in information topology by the fact that the algebra of random variables and conditioning is fundamentally associative, and that consciousness is the qualitative byproduct of the mnemonic activation and consolidation process. Hebb proposed the principle of associative plasticity and learning \cite{Hebb1949} generating cell assemblies and providing the physiological support of consciousness and memory. The theory was refined by Von der Malsburg \cite{Malsburg1981} in his "correlation theory of brain function", proposing that the correlate of cognition-consciousness lies in the patterns of neural activity quantified by correlations, and that simultaneously activated nerve cells represent the basic internal objects. This theory was further pursued from a computational perspective by the studies of synfire chains made by Abeles \cite{Abeles1982}, Diesmann, Gewaltig and Aertsen \cite{Diesmann1999}, examined experimentally from the perspective of the theory of synchrony and binding by Singer, Gray and colleagues \cite{Singer1995} and looked into via studies of cortical songs \cite{Ikegaya2004}. The basic hypothesis is that synchronization of neuronal discharges can serve for the integration of distributed neurons into cell assemblies, and that this process may underlie the selection of perceptually and behaviorally relevant information \cite{Engel1999}. The consciousness aspect of the theory was further clarified by the observation that a single external object stimulating the respective receptive fields of two disconnected neurons induced  synchronous oscillations in the 40-100Hz gamma range frequencies, the signature frequencies of attentional and waking states \cite{Gold1999}. The synchrony theory remains one of the simplest and deepest theories of consciousness, since synchrony unavoidably provides a definition of the space-like subspace in space-time structures and also corresponds to the "stable invariant" subspace of coupled dynamical systems as notably emphasized in the work of Stewart and Golubitsky \cite{Golubitsky2006a,Stilwell2011}. As we will see, homology theory provides a natural ground to define and distinguish patterns and assemblies. Homology measures have been applied to characterize neural activity patterns and assemblies in the work of Curto and Itskov \cite{Curto2008} on hippocampal place cells, to examine persistence in visual activity by Singh and colleagues \cite{Singh2008}, and in neural networks by Petri and colleagues \cite{Petri2014}. As outlined in the mathematical and appendix sections of the paper, homology is the standard and appropriated mathematical theory to formalize what patterns may be. The main theory and applied measure to formalize and quantify those assemblies is probability theory, e.g. Bayesian and information theory. The mathematical section provides an introduction to those theories and underlines, following Kolmogorov \cite{Kolmogorov1983} and Jaynes \cite{Jaynes2003}, that they are indeed a single theory. In what follows, we will briefly review their application and biological meaning in neural coding, further justifying their current status as qualitative theories of the brain (see for review Griffiths \cite{Griffiths2008} and Friston \cite{Friston2012} and references therein).   

\textbf{Functional - black box approach} The classical functional characterization of consciousness considers electrical activity as a function of stimulus. The process of consciousness is considered to be a function which consists in the "cross-correlation" or convolution of the stimulus with the neural response. Characterizing consciousness by functions may appear an inappropriate approach that focuses too much on the final result. However, such an interpretation of function is partial, and it is more relevant to consider functions from a mathematical perspective and instead highlight the "dynamic" aspect of consciousness: a kind of relation between a domain and a codomain (that assigns to any element of the domain a single element of the codomain, creating an ordered pair). Function spaces provide very rich and diverse structures in mathematics, including Hilbert and Banach spaces, and are usually classified according to topological criteria. As further discussed in the mathematical section, information topology relies on an arguably general space of functions, the space of measurable functions, and provides a characterization of its structure.
In biology, these input-output functions provide a "representation" or a coding of the (perceived) stimulus on a given functional basis. From the biophysical point of view, this function is usually characterized using the linear response theory, which studies the fluctuation and dissipation (i.e the return to equilibrium) of a system following the external perturbation generated by the stimulus, as formalized by Kubo and applied to neural responses by Stevens \cite{Kubo1966,Stevens1972}. From an engineering perspective, this function is usually characterized using Volterra or Wiener's kernels methods \cite{Wiener1958, Palm1977,Palm1978} using white noise as input. Figure \ref{scale_function} presents the impulsional response (first order linear kernel) obtained at different organizational scales of the nervous system. At each of these scales, the higher order kernels, representing non-linear components or interactions in the system's function, complete these linear functions. For example, at the "network scale" of V1 responses, the linear kernel accounts for about 20\% of the response at the spiking \cite{Carandini2005} and Vm level \cite{Baudot2006,Baudot2013}. A diversity of biological mechanisms sustains the impulsional response at those different scales, which is extremely different from the biological point of view, involving amino-acid steric interactions, synaptic processes, neural passive and active integration processes, excitatory-inhibitory network processes etc.  
These approaches allow the experimental characterization of memory stored and retrieved by nervous systems and linear and nonlinear representations of elements of consciousness, also called receptive fields in the study of sensory coding at the neuron level \cite{Hubel1959,Hubel1962a,DeAngelis1995,Fournier2014}. Memory duration is the time taken by a system after a perturbation to go back to its resting equilibrium state. Figure \ref{scale_function} clearly illustrates that memory duration increases as one goes from fine to coarse scales of nervous system organization. A caveat of such an approach is that characterization with impulsional noise rapidly becomes inefficient when functions become highly non-linear and are represented in very high order kernels, as is the case when one studies sensory neurons with high cognitive functions, far from low-level sensory areas.   \\  

\textbf{Frequency (rate) and temporal code, from spike coding to Vm coding}: The probabilistic functional approach just reviewed can be used to investigate the code's temporal precision. The book by Rieke and colleagues provides an introduction to spike coding \cite{Rieke1999}. The first, simple code to have been proposed was the rate (or frequency) code, which simply observed that the rate of spiking discharge increases with stimulus intensity \cite{Adrian1926}. The rate code postulates that information is transmitted by the rate of spiking. In practice, the variable is the number of spikes within a time window, normalized by the duration of the window: $\text{rate}=n_{\text{spike}}/\Delta t$ (or equivalently, the variable $X_i$ can take $N_i$ values of rate). It is possible to consider variations of the spike rate using several consecutive time windows, each giving a variable $X_i$ and altogether forming a stochastic process. 
Temporal (or time or latency \cite{Thorpe1996,Gawne1996}) coding postulates that information is transmitted by the precise time of a spike. It corresponds to an instantaneous rate code, e.g the limit of the rate code when the duration of the window tends to be small $\lim_{\Delta t \rightarrow 0}\text{rate}$. There have been debates on whether nervous systems use spike time or rate coding, together with studies of information as a function of the duration of the window-bin \cite{Rieke1999,Strong1998}. Results of experiments show that the nervous system uses a temporal or rate code depending on the stimulus or task; simple stimuli with low information content or relevance evoke rate codes while highly informative, complex time-varying stimuli (for instance with high cognitive content), like natural conditions or stimulus the system has learned, tend to evoke a temporally precise spiking code \cite{Bialek1991,Victor1996,Mechler1998,Baudot2006,Baudot2013}. Synchrony and neural assembly theory presuppose a temporally precise code for which precise temporal coincidence or phase relationships are detected. The naturally fluctuating regime eliciting this temporal spiking code is illustrated in Figure \ref{scale_function}. \\
However, consideration of a spiking code is just a rough simplifying approximation and assumption. Historically, notably due to the vast prevalence of extracellular recordings and for simplicity, the coding unit-event of the nervous system has been considered to be the spike - what has been called spike coding, a binary code. It assumes that spike waveform and initiation and synaptic transmission are all-or-nothing processes. Those assumptions are very rough approximations. Information transmission in neurons is not all-or-nothing: spike waveform and threshold vary significantly and further modulate synaptic transmission in an important part, if not all neurons. As reviewed in Juusola \cite{Juusola2007} and Debanne, Zbili and colleagues \cite{Debanne2013,Zbili2016} and  investigated by Simmons, de Ruyter Von Steveninck  \cite{Simmons2005,RuytervanSteveninck1996} and Rama and colleagues \cite{Rama2015}, \textbf{effective information transmission in real nervous systems is not a binary process and the entire membrane voltage codes}. Moreover, spikes differ from cell to cell and the diversity of spikes' shapes and generating mechanisms, notably sustained by a diversity of ionic channels as shown in the celebrated work of Sakmann \cite{Sakmann1991}, are well known to impact neural individual and collective dynamics. 
Such realistic "analog" coding goes hand in hand with an obvious increase in the considered coding capacity of neural processes compared with digital approximation, an increase which is directly imposed by the increase of the size of the coding alphabet. In practice, studies of graded synaptic transmission such as those by de Ruyter Von Steveninck and Laughlin \cite{RuytervanSteveninck1996} report high information transmission rates (see also Borst and Theunissen's review \cite{Borst1999}). \\  

Turning away from the unrealistic assumption that the code is sustained by ideal impulsional spikes (i.e. binary code) leads to the consideration of the more general electromagnetic "Vm code", which includes spiking events.

\begin{figure} [!h]
	\centering
	\includegraphics[height=14cm]{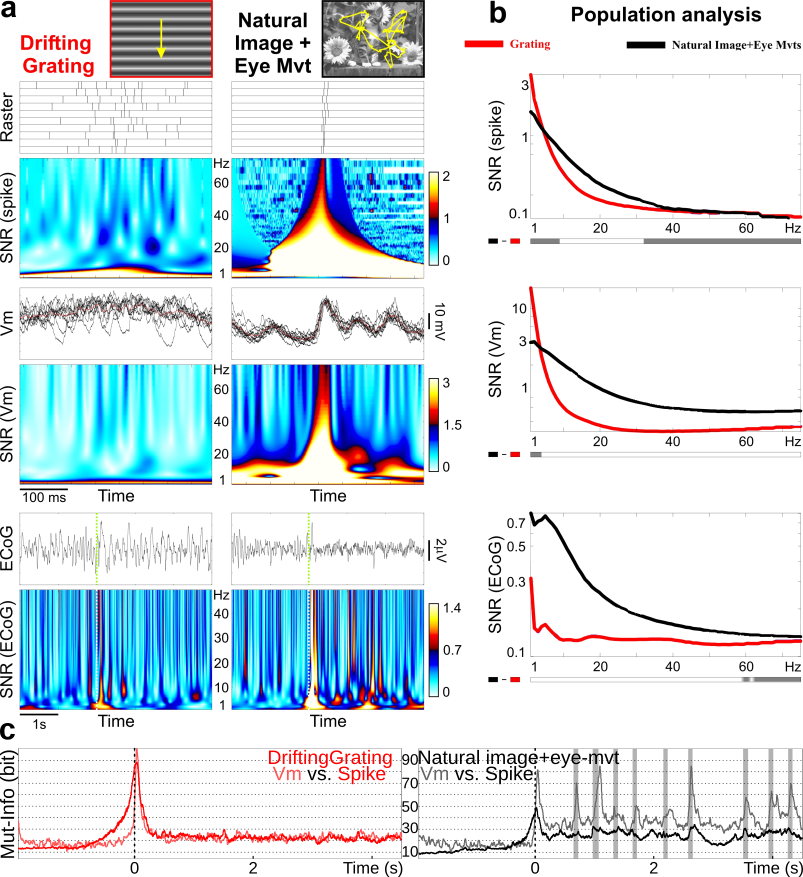}
	\caption{\textbf{Temporal and rate coding with wavelet analysis in the primary visual cortex; SNR and mutual-information rate spectral estimation of spiking, Vm and electrocorticogram (ECoG) responses.} See legend \ref{legend time-frequency}. }
	\label{time-frequency}
\end{figure}

\textbf{Legend of Figure \ref{time-frequency} \label{legend time-frequency}:
Temporal and rate coding with wavelet analysis in primary visual cortex; Signal-to-Noise-Ratio (SNR) and mutual-information rate spectral estimation of spiking, Vm and electro-corticogram (ECoG) responses. a,} comparison of time-expanded epochs of the response of a V1 Simple cell and the simultaneously recorded V1 ECoG (bottom) to an optimal sinusoidal grating drifting at 2 Hz (left) and to natural images animated by eye movements (right). Both epochs illustrate the periods of strongest spike activation for the cell. From top to bottom: i) raster and frequency-time SNR analysis of the spiking response; ii) Vm trials waveforms and SNR analysis. iii) single trial example of ECoG activity and the ECoG time-frequency SNR analysis (2 seconds of spontaneous activity followed by 3 seconds of visual activation).\textbf{ b, population analysis} Comparison of the mean (across cells) average SNR power between various stimulus conditions including grating and natural conditions. From top to bottom: SNR spectra for spiking and subthreshold Vm activity (n=12), and simultaneously recorded ECoG (n=10). Each bar below abscissa expresses the result of a Wilcoxon paired test when comparing two stimuli's conditions for each frequency (color code for “A” minus “B”,  white : “A” significantly higher than “B”; grey : “A” not significantly different from “B”; black : “A” significantly lower than “B”, with p<0.05). \textbf{c,} Temporal modulation of the informational flow of Vm and spiking responses. Comparison of the temporal profile of the estimated mutual-information between Vm and spiking responses averaged across cells for Drifting-grating and Natural Image with eye-movement (saccades are depicted by gray vertical bars). The figure is adapted and modified with permission from \cite{Baudot2006,Baudot2013}.\\

An adequate method for the study of time vs frequency code, avoiding the assumption of a spiking code, is time-frequency wavelet decomposition (or time-energy in physic) \cite{Baudot2006,Baudot2013}, illustrated in Figure \ref{time-frequency} for intracellular recordings and an electrocorticogram of V1 during stimulation with drifting grating (low qualitative content) and natural image animated by eye-movement (high qualitative content). The Signal-to-Noise Ratio (SNR) in time-frequency representation allows one to estimate the mutual-information transmission rate between the stimulus and the recorded response at each time and frequency/rate under Gaussian approximations \cite{Baudot2006,Baudot2013}. Such a method gives a rough estimate that overestimates mutual information. In drifting grating conditions, high SNR values are restricted to a low-frequency band indicating a rate code, and the responses are highly variable from trial to trial. In natural conditions, Vm and spiking responses are highly reproducible (low noise) with high SNR located in high-frequency $\beta - \gamma$ range spectrums and in a time band (see Baudot and colleagues \cite{Baudot2006,Baudot2013}), meaning a temporally localized code and a temporal code. From this electromagnetic point of view, the spike itself is indeed the signature of a temporally localized intermittent code, a fast process which is perhaps consciousness. Note that whereas the spiking code is quite sparse in such conditions, the Vm code is very dense and reveals the unceasing synaptic bombardment expected from a highly recurrent network, and hardly adheres to activity minimization constraints as proposed by Olshausen and Field \cite{Olshausen1996}. Rather, it appears to dissipate free-energy very efficiently in an "efficient dissipation" \cite{Baudot2006}. 
The estimated mutual information, which in this simple two-variable case is equal to the Integrated Information of Tononi and Edelman \cite{Tononi1998} (definitions are given in section \ref{info_functions}) and accordingly quantifies consciousness as regards to the stimulus, is low for the qualitatively low drifting grating stimulus (except at its onset) and evokes EcoG close to resting states. Under natural conditions, the estimated mutual information is higher and Vm conveys higher mutual information rates than the spike train (which can be also interpreted in terms of a neuronal integration process). Such a study confirms the models of integrated information, of temporal coding, and $\beta - \gamma$ frequencies of consciousness. The study of cortical dynamics under natural conditions was pioneered by Vinje and Gallant \cite{Vinje2000,Vinje2002}, and the study on variability and coding in such conditions was reported by Baudot and colleagues \cite{Baudot2006,Baudot2013,Fregnac2005} and then by Butts and colleagues \cite{Butts2007}, Haider and colleagues \cite{Haider2010} and Herikstad and colleagues \cite{Herikstad2011}. \\

\textbf{Population code} Bayesian theory and information theory provide the basic quantification of populational code. It consists in considering the multivariate case where each neuron corresponds to a variable or considering more complex time-dependent generalizations (as in the work of Martignon \cite{Martignon2000}), hierarchical families of probability distributions as in the work of Amari \cite{Amari2001a}, which considers higher order statistics of cell assemblies. For example, Ma and colleagues developed probabilistic population codes \cite{Ma2006} to infer the stimulus from a population activity. The general case of population coding is barely distinguishable from some of the current neural network or machine learning approaches and are reviewed in the next section. The information topology sketched in section \ref{infotopo_synthesis} aims to characterize the structure according to the topological and informational criteria of such multivariate cases.\\

\textbf{Noise, spontaneous-ongoing activity, self and free-will :} it is a leitmotif in biological and neuronal studies to investigate the role of noise, whether it be an un-mastered or "spontaneous" source of variability, and to propose that such a non-deterministic source is responsible for phenomena like consciousness \cite{Dehaene2005,Wyart2009}, or living principle, as in the work of Brown which looks for the "vital forces" in living material \cite{Brown1828}. Many studies have been dedicated to the functional role of noise and have pointed out that noise is "far from being a nuisance" \cite{Vilardi2009,Eldar2014}. Some have formalized noise, for example using stochastic resonance or self-organized criticality formalisms \cite{Bak1987}. Control theory and the conceptualization of a channel of communication in information theory has also made use of such an ad-hoc noise source \cite{Shannon1948}, using the deterministic "0 noise" formalism as a reference. Intrinsic variability has a very important role in human cognition and consciousness, as it allows free-will to be conceivable. As Ruelle remarked in his course on deterministic dynamical systems, critical-bifurcation points, hyperbolic saddle points, are the only places where the choice is given and left possible (see \cite{Baji2005} for a review on this topic; see also the section  of this paper on dynamical systems \ref{dynamical_system}). Recurrent network formalization using statistical physics, pioneered notably by Hopfield networks, introduced a new view on ongoing activity and thermal noise, proposing that it corresponds to the autonomous generative capacity of consciousness, illustrated in the context of the Helmholtz machine as accounting for the wake-sleep sequence and dreams \cite{Dayan1995}, which further gave a conceptual framework for the studies on "cortical or hippocampal replay". The probabilistic approach, as notably proposed by Ma and colleagues \cite{Ma2006} and also in information topology (see the mathematical section of this paper \ref{infotopo_synthesis}), generalizes the noise approach by considering biological and neural computation to be intrinsically probabilistic: the deterministic case is no longer the reference but a peculiar limit subcase. In such a view, any component of a (possibly nervous) system corresponds to a random variable, can be considered as a signal and a noise source and can be both emitter and receiver. A variable or source without noise is deterministic and is the trivial constant "0" information of an informational structure (cf. section \ref{infotopo_synthesis}): information only sees what varies, so to speak. In a sense, such probabilistic studies describe the heterogeneous structures of constitutive ongoing activity and the relative variations of ongoing activity. Indeed, the information topology framework introduces a variational view of probability, which was also proposed by Friston \cite{Friston2006}. Hence, the probabilistic and informational view attributes not only consciousness but also free will to varying observables. We believe it is the fundamental theoretical contribution of probability theory to cognition to allow free will to exist and be an important constitutive part of the system's dynamic. \\       
As a quite consensual conclusion of this section on coding, in agreement with the current neuroscience theories, the code implemented by the nervous system and which sustains consciousness is proposed to be an electromagnetic and probabilistic code; both are consistent with regard to physics. However, it has not yet been established that the expression of statistical physics implemented by information topology can also account for the standard electrodynamic theory. This is a hope for the future which is left here as conjecture.     \\

\subsubsection{Psychophysics - Fechner and Gestalt's heritage}\label{Psychophysic}

In this section, we ask if qualia can be quantified and if it is possible to define the structure of qualia's interactions in order to provide a quantified Gestalt theory.\\  
\textbf{Quantum of qualia - adaptation:} Measures of the nervous system's electromagnetic activity are not sufficient for a complete study of consciousness, as they have to study the "correlation" of those measures with a subjective state-dynamic. Hence, a quantification of subjective states is required, meaning a definition of subjective observable phenomena and an appropriate mathematical definition of "correlation" need to be given. Such an approach defines the entire domain of what is called psychophysics, or, more generally, experimental psychology, wherein the definition of "correlation" is usually the functional black-box approach just reviewed. 
The principles of this domain were laid down by Fechner in 1860 \cite{Fechner1860}. Fechner's main contribution has been what is known as the Weber-Fechner law of adaptation, according to which sensation is proportional to the logarithm of the stimulus (but see Mengoli's 1670 work on the "logarithmic ear" \cite{Mengoli1670}):
\begin{equation}
S(x)=k\ln \frac{x}{x_0}
\end{equation}
where $S(x)$ is the subjective magnitude of the sensation, $x$ is the stimulus intensity and  $x_0$ the “absolute threshold” (the intensity at which the stimuli is no longer perceived). To derive this law, he introduced the concept and measure of "just-noticeable difference", a quantization of the subject 40 years prior to the establishment of quantum physics by Planck \cite{Planck1901}. We now know that the "just-noticeable difference" is the quantum of action (cf. Figure \ref{feynman_diag}).
This law holds in all sensory modalities: vision (light intensity), hearing, taste, touch (weight intensity), smell, but also time estimation \cite{Takahashi2006}. It notably established the use of decibel units for auditory signals. In crossmodal matching, it is replaced by Stevens' power law \cite{Stevens1965} \cite{Krueger1989}. Poincar\'{e} gave a simple, pure, and fundamental topological interpretation of Fechner's law of perception that has been reproduced in the annex of this paper  \ref{topology of psychophysic} \cite{Poincare1905,Poincare1902}.
Laughlin proposed a seminal mutual information maximization framework (infomax) that accounts for Fechner's law \cite{Laughlin1989} in a biophysical adaptation context (cf. Figure \ref{networklearning}a). Among other reports, Kostala and Lansky went further and gave an information theoretic explanation of this law \cite{Kostala2016}, and a fine mathematical introduction with a topological perspective was proposed by Dzhafarov \cite{Dzhafarov2012}.\\

\begin{figure} [!h]
	\centering
	\includegraphics[height=5cm]{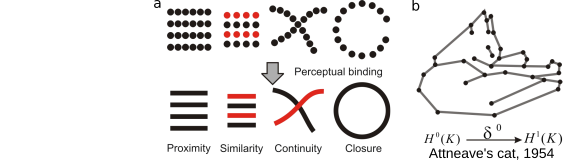}
	\caption{\textbf{Gestalt laws of elementary perception and Attneave's cat. a} Gestalt's four elementary laws of binding \cite{Kohler1947,Wertheimer1924,Wertheimer1923}. \textbf{b,} Attneave's cat, redrawn from Atteneave's 1954 paper \cite{Attneave1954}. The points of a graphical drawing of a sleeping cat were chosen according to a maximal curvature criterion. To underline the topological intuition behind this, we illustrated the binding of incoherent points into a coherent form by a simplicial coboundary (differential) operator from a 0-complex of points to a 1-complex (i.e. the graph of the cat. See the section \ref{What is topology} for a presentation of those topological objects).}
	\label{figure_Attneave-Gestalt}
\end{figure}

\textbf{Gestalt - efficient coding theory: } The Form and geometrical theory of consciousness go back at least to antiquity, notably with Plato and Aristotle's theory of Forms and their debates on its metaphysical or physical nature \cite{Aristotle350B.C.E,Ross1951}. The idea, basic yet at the cutting edge of mathematical neuroscience, was captured by Riemann in the citation given in this paper's introduction \ref{riemanncitation}.
The theory of forms was developed was developed by the school of Gestalt psychophysics, notably K\"{o}hler \cite{Kohler1947} and Wertheimer \cite{Wertheimer1924,Wertheimer1923}, who provide elementary associative laws of binding that govern the construction of complex perceptual forms from basic elementary shapes (cf. Figure \ref{figure_Attneave-Gestalt}). This school was interested primarily in the relationships between objects and even considered that relations form the object, which they summarized by their famous adage “the whole is greater than the sum of its parts”. Wertheimer used the following example: we can perceive trees and not multiple different elements of sensation. And most importantly, we see the same tree even when light conditions and perspective imply a complete change of the elements in question. In other words, it is not the elements but the perceptual structures that are fundamental; these so-called Gestalt qualities are collective mental phenomena. They proposed “the hypothesis of psychophysical isomorphism” between environmental and brain processes and  suggested that brain processing consists in assembling parts into a coherent whole (for a review of this topic see \cite{Rosenthal1999}). Among the perceptual laws they propounded, there are 3 or 4 elementary laws of binding and segmentation: proximity, similarity, continuity, and sometimes closure (cf. Figure \ref{figure_Attneave-Gestalt}). 

\begin{figure} [!h]
	\centering
	\includegraphics[height=11cm]{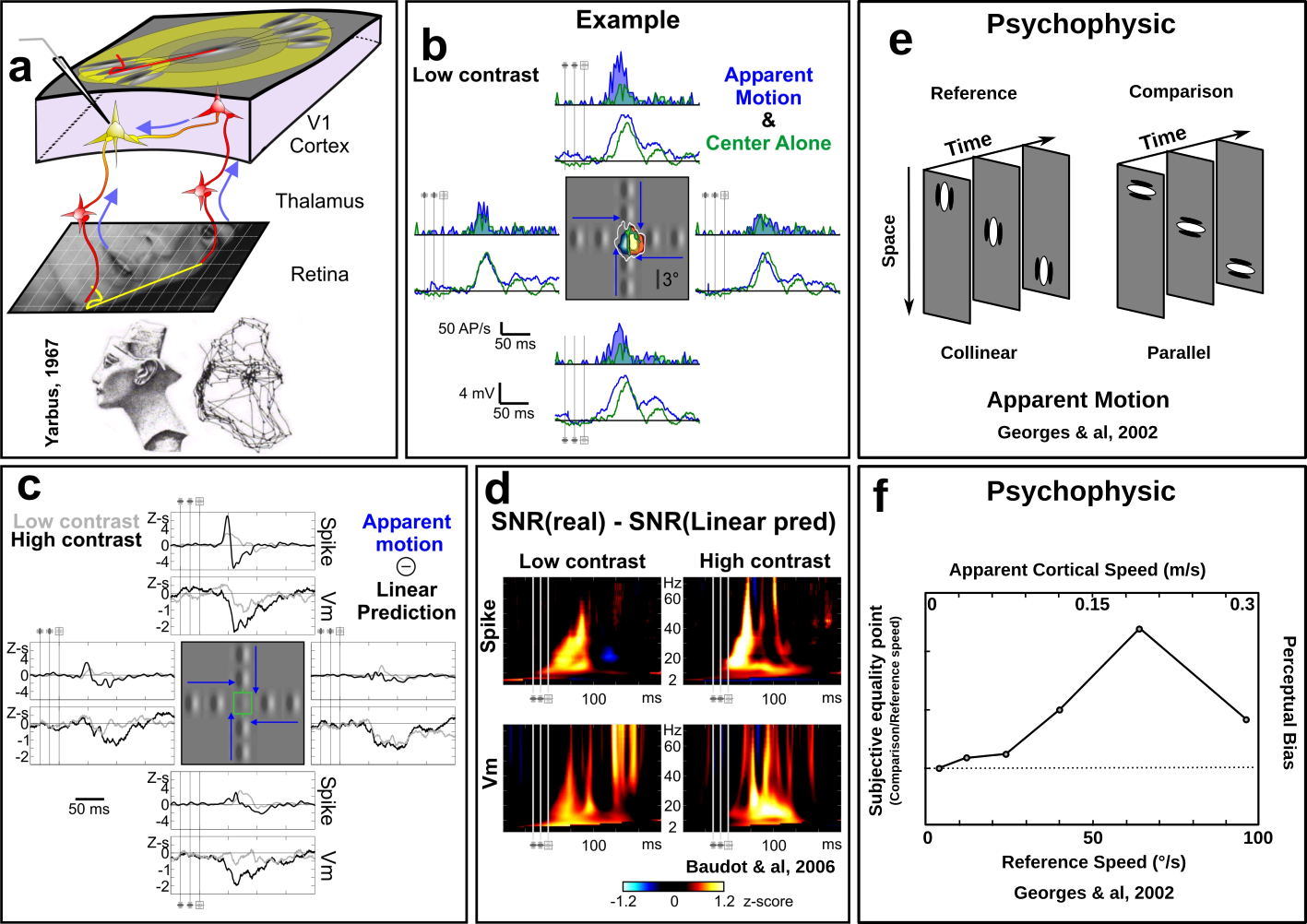}
	\caption{\textbf{Subjective speed of apparent motion and its neuronal substrate}. See Legend \ref{legend apparent_motion}.}
	\label{apparent_motion}
\end{figure}

These laws of binding operate both in space and time; for example, in the temporal domain, the law of continuity was called “common fate”. Of course, their definition lacked mathematical precision, but they have a remarkable geometric and topological flavor. Moreover, the binding problem (generalization) is today considered together with its dual segregation (discrimination - segmentation) task both in psychophysics and neuroscience. The results of modern psychophysics are much more precise and complex. For example, Field et al \cite{Field1993} and Polat and Sagi \cite{Polat1993} gave a description of the spatial, orientation and contrast dependency of visual contour integration, the so-called “association field”. Such results were investigated by Georges and colleagues in the spatiotemporal case, using tasks of speed estimation of various apparent motion configurations \cite{Georges2002}. The principle of dynamical space-time “association field” is depicted in Figure \ref{apparent_motion} a and e. The perceptive bias in speed estimation for coherent Gestaltic configurations is represented in Figure \ref{apparent_motion} f.  \\

The correspondence of such psychophysical experiments with electro-physiological activity has been reported by several studies at all scales of the nervous system and with many kinds of recording methods. The converse has also repeatedly been shown, starting with the experiments of Penfield \cite{Penfield1958}; electrical stimulation of single neurons, as in the study of Salzman and colleagues \cite{Salzman1990}, or of a neuronal population, can bias and modify perceptions, as reviewed by Parker and Newsome \cite{Parker1998} and Cicmil and Krug \cite{Cicmil2015}. Here, we simply present particular examples of single neuron recording results that provide the neural basis for the visual association field. The work of Gilbert and Wiesel \cite{Gilbert1989} and Schmidt and colleagues \cite{Schmidt1997} provide the neural correlate and mechanisms for the static psychophysical association field and support the view that the binding of visual contours onto perceptually coherent objects involves long-range horizontal connections between V1 cortical neurons as represented in Figure \ref{apparent_motion}a. 
 
\textbf{Legend of Figure \ref{apparent_motion}\label{legend apparent_motion}: qualitative perception of the speed of apparent motion and its neuronal substrate.} Center-surround directional selectivity for saccadic-like apparent motion and the temporal and SNR modulation of the response of primary visual cortex neurons. \textbf{a,} Schematic representation of a visuo-oculomotor model of V1 processing, sequentially integrating visual information along spatial long range (saccade) and short range (fixational movement) eye-movements of coherent shapes (gestaltic). During saccadic high-speed long-range movement, V1 neurons, by means of their spatiotemporal association field, selectively integrate the visual information iso-oriented to saccadic motion along their collinear axis (co-aligned with the motion path), whereas during fixation they integrate the visual information on low spatial scales and at low speeds corresponding to their classical direction selectivity (classical direction preference axis across the discharge field width). Furthermore, the eye-movement scan-path is correlated to image features, notably the contours for saccades' path in this image exploration. The bottom cartoon is Yarbus's original illustration \cite{Yarbus1967} (1967) and illustrates the eye-movement pattern of a human observer (right panel) along the corresponding photograph (left panel).\textbf{ b,} an example of a simple cell response to apparent motion stimuli (blue color) and center only control (green color), for low contrast center conditions,
exemplifying a collinear surround facilitation. Picture in the middle represents the four tested axis of apparent motion superimposed with the RF map obtained with sparse noise (ON responses in red scale color, OFF responses in blue scale, depolarising field extent white line). Gabor patches were sequentially flashed from the surroundings to the center. \textbf{c,} The biphasic temporal profile of center-surround apparent motion nonlinearity, and its directional collinear selectivity and modulation by contrast (population analysis $n = 23$). The temporal waveforms of nonlinearity are calculated for each cell by subtracting the linear predictor (Center alone + surround alone responses) from the real response observed to the full apparent motion sequence, both at the spiking levels (top panels) and at the Vm level (bottom panels). Here, we present the average cross-cell temporal waveforms of nonlinearity expressed as a z-score of the spontaneous activity. The temporal profile of the nonlinearity is given for the low contrast center (grey color) and the high contrast center (black color). \textbf{d,} apparent motion nonlinear modulation of the SNR of the responses. To measure the center-surround SNR modulation gain, each trial of the center alone condition are summed with those of the surround alone condition to obtain a pool of linear predictor trials, on which we could apply the SNR time-frequency analysis. The time-frequency apparent motion nonlinear SNR gain is then obtained by subtracting the apparent motion SNR from the linear predictor SNR, expressed as a z-score of spontaneous activity (significant threshold calculated independently for each frequency z-score p>0.001), and averaged across cells (adapted and modified with permission from Baudot \cite{Baudot2006}). \textbf{e,} the psychophysics experiment to quantify the bias in the perceived speed of apparent  motion relies notably on two stimuli : i) a reference spatio-temporal sequence with collinear Gabor patches (dynamic Gestaltic association field configuration) and ii) a control sequence with parallel Gabor patches (non Gestaltic configuration). \textbf{f,} The result of the perceived speed bias quantified by the subjective equality point (ratio of comparison/reference speed) as a function of the speed of the stimuli or of the corresponding cortical speed. The maximal bias, consisting in an overestimation of the speed for Gestaltic configurations, is found for speeds in the range of saccades' speeds and of horizontal propagation speed in the cortex (adapted and modified with permission from Georges, Series, Fregnac and Lorenceau \cite{Georges2002}).  \\

The neural correlate of the dynamic association field sustaining the apparent motion percept has been investigated in the work of Chavane and colleagues,\cite{Chavane2000} Baudot and colleagues \cite{Baudot2006} and Gerard-Mercier and colleagues \cite{Gerard-Mercier2016}, and is illustrated in Figure \ref{apparent_motion} which summarises most of the electrophysiological paradigms of consciousness: shapes and binding, temporal coding, $\beta−\gamma$ oscillations, "feedback" and non-linearities (here we consider feed-forward inhibition and lateral intracortical excitation as formal feedback from the cybernetic point of view), active sensing (see next chapter) and reliability or mutual information (transient) increase. The space-time sequence of surrounding stimulation in a coherent gestaltic configuration, optimally fitted to recruit horizontal cortical connectivity, improved the temporal precision of spiking responses, the SNR of Vm and spike responses, $\beta−\gamma$ band activity, and decreased the latencies of the responses. As shown in Figure \ref{apparent_motion}c, the non-linearities induced by surround stimulation present a space-time profile in agreement with saccadic eye-movements' kinematic characteristics and sustains the impulsional and reliable code. They exhibited a biphasic temporal profile resulting from an excitatory-inhibitory temporal phase shift analog to what was observed in the auditory cortex and in the olfactory bulb by Wehr and Zador \cite{Wehr2003} and Laurent’s team \cite{Collins2007}. It hence underlines a generic cortical (or cortical-like) mechanism for spike sparsening, temporal precision refining, and for the amplification of $\beta−\gamma$ oscillations. Such studies also lay a path towards an answer to Olshausen and Field's question, "What are the other 85\% of V1 doing?", \cite{Olshausen2004} referring to the low explanatory power of the linear simple RF; the dynamic of eye movement which is the major component in natural condition statistics and their adapted non-linear interactions have been missing in classical studies. V1 neurons code for much more sophisticated space-time shapes than the linear component displays.

\textbf{Gestalt and efficient coding:}\label{barlow} The principle of efficient coding, that the goal of sensory perception is to extract the redundancies and to find the most compressed representation of the data, was first stated by Attneave in 1954 \cite{Attneave1954} followed by Barlow \cite{Barlow1961}. Attneave notably claimed that any kind of symmetry and invariance are information redundancies and that Gestalt principles of perception can be defined in terms of information. Attneave's illustration of this principle is reproduced in Figure \ref{figure_Attneave-Gestalt}, in which his cat, drawn out of only 38 "informative" points, has an intuitive cohomological interpretation as a cochain complex (this illustration is only included to facilitate intuitive understanding a la Attneave). We propose in the section mathematical section of the paper \ref{Math} that homology is an adequate formalism both to achieve the assimilation of symmetries and invariance as information quantities and to provide mathematical and even logical Gestalt laws of perception. Such an idea is already present in essence in the work of Thom and Petitot on semiophysics, structural stability and morphogenesis \cite{Thom1977,Petitot1983,Thom1990}, although their presentation of the idea is rooted in probability and information theory rather than catastrophes. Following the seminal work of Kendall \cite{Kendall1984} defining shape spaces, statistical shape space analysis undertaken by Dryden and Mardia \cite{Dryden1998} and a series of works by Mumford and Michor \cite{Michor2006}, a whole domain of statistical and informational investigation of shapes appeared, for example the pattern theory of Mumford and Desolneux \cite{Mumford2010} among many other works in this active field. The relation or expression of shape space analysis to information topology is not yet known and would require its generalization to continuous symmetries.

\subsection{Active consciousness - plasticity, adaptation homeostasis - Dynamic and persistence of qualia}

\subsubsection{Action-perception inseparability}

In this section, we investigate how consciousness dynamics are related to action. 
Following James \cite{James1890}, Tononi and Edelman proposed that  consciousness is not a thing or a state, but a process or a flow fluctuating in time, taking a quite continuous view of the subject. The model of Changeux-Dehaenne may seem opposed to such view in the sense that it is "discontinuous" and involves a transition from unconscious to conscious, two separate phases. However, as claimed previously, such discontinuity should simply be considered a critical point between two different "conscious phases", one being unaware of the other, a perspective which highlights the fact that taking the diversity of consciousness into consideration renders these two theories quite indistinguishable. As illustrated by the involvement of eye-movement in visual responses and coding in the previous section, consciousness is a dynamic and active process. This central statement arises from the general action-perception conceptualization of cognition in neuroscience (see Llinas \cite{Llinas2002}, Berthoz \cite{Berthoz2000,Berthoz2003} and Jeannerod \cite{Jeannerod2006} for general introductions to this subject), from the phenomenology of Merleau-Ponty \cite{Merleau-Ponty1945} and active sensing in electrophysiological studies \cite{Saraf-Sinik}. The formalization of the action-perception paradigm was pioneered by the tensor networks model of Llinas and Pellionisz \cite{Pellionisz1979}. A remarkably clear and relevant presentation of "active consciousness" was given by O'Reagan and Noe \cite{ORegan2001}. Borrowing their metaphor, conscious perception is like a blind man who scans his environment with his cane to perceive its structure and extract its information. Visual qualitative perception exemplifies this statement; our eyes continuously scan the environment by saccadic and fixational eye-movements (drifts and tremors), as represented in Figure \ref{apparent_motion}a. Whenever the eyes or the retinal image are artificially kept immobile, the visual qualitative perception fades within hundreds of milliseconds (and the cortex goes back to its spontaneously fluctuating ongoing states). It is misleading to consider such movement as unconscious (as we do not see them or see our visual world moving with them) or even to think that a stabilizing process is required to compensate them. As underlined by O'Reagan and Noe \cite{ORegan2001}, they indeed construct our perception and are an intrinsic, constitutive, necessary component of our conscious flow that constructs our subjective space-time, further probing the integrative action-perception process. This is the striking conclusion, and also Piaget's opus \cite{Piaget1964,Phillips1981}: what we consider as external fixed reality, our perception of space and time, is a sophisticated construction of our nervous system which learns spatial and temporal relations. Moreover, the inseparability of action and perception has wide-ranging ramifications in physics, as can be seen from the fact that the duality of Hamiltonian (perception-like) and Lagrangian (action-like) approaches are encompassed by the consideration of the Legendre transformation.       
This inseparability of action and perception is a hallmark of the adaptive and dynamic nature of consciousness, which is a perpetually adaptive process; we can go a step beyond the dynamical nature of consciousness. Consciousness is variational by essence - what is constant, we are not conscious of; we all have a blind spot corresponding to  missing photoreceptor on the retina occupying an important part of our visual field, but none of us have ever seen it. This brings to light the importance of the adaptive or learning mechanisms in the nervous system that are the main topic of neuroscience and which we review in the next section, in which we also introduce the thermodynamic and informational formalization of the process of learning in consciousness. 

\subsubsection{Plasticity - (machine) learning - Informational and Statistical physic of consciousness dynamic}
\begin{figure} [!h]
	\centering
	\includegraphics[height=9cm]{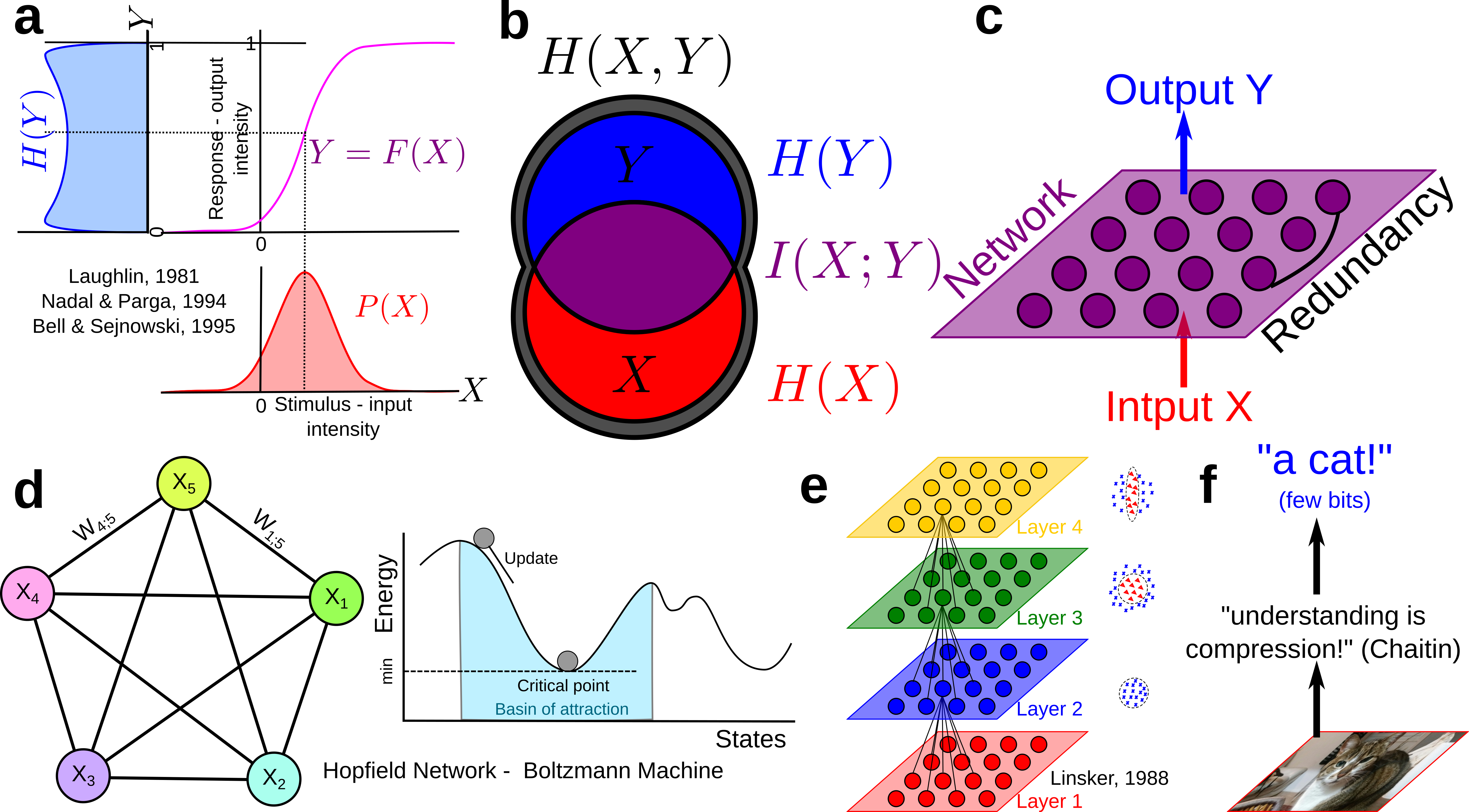}
	\caption{\textbf{Statistical and informational models of learning-adaptation. a,} the information maximization principle (adapted and modified with permission from Laughlin \cite{Laughlin1981}, Nadal and Parga \cite{Nadal1999,Nadal1994} and Bell and Sejnowski \cite{Bell1995}): the stimulus or input $X$ has a probability law $P(X)$ (red). The system is modeled as a black box with function $Y=F(X)$ (invertible continuous deterministic \cite{Bell1995}, the cumulative of $P(X)$ in \cite{Laughlin1981}, in purple) and implements a gain control. The response or output of system $Y$, and its entropy $H(Y)$. The maximization of mutual information  $I(X;Y)=H(Y)-H(Y/X)$ (\textbf{b}) between the input and the output comes to maximize $H(Y)$ since the system is deterministic ($H(Y/X)=0$), and hence removes the redundancy in $Y$ or produces "factorial" code that can be used for Independent Component Analysis (ICA), as depicted in \textbf{c}. \textbf{d,} an illustration of a Hopfield network of 5 McCulloch and Pitt binary neurons or a  Boltzmann Machine with 5 binary variables (vertex of the graph); the edges of the graph represent the synaptic weights $ w_{i,j}$ of the network's connectivity matrix. On the right is illustrated a 1-dimensional free-energy landscape with several minima learned by a network; see Hopfield and Ackley et al for details \cite{Ackley1985}\cite{Hopfield1982}. \textbf{e,} the infomax self-organizing mutilayer network of Linsker \cite{Linsker1988} that learns static, approximately realistic neuronal receptive fields. This seminal model, together with multilayer perceptrons, can be understood as an effective informational version of Marr's primal sketch \cite{Marr1982} and the precursor of current deep learning machines. \textbf{f,} The cognitive, very basic principle of redundancy removal, expressed by Chaitin as "understanding is compression!" . It is  illustrated by the fact that humans can extract a complex image from a very short linguistic representation ("a cat!") in about 150ms \cite{Thorpe1996}. }
	\label{networklearning}
\end{figure}

In this section we investigate the relation of the dynamics of our qualitative experience with the plasticity of the nervous system and the way this is usually formalized.  
The formalization of plasticity, the dynamic of consciousness, has formed the crossroads between information theory, statistical physics and data analysis (machine learning). A basic principle ruling nervous system dynamics and learning was inferred by Hebb \cite{Hebb1949} in what is now called the Hebbian rule of plasticity, stating that if a presynaptic cell $X$  tends to repeatedly excite and take part in the firing of a postsynaptic cell $Y$, then the efficiency of $X$ on $Y$ will be reinforced. It notably found a biological verification in the study of L\o{}mo and Bliss, who demonstrated Long Term Potentiation (LTP) at the hippocampal synapses \cite{Lomo1966,Bliss1973}. The principle of efficient coding proposed by Attneave \cite{Attneave1954} and Barlow \cite{Barlow1961} and restated in section \ref{barlow} can be reformulated as an optimization problem, aiming to maximize mutual information between the input and the output of a system that provides a decorrelated or factorial, informationally efficient representation of the input, as illustrated in Figure \ref{networklearning} a,b,c. From the cognitive point of view, the idea was resumed by Chaitin \cite{Chaitin2006} and Grassberger (private communication, cf. Figure \ref{networklearning}f): "understanding is compression!". This could also be stated as finding all the redundancies that characterize the structure of the environment. Maguire and colleagues have constructed a whole theory of consciousness based on such compression principles \cite{Maguire2016}. Linsker's seminal work showed that the "infomax principle" applied in feed-forward linear networks is equivalent to considering that synaptic weights follow a Hebbian covariant rule and achieve a certain kind of Principal Component Analysis (PCA), with neurons developing  static oriented simple-like receptive fields  \cite{Linsker1986,Linsker1988} (cf. Figure \ref{networklearning}e). On a biological side in 1981, Laughlin and Srinivasan formulated the information maximization principle, showing that it implements predictive coding in vision and gaining experimental control of the interneurons of the eyes of flies. \cite{Srinivasan1981,Laughlin1989} (cf. Figure \ref{networklearning}a). Nadal and Parga \cite{Nadal1999,Nadal1994} and Bell and Sejnowski \cite{Bell1995} further generalized the approach, showing that  maximizing the mutual information $I(X;Y)=H(Y)-H(Y/X)$ between the input $X$ and output $Y$ of a network (see figure \ref{networklearning} and legend), imposing low noise or deterministic system conditions ($H(Y/X)=0$), leads to redundancy reduction, a factorial code that can be used to achieve Independent Component Analysis (ICA). In real neuroscientific experiments, as shown in the column on variability in Figure \ref{scale_function}, the effect of maximizing mutual information, at all scales of nervous system organization (and reference therein), is to reduce the noise-variability of the responses and for the system to become close to deterministic given the stimulus - what is usually called 'reliable'. This ensures a consistency of consciousness, ensuring that two individuals experiencing the same environment will develop and share the same conscious experience (and hence can communicate about it consistently). This fact is all the more clear in the fMRI experiments of Hasson and colleagues, where the readings are taken from humans \cite{Hasson2009,Hasson2004}. In simple terms, infomax accounts for the fact that our experiences can be shared, most notably human or animal communication. 

A historic breakthrough in learning theory, arguably the first in Artificial Intelligence (AI) after Turing, was achieved by Hopfield in 1982 \cite{Hopfield1982}, who showed that an Ising system could be used to formalize the associative memory learning of a fully recurrent network of binary neurons with the Hebbian plasticity principle (cf. Figure \ref{networklearning}d). Ackley, Hinton and Sejnowski \cite{Ackley1985} generalized recurrent networks by considering neurons as random variables and imposing the Markov Field condition, allowing the introduction of conditional independence and the further construction of "deep" network structures with hidden layers \cite{Hinton2012}. The result, the Boltzmann or Helmholtz machine \cite{Dayan1995}, relies on the maximum entropy or free energy minimization principle, and originally relied on minimizing the relative entropy between the network and environmental states \cite{Ackley1985}. These studies presented the first formalization of learning within statistical physics, explicitly in terms of entropy and free energy functions, but also in terms of information functions. Friston et al introduced a variational Bayesian formalization of the minimum free energy principle and proposed  a theory of embodied perception based on it \cite{Friston2006}. Recently, the restricted Boltzmann machine, originally introduced by Smolensky in his Harmony theory \cite{Smolensky1986}, found an enlightening reinterpretation through the variational renormalization group method, thanks to the work of Mehta and Schwab \cite{Mehta2014}. Their  results clarify the principles involved in deep networks, part of the developments of which can appear alchemical \cite{Rahimi2017}, and further reinforce the electrodynamical nature of consciousness.  \\
\textbf{The biological relevance of recurrent models and epigenetic generalization:} Following the model of convergence of thalamo-cortical connections on simple and complex cells proposed by Hubel and Wiesel \cite{Hubel1962a}, recurrent models have been given the status of abstract models which are mostly relevant for artificial intelligence and machine learning. However, following the work of Fregnac, intracellular studies, revealing synaptic excitatory, inhibitory and neural integration, that is to say, electrical code, have revealed that this sparse convergence model is highly misleading (see \cite{Baudot2006a} and \cite{Fregnac2015} for a review of this topic). Using Hebbian conditioning protocols, Debanne and colleagues were able to turn simple cell responses into complex cell responses \cite{Debanne1998}. Orientation selectivity was shown to arise from a diverse combination of excitatory-inhibitory cortical recurrence balances by Monier and colleagues \cite{Monier2003}, and recordings in natural visual conditions of eye movements revealed very dense Vm activity \cite{Baudot2006a,Baudot2013}, as shown in Figure \ref{time-frequency} and \ref{scale_function} and quantified in \cite{Baudot2006a}. Markov and colleagues were able to quantify that the proportion of feed-forward connections in the visual cortex only represent few percent of the full number \cite{Markov2013}. Hence, what has been considered as artificial intelligence may be much closer to biologically realistic computation, at least in principle: high cortical recurrence allows statistical computation and can provide a robustness of cortical functions to single connection variability. 
  
However, from a biological point of view, the picture painted by recurrent networks, with Hebbian or anti-Hebbian, LTP and LTD and instantaneous transmission is far too simplistic. As we saw in Figure \ref{apparent_motion}, propagation delays are part of computation, codes and percepts. Roxin and colleagues were able to show that reintroducing conduction delays (space-time) in recurrent network model increases the multistability and richness of the phase diagram, with oscillatory bumps, traveling waves, lurching waves, standing waves (...) \cite{Roxin2005}.Hebbian plasticity in excitatory cortical and hippocampal neurons critically depends on the relative timing of pre- and post-synaptic spikes as shown by Debanne and colleagues \cite{Debanne1994} and Bi and Poo \cite{Bi1998}. This "spike-timing dependent plasticity" (STDP) with a Depression-Potentiation waveform is  illustrated in Figure \ref{scale_function}. Since those seminal studies, such a waveform has been found to be highly dependent \cite{Abbott2000,Graupner2010} on i) developmental stage, ii) SNC region and iii) neuromodulatory context, exhibiting waveforms with only depression or potentiation, or with Depression-Potentiation-Depression profiles. Graupner and Brunel proposed a model based on Calcium concentration to account for such a modulation, for which each synapse has a fixed point ruling the balance between Potentiation and Depression \cite{Graupner2012}. Such individual synapse dynamics further enrich the dynamic and memory stage capacity of the whole network. As already highlighted in the electromagnetic view, the standard, mostly synaptic efficacy view of learning introduced above should be completed by other plastic mechanisms, possibly at other scales, such as morphological, developmental etc. Notably, current biological studies tend to show that developmental mechanisms and learning processes are "entangled" and form a continuum \cite{Galvan2010}, yet they can consistently be formalized into a general epigenetic adaptive formalism within an information topology framework, as proposed in Baudot, Tapia and Goaillard \cite{Baudot2018}. Indeed, the electromagnetic proposition is blind to an a priori ontological distinction between development and learning, and this epigenetic generalization simply consists of considering "analog" (discrete in the first approximation) multivalued variables rather than binary variables, as discussed in section \ref{infotopo_synthesis} \cite{Baudot2018}. In this topological analysis we further introduce the idea that the total free energy function is equivalent to the Integrated Information of Tononi and Edelman, but applied to genetic expression. Hence, genetic regulation and differentiation are also components of our consciousness which could be assumed to be slow components, given the "impulsional" response characteristics of gene regulation mechanisms. According to operon lactose kinetic studies the magnitude of typical time of the impulsional response of gene regulation is of the order of several hours \cite{Jacob1961} \label{jacob} (compare to Figure \ref{scale_function}). Hence, we propose that such epigenetic plasticity corresponds to modulation of our conscious experience on the scale of hours. Of course, such a slow process could have a direct impact on fast neuronal electrophysiological dynamics, as previously stated (for an example see Soden and colleagues \cite{Soden2013}). \\

\subsubsection{Homeostatic plasticity - consciousness invariance}\label{Homeostatic plasticity}
\begin{figure} [!h]
	\centering
	\includegraphics[height=12cm]{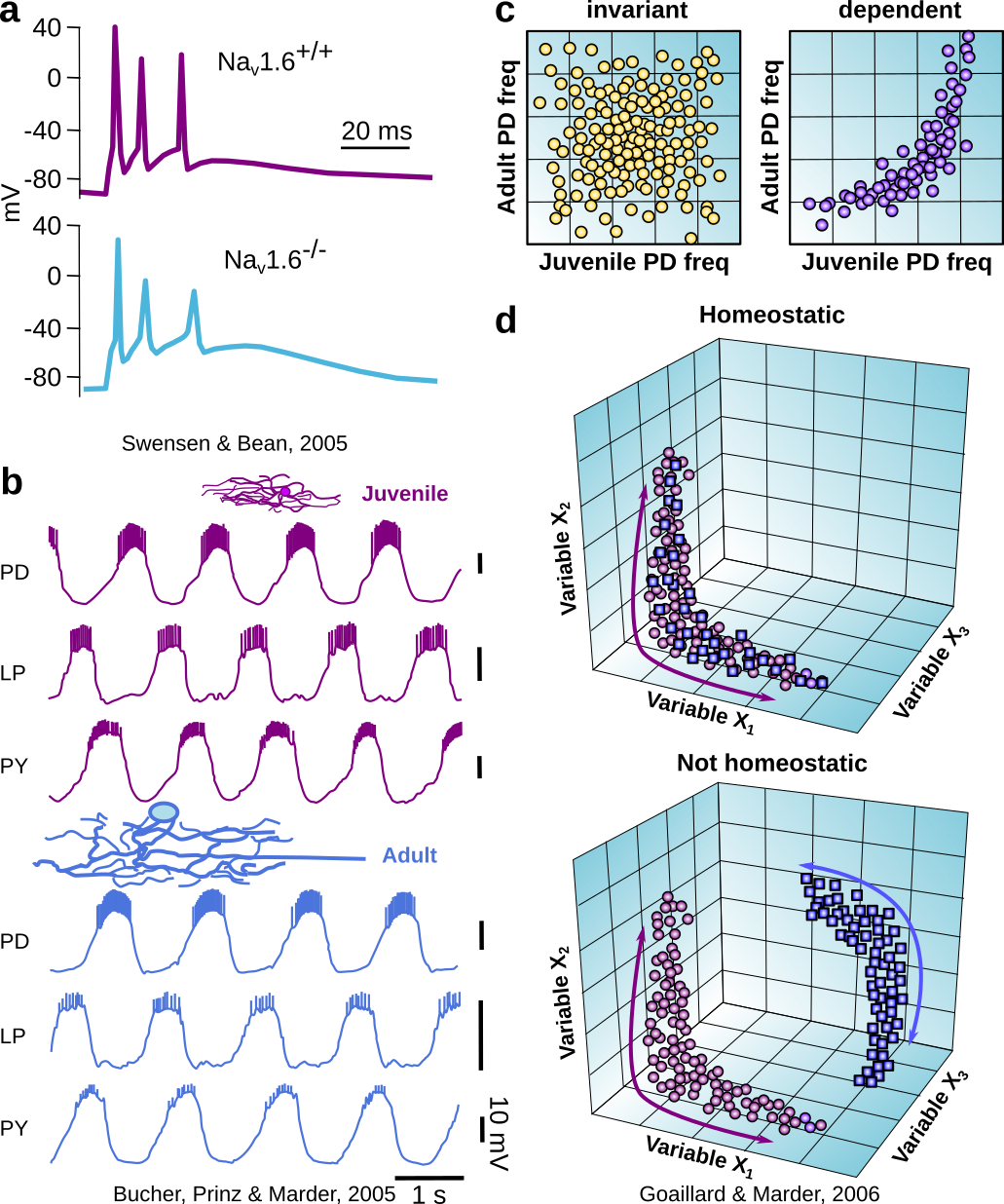}
	\caption{\textbf{Principles of homeostasis and invariance: a,} an example of homeostatic plasticity of bursts of action potentials recorded from Purkinje cell in wild animals ($Na_V1.6^{+/+}$, purple) and in animals knocked out for the sodium channel $Na_V1.6$ subunit ($Na_V1.6^{-/-}$, blue). A transient application of TTX that blocks sodium conductances eliminates the bursting pattern, revealing that a compensation occurred in KO animals to maintain the burst pattern (adapted and modified from Swensen and Bean \cite{Swensen2003}). In such a case, the basic information analysis presented here cannot be applied since the cases of the KO and the wild animals are exclusive and do not provide a joint-distribution but rather two distinct probability laws. In such a case, the more general relative entropy (Kulback-Leibler divergence) has to be applied. \textbf{d,} (top) drawing dye fills of identified pyloric dilator (PD) neurons from a lobster (homarus americanus). (Bottom) Simultaneous intracellular recordings from the PD, lateral pyloric (LP) and pyloric (PY) neurons in a juvenile and an adult preparation, showing similar waveforms and motor patterns, despite the important change of size. Such invariance or independence of the frequencies and relative phase of the patterns to the cell size imply the presence of an active mechanism tuning capacitance (adapted and modified from Bucher, Prinz and Marder \cite{Bucher2005}). \textbf{c,} schematic data for the case of invariance and dependence upon aging of the rhythm generated by neurons or a conductance pattern (see text). \textbf{d,}  Schematic data corresponding to 3 measured variables $X_1,X_2,X_3$ of a system (e.g. 3 conductances of neurons from c) are plotted, for the case of homeostatic process (top) and non homeostatic process (bottom), for two measured populations represented by purple circles and blue squares (e.g. juvenile and adult). The homeostatic case, which is proposed to correspond to the observed case in b; the two populations present the same dependences and produce the same pattern of activity. A modification of the variable in this region does not produce qualitative changes in behavior (purple arrow). The population variability is important as displayed by the widely different values of all three variables for different neurons. Homeostatic tuning rules that maintain a constant type or activity pattern can in principle beneficiate the broad range of variation of its variables to tune the individual neurons. The study by Taylor et al \cite{Taylor2006a} of conductances of a large database of model neurons and by Tapia et al \cite{Tapia2018} and Baudot et al \cite{Baudot2018} of genetic expression reveal that neurons pertaining to the same type are "connected" in the data space. The bottom panel presents a case that would not be homeostatic, for which the two populations would present distinct forms of dependences and would correspond to distinct clustered patterns. Passing from one of these patterns to the other is conjectured to be associated with a qualitative change in activity. (Adapted and modified with permission from Marder and Goaillard \cite{Marder2006})}
	\label{homeostasis}
\end{figure}

The biological complexity of learning mechanisms and rules, requiring refined statistical and dynamical models as just discussed, has introduced a key concept in biology and cybernetics \cite{Cowan1965}: homeostasis. Danchin used the metaphor of the Delphic boat to illustrate the astonishing stability of biological structures and functions despite their constant replacement of their constitutive components, and concluded that it is the  relationships between the 'planks' (that is to say, the genome, the genes, for the nervous system, the neurons...) that define the boat \cite{Danchin2003}. In terms of consciousness, such stability corresponds to the persistence of the self-being, the subjective I.
Marder and Goaillard's review provides explicit electrophysiological examples of such persistence \cite{Marder2006}. While learning principles suggest changes in  function and structure or a displacement of an equilibrium, homeostasis, the other fundamental aspect of plasticity, also ubiquitous at all organizational scales, implies the maintenance or regulation of the stability or equilibrium of the organism, or of some its sub-processes. Such a principle clearly holds at the elementary level of atomic or sub-atomic structures: the atom stays the same while its electrons are continually being exchanged, which further suggests that what we call "self" is an energetic relational structure.  
Biological studies have revealed that even what could be considered static and stable states are indeed maintained by active processes, as illustrated by the invariance of electrical activity patterns in gene knockouts, as seen in Swensen and Bean \cite{Swensen2003}, or by neuronal size increase during development, as shown in Bucher and colleagues \cite{Bucher2005} in Figure \ref{homeostasis}a,b. Note that the electrical activity pattern of lobsters' pyloric systems expresses strong synchrony and integrated information (although it has not been quantified), and hence should be considered as conscious according to usual theories. 
A usual formalization of the relationship between homeostasis and learning involves several scales of organization, as already underlined in Fusi and Abbott's cascade model: fast adaptive-learning models at small scales induce fast temporal fluctuations which are "renormalized" or compensated for by a slow homeostatic process at a large scale. For example, at the synaptic plasticity level, homeostasis is expressed as a slow synaptic scaling process that keeps a neuron's synapses' efficacies in a "stable" physiological range, as studied in the work of Turrigiano and colleagues \cite{Turrigiano2008,Watt2000}. This synaptic weight homeostasis is formalized as a metaplasticity rule playing on the balance between potentiation and depression \cite{Watt2010}, that can be also accounted for by the model of Graupner and Brunel just discussed \cite{Graupner2012}.
In terms of information theory, the formalization of invariance is straightforward, as it corresponds to the general definition of action invariance detailed in section  \ref{infotopo_synthesis} and appendix \ref{Geometries-Homeostasis}, that is a "robustness to noise" or to any variation $X$, a very basic and typical definition of invariance. For example, in the context of physiology, Woods and Wilson proposed that minimizing noise is a key factor driving the evolution of homeostasis \cite{Woods2013}, and we simply relativize their proposition by defining invariance as the minimization of "shared noise". Such a definition of invariance can be formally expressed as a statistical independence requirement: 

\textbf{Invariance (definition):} \label{Invariancedef} a system or process $X$ is invariant with respect to the variations of a system or process $Y$  if conditioning by $Y$ does not change the entropy of $X$, that is if $H(X/Y)=H(X)$, or equivalently $I(X;Y)=0$, or equivalently if $X$ and $Y$ are mutually independent.\\

This definition of invariance appears more generally and intuitively in the fundamental chain rule of information (introduced and explained in the mathematical section of this paper \ref{infotopo_synthesis}):
$I(X_1;.;X_k;..;X_n)=I(X_1;.;\hat{X_k};..;X_n)-X_k.I(X_1;.;\hat{X_k};..;X_n)$, where the hat denotes the omission of the variable. It can be read literally as: the $k+1$ dependencies quantify the default of invariance to the conditioning of a k dimensional system.
In the commutative setting presented here, we can see that such a definition of invariance is symmetric: if $X$ is invariant to $Y$, the converse also holds. In the appendix pertaining to the geometry of consciousness we justify such a definition by the fundamental role of invariance in an action in geometry \ref{Geometries-Homeostasis}: it defines classically a geometry, or a structural stability in topology, and there is hence not much choice in such a definition unless mathematics radically changes. According to this definition, searching for the invariance of a system against the variation of its environment is opposite to learning defined as the infomax principle: the first maximizes independence, the second dependence. This basic principle of invariance is illustrated in Figure \ref{homeostasis}c. Crucially, such study requires that the measures of juveniles and adults are made on the same identified neurons (in order to obtain the joint distribution). In cases where the rhythm generated by neurons and the conductance patterns, is invariant across development, the variability observed in the juvenile would be independent of the variability observed in the adult. In cases where the rhythm generated by neurons depends on the developmental sequence, the variability observed in the juvenile would be dependent on the variability observed in the adult, indicating that a learning process occurred during development. In the latter case, the pattern has memorized something from the environment, although the marginal variability in the juvenile and the adult may stay unchanged or "identical".

The definition of homeostasis is more refined than that of invariance just introduced. The general definition considers  homeostasis to be the maintenance or regulation of the stable condition of an organism or its equilibrium \cite{Martin2008}. It hence requires the introduction of a complex organism and of the concept of equilibrium or stability. As illustrated in Marder and Goaillard \cite{Marder2006}, homeostasis refers to the invariance of a whole, possibly complex, multi-dimensional pattern. According to this view, homeostasis is "a consequence of the potential for compensation among components with different properties", "achieved simultaneously through overlapping functions and compensation". As illustrated in Figure \ref{homeostasis}, a homeostatic process preserves the structure of dependencies among variables, and we propose to define it formally as follows:\\
\textbf{Homeostasis (definition):} \label{Homeostasisdef} a system or process $X_1;X_2;...;X_k$ is homeostatic with respect to the variations of a system or process $X_{k+1}$ if conditioning by $X_{k+1}$ does not change the information of $X_1;X_2;...;X_k$, that is if $I(X_1;X_2;...;X_k)=I(X_1;X_2;...;X_k,X_{k+1})$, or equivalently $I(X_1;X_2;...;X_k/X_{k+1})=0$, or equivalently if $X_1;X_2;...;X_k$ are conditionally independent given $X_{k+1}$.\\

This definition of homeostasis could also read $I(X_1;X_2;...;X_k;... ;X_k)=I(X_1;X_2;...;\widehat{X_k};...;X_{k+1})$, literally, that the removal of a variable $X_k$ (noted $\widehat{X_k}$) from the whole system or process does not change the information. Homeostasis is stronger than invariance in the sense that invariance implies homeostasis but not the converse ($I(X;Y)=0 \Rightarrow I(X;Y/Z)=0$). In the section dedicated to information topology \ref{infotopo_synthesis}, we will see that this definition is related to the usual definition of equilibrium in thermodynamics. Another way to quantify homeostasis that is less rigorous and general compares the mutual information $I(X_1;X_2;...;X_k)$ with the probability densities estimated in the points of the reference type (e.g. adult, or wild type, $P$) and with all the points ($Q$): an increase or constancy of mutual information, $I(X_1;X_2;...;X_k;Q)\geq I(X_1;X_2;...;X_k;P)$, would indicate a homeostatic process.  

\section{Mathematic, cognition and consciousness}\label{Math}

\subsection{Mathematical nature of consciousness}

\subsubsection{Empirical proof of the mathematical nature of consciousness: subjective axioms}
\begin{figure} [!h]
	\centering
	\includegraphics[height=9cm]{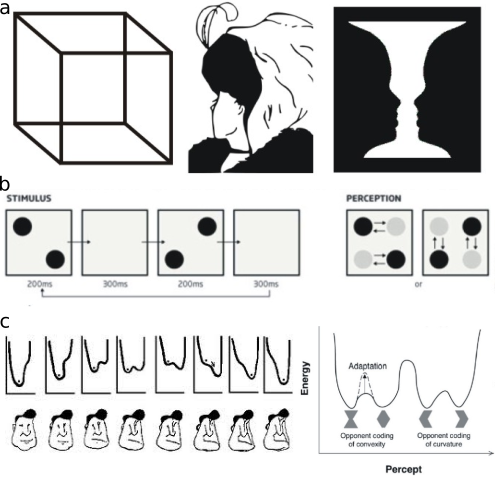}
	\caption{\textbf{Consistency and diversity of the subject's qualia, psychophysical evidence. a} 3 paradigmatic examples of bistable images and perception. From left to right, the Necker's cube \cite{Necker1832}, Hill's "My wife and my mother-in-law" \cite{Hill1887–1962}, Rubin's "Face-Vase" \cite{Rubin1915}. \textbf{b,} The elementary dynamical bistable stimuli of apparent motion designed by Ramachandran and Anstis \cite{Ramachandran1985} (reproduced with permission from Carter et al. \cite{Carter2008}). \textbf{c,} The energy landscape interpretation of the transition from one percept to another (adapted and modified with permission from Fisher  \cite{Fisher1967}). Right, the bistable case, left the multi-stable case designed by Suzuki and colleagues as a tetra-stable stimulus \cite{Suzuki2002} (reproduced with permission from Suzuki et al. \cite{Suzuki2002}).   }	
	\label{bistability}
\end{figure}

In this section we ask if the language of consciousness could be mathematical.  
In the early stages of mathematical formalization, Aristotle stated the principle of non-contradiction, now known as consistency. For simplicity, we use this axiom: $X \wedge \neg X = 0 $; or, literally, "there is no proposition $X$ such that $X$ and $\neg X$". A more formal modern definition is: a set of formulas - a theory $T$ in first-order logic - is consistent if there is no formula $X$  such that $T \vdash X$  and $T \vdash \neg X$. Non-contradiction is still a fundamental principle in almost all mathematical theories (some researchers having developed para-consistent logic, by which any theorem is true without requiring proof). The original statement of Aristotle is "it is impossible that contrary attributes should belong at the same time to the same subject." (\cite{Aristotle350B.C.E}, III.3). Aristotle's formulation is both a cognitive principle and a mathematical axiom, and it can be illustrated by a standard psychophysical experiment of perceptual bistability or multistability, as portrayed in figure \ref{bistability}. We hence consider consistency as the first axiom of cognitive theory and propose that experiences such as the one illustrated in Figure \ref{bistability} provide empirical evidence for such a statement. Moreover, we do not consider the excluded-third $X \wedge \neg X$ (or excluded middle, which could also be expressed as "for every proposition
$X$, either $X$ or $\neg X$") as an axiom or theorem of such a cognitive theory. The reason for this will become more clear in the following sections, and this exclusion is in practice motivated by the fact that we require a multivalued logic that corresponds intuitively to the usual conception of probability valuation. As illustrated in Figure \ref{bistability} and for example in the work of Suzuki and colleagues \cite{Suzuki2002}, stimuli and perception are possibly multi-stable. The logical aspects of topos \cite{Proute2008,Bauer2017}, which we will briefly introduce in the next section, account for this bistability and multistability by the fact that the internal logic of the subject perceiving that $X$ or $\neg X$ does not negate the excluded third while the external observer will report that $X$ has an intermediate truth value, e.g. the probability that the observer saw $X$ \cite{Suzuki2002}. Brouwer noted that the excluded third is a kind of "omniscient principle", allowing one to prove $X \wedge \neg X$ even in the case where $X$ is Fermat's Last Theorem \cite{Proute2008}. Here, the excluded third is independent in constructive logic \cite{Proute2008,Bauer2017}; it cannot be proved or disproved. These two considerations lead us to propose that cognitive logic is constructive-intuitionist. In what follows we present the modern expression and results of such a model, taking into consideration that a pertinent formalization of perception is measure theory. Hence, we review what a constructive measure and probability could be.  

Figure \ref{bistability} illustrates the ongoing challenges in a mathematical theory of consciousness or cognition: a consistent theory, for which every thought is true and non-trivial, yet one which encompasses the obvious multiplicity of thoughts: a tolerant theory that allows for the construction of all subjective thoughts. What follows aims to provide axioms and mathematical foundations for such a theory, notably accounting for Edelman and Tononi's model and quantification of cognition and consciousness, the integrated information theory (IIT) \cite{Tononi2016,Oizumi2014,Tononi1998,Edelman2000}. Such foundations are the mathematical foundations of measure, integration and probability theory.     

\subsubsection{Mathematic, a model of cognition}

Following Boole, some mathematicians have considered the logical foundations of mathematics and the logical foundation of a theory of mind as equivalent problems. Cantor gave the following definition: \textit{"a set is a gathering together into a whole of definite, distinct objects of our perception [Anschauung] or of our thought, which are called elements of the set"} \cite{Cantor1895}. As stated by one of its main conceivers, set theory is a cognitive theory. Whitehead and Russel completed a work on the foundations of mathematics with the Principia Mathematica \cite{Whithead1910}. Whithead went even further in the cognitive conception of mathematics, developing at length a theory of perception in which any entity of the universe, even a particle, perceives, and considering events as elementary constituents in place of masses. He stated: \textit{"there is urgency in coming to see the world as a web of interrelated processes of which we are integral parts, so that all of our choices and actions have consequences for the world around us"} \cite{Whithead1929}. Frege developed a fully cognitive  and linguistic interpretation of logic. In cognition, it gave birth to the whole field of analytical philosophy of mind and language. Hilbert formulated the project of arithmetization of mathematic, such that any theorem could be derived automatically, stated the isomorphism between the mind and logical calculus, and notably claimed \textit{"My theory of demonstration only simulate the internal activity of our understanding and record the proceedings of the rules that govern the functioning of our thoughts"}\cite{Hilbert1930}.
G\"{o}del promptly discovered the theorems of completeness and incompleteness, creating an obstruction to Hilbert's program on the basis of ZFC axioms (a particular model of Arithmetic; using another model such as Presburger arithmetic, for example, G\"{o}del's obstructions do not hold and the theory is complete) and using an arithmetic coding procedure \cite{Godel1931}. It is also such an equivalence that guided Turing in defining computational classes and machines on cognitive reasoning principles \cite{Turing1937}: \textit{"computable  numbers  are  those  whose  decimals are calculable by finite means... the justification lies in the fact that  the human memory is  necessarily limited"}. Such an equivalence can be legitimated using the following reasoning inspired by Penrose \cite{Penrose1989}, in which the inclusion sequence is justified: $\text{cognition}\subseteq \text{physic} \subseteq \text{mathematic}$. This is a purely theoretical and classical sequence; the complement of physics with mathematics is called metaphysics, and the complement of cognition with physics is called in-animated matter. However, it is also possible to argue about the inverse inclusion short sequence in practice: $\text{cognition}\supseteq \text{physic} \supseteq \text{mathematic}$. Mathematics and physics are subsets of human cognitive activities, and mathematics is a subset of activity pertaining to physics (since there are purely empirical studies in physics, which are purely descriptive without theoretical, formal or modeling aspects). This gives a reasonable basis for thinking that mathematics, physics and cognition are not so far from being equivalent, the isomorphism postulated by Hilbert between the mind and logical calculus. Simply put, the conclusion is that biological systems can be understood as "living theorems" following the path of their proof (just as in Villani \cite{Villani2013}): mathematicians are not the only great geometers, so that any stone or insect can enter the monument of academia, while still respecting the severe policy written at the entrance: "Let no-one ignorant of geometry enter". 
It could be possible to construct a mathematical paradise that would not be a tiny room, but a grand hotel, and to reconcile Hilbert's formalism with Brouwer's intuitionistic approach; the problem is to construct a hotel such that it can safely, harmoniously and even comfortably welcome all its legitimate members. So let's try to give a free unlimited VIP pass to Hilbert's paradise, and remind ourselves that there will be no queue or competition to enter (we indeed start in what follows with very tiny elementary hotel).\\

Set theory is no longer considered a sufficient foundation for mathematics by an important segment of the mathematical community; the theory of category has begun to replace it (see Mac Lane's book \cite{MacLane1998} and Lawvere \cite{Lawvere2005}). Here, for didactic and fluency purposes, we will mainly use the old, classical language of set theory, although categorical formulations as preliminarily developed, for example in \cite{Baez2014, Baez2011, Baudot2015a, Vigneaux2017}, are more appropriate (and perhaps even necessary in order to fully legitimate the generalization of the boolean logic highlighted here). Logic studies have made great improvements since set theory and G\"{o}del's work; most notably the introduction of and reference to a meta-logic can now be avoided, and logic has became contextual: a theorem or statement is considered in the context of a given theory with its axioms \cite{Proute2013}.
To conclude this section, consciousness is proposed to be by nature mathematical. One obvious reason for this is that mathematics is the only scientific domain that can guarantee currently and consistently the unity required for a theory of consciousness, while physics and biology are commonly considered, with regard to one another, as divided and non-unified.  \\

\subsection{Constructive logic, information topos and e-motivs}

This section asks the questions of what could be the mathematical axioms of a theory of consciousness, whether any given thought can be conceived of as a theorem that can be derived from a set of axioms, and whether it can be identified by its information. Rather than providing a complete definite theory, the results presented here point out all the way left in order that we understand what information and consciousness is. 

\subsubsection{Measure Theory: a mathematical theory of subjective and objective experience-observation}\label{measure}

This section investigates the mathematical axioms of a theory of consciousness as the axioms of measure and probability, which would further avoid paradoxical decomposition induced by the Axiom of Choice.

\textbf{Measure with countable choices, Quantified volumes of e-motion, consciousness integration, constructive axioms of perception-action?} 

In 1854, Riemann defined the integral of a function as the asymptotic limit of the sums of areas of small rectangles approximating the function \cite{Riemann1854b}. With his method, a large class of derived functions had no definite integral. Lebesgue in 1901 proposed a formalism such that integration operation could be the inverse of derivation operation, that is, for any function $f$ continuous on $[a,b]$ and differentiable over $]a,b[$, we have $f(b)-f(a)=\int_a^b f'(x)dx$. The work of Lebesgue \cite{Lebesgue1901} and Borel (\cite{Borel1898} Chap III) also showed that their integration theory relies on elementary and general definitions and axioms within set theory. In his “lessons”, Borel explicitly assigned a measure to subsets of $[0,1]$ generated from the sub-intervals by the operations of countable unions or by taking the complementary. Borel also explicitly stated that the measurement of these subsets (later termed 'measurable' by Lebesgue) satisfies the additivity property: if $X_n$ is a finite or countable family of such pairwise disjoint sets, then the measure of their union is equal to the sum of their measures $\mu(X_1\cup...\cup X_n)=\mu(X_1)+..+\mu(X_n)$. Moreover, he claimed that the only measure a sub-interval has is its length. Borel proved the existence of such a measure, and Lebesgue its uniqueness. Borel moreover stated that the measure of a set is always non-negative. These axioms of measure provide a formal and reasonable basis for a theory of experimental measurement and subjective observation, that is, for the properties one should reasonably expect of measures (data, experience) in general. The mathematization of such subjective measures ensure their intersubjective communication and understandability: to reliably share subjective perception with another subject, mathematics is the less ambiguous language. The central axiom is additivity, known as extensivity in physics: for two non-intersecting sets $X,Y$ we have  $\mu(X\cup Y)=\mu(X)+\mu(Y)$. Generally, for arbitrary, possibly intersecting sets, we have the inclusion-exclusion principle: $\mu(X\cup Y)=\mu(X)+\mu(Y)-\mu(X\cap Y)$ and for $n$ sets $\mu(\bigcup_{i=1}^n X_i)=\sum_{i=1}^{n}(-1)^{i-1}\sum_{I\subset [n];card(I)=i}\mu(X_I)$. To properly formalize this in set theory, Borel and Lebesgue proposed operations that generate measurable sets, of which there are two: countable union and taking the complementary. They define an algebra known as $\sigma$-algebra (sigma here denoting additive), and its axioms are:\\
Let $\Omega$ be a given set and $2^|\Omega|$ be its power set. A subset $\mathcal{F}$ is then called a $\sigma$-algebra if it satisfies the following three axioms: 
\begin{itemize}
	\item $\mathcal{F}$ is non-empty
	\item $\mathcal{F}$ is closed under complementation
	\item $\mathcal{F}$ is closed under countable union
\end{itemize}
From the 2nd and 3rd axioms, it follows that a $\sigma$-algebra is also closed under countable intersections, and one can remark that in the finite $\Omega$ case it is also a Boolean algebra and a topological space. The axioms of a measure are: let  $\Omega$ be a set and  a $\sigma$-algebra over  $\Omega$. A function $\mu$ from the extended real number line is called a measure if it satisfies the following properties:
\begin{itemize}
	\item   $\forall X \in \mathcal{F},\mu (X)\geq 0$
	\item   $\mu (\emptyset)= 0$
	\item  Countable additivity (or sigma-additivity): For all countable collections  of $X_{i,i\in I}$ of pairwise disjoint sets in $\mathcal{F}$:  $\mu(\bigcup_{i\in I} X_i)=\sum_{i\in I} \mu(X_i)$.
\end{itemize}	

As a conclusion to this section, the importance of the axioms of measure arises from the fact that they are at the foundations of mathematics and are an obvious minimal requirement for formalizing an observed quantity in physics, as well as any objective or subjective measure, if a such distinction makes sense.\\
Just as for dynamical systems and physics, for which initial conditions can dictate the dynamic in the long run, the axioms of a mathematical theory dictates what theorems are available within the theory. As exemplified by the independent 5th axiom of Euclid, which hid the existence of non-euclidean geometries, Occam's razor is also a guiding principle in the choice of axioms: considering spurious, unnecessary axioms can lead to theories being too restricted to account for peculiar physically-observed phenomena. What is worse, the axioms of a theory can contain contradiction in their very germ.  While modern model theory simply defines set theory by an object, a set $\Omega$ together with an operation of inclusion $\subseteq$, a classical construction of the theory of sets, relies on Zermelo-Fraenkel axioms together with the Axiom of Choice (AC), forming the ZFC model \cite{Ciesielski1997}. The AC roughly proposes that, given any, possibly infinite, collection of bins, each containing at least one object, it is possible to make a selection of exactly one object from each bin, or equivalently, that any set may be well-ordered (see Howard and Rubin for the various expressions of AC \cite{Howard1998}). Fraenkel   proved in 1922 the independence of AC from a model/theory of set with atoms $T(A)$ \cite{Bell2015}. His proof,  reproduced in the article of Bell \cite{Bell2015} and further generalized by Cohen using his forcing method (without use of atoms, and holding for real numbers), relies on the fact that permutation of the set $A$ of atoms induces a structure-preserving permutation, an automorphism, of the Theory $T(A)$ of sets built from $A$, allowing to construct an equivalent Symmetric model $Sym(T)$ of set theory in which it is easy to prove that a set of mutually disjoint pairs of elements of $A$ has no choice function. The proof given by Fraenkel ensures that in a mathematical logic based on Galoisian group will not dispose of the infinite choice axiom. The relation of the axiom of choice to intuitionist logic is straightforward: The axiom of simple choice (a finite subcase of AC) is equivalent to the principle of excluded third, as shown in few lines by Ageron \cite{Ageron2002}.  
The AC caused severe problems and debates, notably concerning measure theory, as it implies the paradoxical existence of non-measurable sets, leading for example to the Banach-Tarski paradox \cite{Wagon1986}. This paradox states that $B^3$, the solid ball in $\mathcal{R}^3$, is $G^3$-paradoxical: considering (countably) infinite choices, the ball could be decomposed into a finite number of point sets and reassembled into two balls identical to the original, or a new sphere of any size. The physical interpretation of the Banach-Tarski paradox allows matter, or indeed gold, to be created ex nihilo \cite{Dewdney1989a}, a first principle failure whenever one would wish to axiomatize thermodynamics in logic (as we wish to do here). It is hence legitimate and usual to consider non-measurable set as metaphysical sets. The important result was found by Diaconescu: he showed that AC implies the excluded-third \cite{Diaconescu1975} (see Bauer for short proof \cite{Bauer2017}); hence, in this sense, the ZFC model is not constructive. Another equivalent expression of the excluded-third is "subsets of finite sets are finite". If this statement seems at first glance reasonable, we shall see that it imposes a notion of "point" or "atom" far stronger than Euclide's definition “that which has no part” and avoids, in a sense, any "spatial extent" of a point or "atom". 
What would be a constructive (with something like a finite choice version of AC) version of the decomposition proposed by Tarski, avoiding those metaphysical sets? An answer arose from Dehn's solution to Hilbert's 3rd problem and is called dissection or scissors congruence. Two polyhedra in Euclidean 3-space are scissors congruent if they can be subdivided into the same finite number of smaller polyhedra such that each piece in the first polyhedron is congruent to one in the second. For example, Bolyai-Gerwain's theorem states that two polygons are scissors congruent if and only if they have the same area. However, in higher dimensions, this theorem no longer holds, and one has to add Dehn's invariant. For instance, Dehn proved that a regular tetrahedron in $\mathbb{R}^3$ is not scissor congruent with a cube of the same volume \cite{Dehn1901}. The Banach-Tarski paradox nevertheless states that they are equidecomposable. Hence, Dehn's finite dissections appear finer than Tarski's infinite decomposition. This was formalised by Wagon, who established that if two polygons are equidissectable, then they are equidecomposable \cite{Wagon1986}. Scissor congruences, defining groups, were generalized to arbitrary dimensions and geometry, and their homology extensively studied, notably by Dupont and Sah \cite{Dupont1982,Dupont2001}. The axiomatization and formalization of those groups was notably further pursued by Denef and Loeser, in the domain known as motivic measure and integration, which explicitly provides a field-independent measure of mathematical formula, a modern version of Leibniz's analysis situs \cite{Hales2005}. This brief presentation of dissections and decompositions is sufficient to conclude that consideration of AC implicitly involves the inability to distinguish elementary distinct geometrical objects that finite dissections discern. A possible way to circumvent the problems raised by AC is to consider the cheap nonstandard analysis obtained by the consideration of the Frechet filter, as explicated by Tao \cite{Tao2012}.   
Another possible way to give a more precise axiomatization for a cognitive theory without non-measurable sets is to follow Solovay's work \cite{Solovay1970}. In Solovay's construction, that is, classical Zermelo-Fraenkel axioms with Dependent Choice (DC, countable-finite weakening of the axiom of choice) and the existence of an inaccessible cardinal IC, any set is measurable. His axioms provide a construction of "real" numbers, called "random reals" which are in bijection with additive homomorphisms. This is, in our opinion, one of last and greatest achievements in Hilbert's arithmetization program.\\ 
To conclude, even at the elementary level of the logical axiomatization of a mathematical theory, the formalization of what kind of decomposition-dissection-division is allowed appears crucial, and a slight change in the axiom, e.g. from AC to DC, can avoid important "complications" that appear paradoxical from the physical point of view.at least from basic physical principles. Arithmetic and number theory provide the guiding principle for such a division procedure.

\subsubsection{Probability, the logic of thoughts, the geometry of beliefs}\label{logic of thoughts}

Measure theory allowed the Kolmogorov's axiomatization of probability \cite{Kolmogorov1933a}. Considering probability theory as a cognitive theory has been an obvious option since the early stages of probability theory. The title of Boole's major opus is sufficiently explicit to illustrate our statement: \textit{"An Investigation of the Laws of Thought on Which are Founded the Mathematical Theories of Logic and Probabilities"} \cite{Boole1854}. His work also provided the basis for the development of information theory, as exposed in the book by Nahin \cite{Nahin2012}, and Boole should hence be considered one of the important founders of the theory of consciousness and cognition. Boole's original text is sufficiently limpid for there to be no need of commenting it, and as such we simply cite it in the present work (\cite{Boole1854} Chap. III, Derivation of the laws of the operations of the Human mind): 
\begin{itemize}
	\item Proposition 1: To deduce the laws of the symbols of logic from a consideration of those operations of the mind which are implied in the strict use of language as an instrument of reasoning.
	\item Proposition 2: To determine the logical value and signifiance of the symbol 0 and 1. [...] The Symbol 0, as used in algebra, satisfies the following law,$0\times y=0$ or $0y=0$, whatever number y may represent. [...] Secondly, the Symbol 1 satisfies in the system of numbers the following law, 1*y=y, or 1y=y, whatever number y may represent. [...] Hence, the respective interpretation of the symbols 0 and 1 in the system of Logic are nothing and Universe.
	\item Proposition 3: If $X$ represent any class of objects, then will $1-X$ represent the contrary or supplementary class of objects, i.e. the class including all objects which are not comprehended in the class $X$.
	\item Proposition 4: The axiom of metaphysicians which is termed the principle of contradiction, and which affirms that it is impossible for any being to possess a quality, and at the same time not to possess it, is a consequence of the fundamental law of thought, whose expression is: $X^2=X$ . Whence we have $X.(1-X)=0$. Both these transformations being justified by the axiomatic laws of combination and transposition (II.13). Let us, for simplicity of conception, give to the symbol $X$ the particular interpretation of "men", then $1-X$ will represent the class of "not-men" (prop III.). Now the formal product of the expressions of the two classes represents that class of individuals which is common to them both (II.6). Hence $X.(1-X)$ will represent the class whose members are at once "men" and "not men", and the equation (2) thus express the principle, that a class whose members are at the same time men and not men does not exist. In other words, that it is impossible for the same individual to be at the same time a man and not a man [...] which is identically that principle of contradiction which Aristotle has described as the fundamental axiom of all philosophy.
\end{itemize}   
This "law of duality", or principle of non-contradiction, here made algebraic, will henceforth be called the \textbf{idempotence} property of a composition law. We will see that, like the join and meet of probability of events ($P(X \vee X)= P(X)$ and $P(X \wedge X)= P(X)$), joint of random variables and partitions is idempotent. 
After Boole, Hume founded cognitive sciences in his treatise on human nature by notably stating that "all knowledge degenerates into probability" \cite{Hume1738}, and since Leibniz had established binary calculus and monads, the probability theory of cognition demonstrated an impressive robustness. More than one and a half centuries after Boole, a long list of works and articles still propose that probabilistic or Bayesian theory is the relevant formalism for a theory of the brain (for a review on this topic see Griffiths \cite{Griffiths2008}, Friston \cite{Friston2012} and the references therein). The question of what probability is, its axiomatization, and the foundations of cognition are investigated in depth in a series of works by Gromov, which partially motivated the work presented here \cite{Gromov2015,Gromov2015a,Gromov2013,Gromov2018}. \\ 
Kolmogorov based his axioms of probability on the Lebesgue measure, and it is these axioms that we here consider as still pertinent for a consciousness-cognitive theory; hence we faithfully reproduce his axiomatization (with the exception of the symbols of joint and meet probability, which have been changed such that they are consistent with the preceding logical notations \cite{Kolmogorov1933a}): "Let $\Omega$ be a collection of elements $ \xi,\eta, \zeta,...$, which we shall call elementary events, and $\mathcal{F}$ a set of subsets of $\Omega$; the elements of the set  $\mathcal{F}$ will be called random events.
\begin{itemize}
	\item $\mathcal{F}$ is a field of sets. 
	\item $\mathcal{F}$ contains the set $\Omega$.
	\item To each set A in  is assigned a non-negative real number $P(A)$. This number $P(A)$ is called the probability of an event $A$.
	\item $P(\Omega)=1$
	\item If $A$ and $B$  have no element in common, then $P(A \vee B)=P(A)+P(B)$
\end{itemize}     

A system of sets $\mathcal{F}$, together with a definite assignment of numbers $P(A)$, satisfying axioms 1-5, is called a field of probability. Conditional probability: If $P(A)>0$ , then the quotient $P(B/A)= P(A\wedge B)/P(A)$ is defined to be the conditional probability of the event $B$ under the condition $A$"\cite{Kolmogorov1933a}.\\
Three remarks can be drawn from his axiomatization:
\begin{itemize}
	\item Forgetting the 4th axiom ($P(\Omega)=1$), we obtain the preceding axioms of measure; hence, a probability is a normalized measure such that the “total measure” is a unit.
	\item Probability and randomness are simply the definition of an abstract geometric volume in arbitrary space and rely on additivity: there is nothing more "deterministic" in common sense than a geometric volume of space and addition; the usual opposition between common notions of determinism and non-determinism fails (while the formal definition of determinism as events with a probability of 1 or 0 stays consistent). Notably, famous statements in physics of the kind \textit{“god does not play dice”} \cite{Hawking,Natarajan2008}, where "god" is considered as an abbreviation for the "geometry of space-time", could be interpreted as meaning that space-time has no volume, which is a nonsense.
	\item As stated by Kolmogorov, these axioms of finite-discrete probabilities, which are usually handled as empiric probability, the ratio $n/m$ in the discrete rational case, define the \textbf{"generalized fields of probability"}. To handle continuous probabilities a 6th axiom of "infinite" is required (just as in set theory according to Bourbaki \cite{Bourbaki1968}). We note that, as Kolmogorov, who was one of the main founders of constructive logic, probably wished, the discrete rational empirical "generalized fields of probability" respects constructive requirements (as no infinite choice is required), while in the case of real-valued probabilities, it depends on the precise construction of real numbers (the field being constructed using Solovay's model and random reals to fulfill our measurability completeness requirements).
\end{itemize}

\begin{figure} [!h]
	\centering
	\includegraphics[height=10cm]{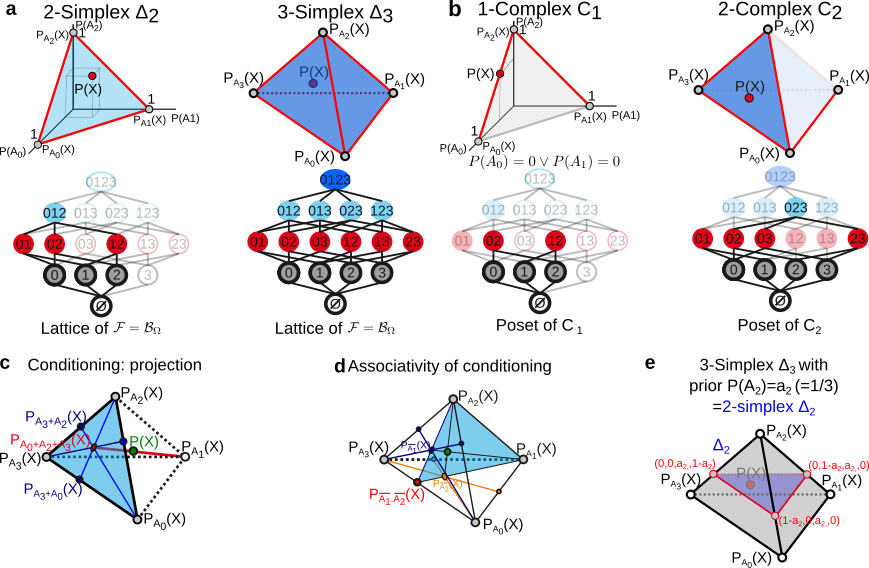}
	\caption{\textbf{The geometry of probability. a,} examples of a 2-simplex and a 3-simplex probability together with their associated Boolean complete lattices (or $\sigma$-algebra, bottom). A sample space of $n$ atomic events $\{A_0,..,A_{n-1}\}$ defines a $(n-1)$-simplex of probability.  A probability $P(X)$ lies in the convex hull depicted in blue which is the $(n-1)$-simplex with a vertex at the units of $\mathbb{R}^{n}$ (more exactly $\mathbb{R}^{\otimes_k n}$). (Left) The example of a 2-simplex $\Delta_2$ which can be illustrated by a coin toss possibly biased but with the coin having three faces, with a sample space composed of 3 atomic-elementary events "face0" ($face0=A_0$), "face1" ($face1=A_1$) and "face2" ($face2=A_2$), and $\Omega =\{ A_0,A_1,A_2 \}$, the $\sigma$-algebra considered in the finite context as a Boolean algebra $\mathcal{B}_{\Omega}$ of all possible, not necessarily atomic-elementary outcomes, is $\mathcal{F}=\mathcal{B}_{\Omega}=\{ \emptyset,\{A_0\},\{A_1\},\{A_2\},\{A_0,A_1\},\{A_0,A_2\},\{A_1,A_2\},\{A_0,A_1,A_2\} \}$. The probability $P(X)$ is given by the theorem of total probability $P(X)=P(A_0).P_{A_0}(X) + P(A_1).P_{A_1}(X) + P(A_2).P_{A_2}(X)$ where $P(A_i), i \in {0,..,n-1}$ provides barycentric coordinates, since we have $\sum_{i=0}^{n-1} P(A_i)=1$. For the 3-simplex, there are 4 possible outcomes (right). \textbf{b,} Examples of a 1-complex and a 2-complex constructed as a sub-complex of the previous simplex with their lattice. The exclusion rule that generates the 1-complex is $P(A_0)=0 \vee P(A_1)=0 $. \textbf{c,} Conditioning is the projection on the lower dimensional opposite $(n-2)$-subface of the simplex. Rigorously, the conditioning follows the inverse path of the simplicial projections; in the example from $P_{A_3}(X)$ to $P_{A_3\vee A_0 }(X)$ to $P_{A_2 \vee A_3 \vee A_0}(X)$ to  $P(X)=P_{\Omega}(X)=P_{A_0\vee A_1 \vee A_2 \vee A_3}(X)$. \textbf{d,} Example of adding a prior, here the elementary constant function $P(A_2)=1/3$ in the 3-simplex.}	
	\label{figure_probability_simplex_complex}
\end{figure}

One of the remarkable aspects of Kolmogorov's axiomatization is that it has a direct and simple geometrical expression, usually named the probability simplex. Although the origin of the probability simplex is unknown to us, it has been in a part of mathematical folklore for a long time; in 1855, Maxwell constructed the simplex of colors in his study "Experiments on Colour as Perceived by the Eye, with Remarks on Colour-Blindness" \cite{Maxwell1855} (he also claimed \textit{"the true logic of this world is the calculus of probabilities"}). As an additional example,, a more modern version of a probability simplex was presented and used extensively in 1982 in Cencov's seminal work \cite{Cencov1982}. 
A simplex is defined as the set of families of the numbers $P_{\omega}, ~ \omega \in \Omega$, such that $\forall \omega, 0\leq P_{\omega} \leq 1$. It parameterizes all probability laws on $\omega$. In more explicit geometrical terms, the fourth and fifth axioms of probability are equivalent to imposing that geometry is affine, and the axiom of positivity (axiom 3) dictates that it is convex. The more general expression of conditional probability using projective space is studied in Morton \cite{Morton2013}. This is depicted in figure \ref{figure_probability_simplex_complex} for a 2 and 3-simplex of probability.
Notably, the theorem of total probability \cite{Kolmogorov1933a} states that given elementary events $A_1 \cup...\cup A_n = \Omega $, we have $P(X) =P(A_1).P_{A_1}(X)+ ...+P(A_n).P_{A_n}(X) $, allowing the consideration of $\{P(A_1),...,P(A_n)\}$ as the set of barycentric coordinates of the probability $P(X)$ in the $(n-1)$-dimensional simplex. It is possible to construct a subcomplex of these probability simplexes by a process of exclusion of faces, utilising an exclusion rule, that traduces the cancellations of a probability \cite{Baudot2015a}. An example of 1-complex and 2-complex of probability, together with their associated set of exclusion rules, is given in figure \ref{figure_probability_simplex_complex}. Conditioning by elementary events is a projection on the complement $(n-2)$-subsimplex (the opposite $(n-2)$-face) and is associative, as shown in figure \ref{figure_probability_simplex_complex}. Addition of priors usually consists of selecting a subspace of the $(n-1)$-simplex by imposing arbitrarily complex functional constraints on the elementary probabilities. This geometrical formalization of probability is not the geometry of the space itself, but the geometry of the volumes within the space. 

\subsubsection{Topos: the consistent diversity of truths and beliefs}\label{topos}

In this section we ask what a probabilistic and informational logic could be in practice.

\textbf{Multivalued logic and probability.} Since Kolmogorov, the axiomatization of probability has been repeatedly questioned, something which has been motivated by the idea that the logic implemented by biological or cognitive calculus could differ from classical logic. There have been many attempts to propose alternatives to Boole's original work on logic and probability \cite{Boole1854} and Kolmogorov's work \cite{Kolmogorov1933a}, for example the definition of a peculiar set of Bayesian axioms and logic that gives a fundamental role to Bayesian sum and product rules following Cox and Jaynes's work \cite{Cox1961,Dupre2009}, or fuzzy logic \cite{Zadeh1965,Hajek2013,Wierman2010}. The basic motivation guiding such research is that, where classical Boolean logic and set theory admits only two valuations, "true" or "false", probability theory provides an obvious multivalued logic.
In a series of works based on Lattice theory and pointing out the relation to factorization in number theory, Knuth proposed to derive the basic principles of probability and information accounting for Cox and Kolmogorov foundations  \cite{Knuth2005,Knuth2009}. The principles proposed by Knuth are basically the same as what is presented in this review that underlines a more usual mathematical expression. 
Carath\'{e}odory and Kappos proposed an alternative, indeed equivalent axiomatisation, but one that is more directly along the lines of intuitionistic logic, which according to Cartier \cite{Cartier2001} postulated: \textit{“instead of requiring the valuation $v(A)$ of a proposition to assume only the values 0 and 1, one may postulate more generally that $v(A)$ is a real number between 0 and 1.”}. With the aim of providing foundations for Laplacian or Bayesian probability, Cox proposed three desiderata-”axioms” \cite{Cox1961}: 
\begin{itemize}
	\item  representations of plausibility are to be given by real numbers
	\item  plausibilities are to be in qualitative agreement with common sense
	\item  plausibilities are to be consistent, in the sense that anyone with the same information would assign the same real numbers to the plausibilities.
\end{itemize}    
As far as we understand these desiderata, they appear consistent in all points with Kolmogorov's axioms (but their "fuzziness" does not allow the proving or the disproving of any equivalence), leading to the conclusion that subjective vs. objective, Bayesian vs. frequentist probability, at least at the axiomatic level, are simply two different interpretations of a single theory: probability. While the Bayesian interpretation remains relevant concerning the theory of mind, this identity enriches the Bayesian interpretation by underlining its obvious pertinence in the domains of physics and empirical science. \\

\textbf{Topos: a bridge between the subject and the object (\textit{'objectifies the subjective'} Lawvere \cite{Lawvere2014}).} The multivaluation of logic found its main mathematical expression in the construction of topos theory. Topos were developed by Grothendieck, Verdier and Giraud \cite{Artin1964}, predominantly on the geometrical ground of sheaves. Grothendieck resumed this work in the following terms: \textit{"This is the theme of the topos which is the "bed" where come to marry geometry and algebra, topology, and arithmetic, mathematical logic and category theory, the world of the continuum and the one of "discontinuous" or "discrete" structures. It is the largest construction I have designed to handle subtly, with the same language rich geometric resonances, a common "essence" to some of the most distant situations."} \cite{Grothendieck1985} (p.59). A simple introduction to ambiguity and Topos, with some cognitive aspects, can be found in Andr\'{e}'s book \cite{Andre2007a} (chap.1). \\
The logical aspects of topos and the fact that it provides an algebraic logic were notably recognized in the work of Lawvere and Tierney (see \cite{Lawvere2014} for a review of this topic). This logical view provides a quite simple definition of a Topos: a "category with a power set". According to Reyes \cite{Reyes1977}, the analogy that has been constructed identifies: \\
\begin{tabular}{|c|c|}
	\hline 
	Topos Theory  & Model Theory\tabularnewline
	\hline 
	\hline 
	Site & Theory\tabularnewline
	\hline 
	Fiber (on the site) & Model (on the theory)\tabularnewline
	\hline 
	Sheaf & Concept (represented by a formula)\tabularnewline
	\hline 
\end{tabular}\\

A topos $T$ is a category with the 3 axioms \cite{Lawvere1972,Diaconescu1975}: 
\begin{itemize}
	\item $T$ has finite limits, i.e. finite products, intersections and a terminal object $1$. 
	\item $T$ has a universal monomorphism $I \xrightarrow{\text{true}}\Omega$, i.e. for any monomorphism of $T$, $A' \xrightarrow{\text{m}}A$ there exists a unique characteristic function
	such that the following diagram is a pull-back:
	\[
	\xymatrix{A'\ar[d]_{m}\ar[r] & 1\ar[d]^{true}\\
		A\ar[r]^{\chi_{m}} & \Omega
	}
	\]
	\item $T$ has for each object $X$ its power set $\Omega^{A}$; this
	is characterized by the fact that the morphisms $X\rightarrow\Omega^{A}$
	are precisely the subobjects of $X\times A$. In particular, its global
	sections $1\rightarrow\Omega^{A}$ are the subobjects of $A$. 
\end{itemize}

Stated in more homological terms, let $C,\mathcal{E}$ be two categories. A topos $\mathcal{T}(C;\mathcal{E})$ of $\mathcal{E}$-valued pre-sheaves on $C$ is the set of contra-variant functors from $C$ to $E$ that forms the set of objects of a category whose arrows are the natural transformations. 
A category $C$ embeds naturally in this topos if we associate the functor $Y \rightarrow C(Y, X)$
to $X$. This definition is sufficient in a finite context, since for discrete topology that provides a discrete site, every pre-sheaf is a sheaf. The complete notion of topos asks for a Grothendieck topology on a category and considers pre-sheaves \cite{Artin1964}. The three most common examples of topos are categories of sets, categories of functors $T(C^{op})$ for any small category $C$ and categories of sheaves on topological spaces. The generalization of topos with respect to usual set theory can be seen from the fact that the topos of sets are topos with two values of truth and the axiom of choice \cite{Lawvere2014}. Moreover, a topos satisfying AC is Boolean (\cite{Barr1985}.
One of the main consequences of the axioms of Topos is that the structure that generalizes the truth tables is a \textbf{Heyting algebra}, a constructive generalization of Boolean algebra. Heyting algebra replaces the Boolean complement by a constructive pseudo-complement. A Heyting algebra $\mathcal{H}$ is a bounded lattice such that for all $X$ and $Y$ in $\mathcal{H}$ there is a greatest element $Z$ of $\mathcal{H}$ such that:

\begin{equation}\label{Heyting Algebra}
X \wedge Z \leq Y 
\end{equation}

The element $Z$ is called the relative pseudo-complement of $X$ with respect to $Y$ and is denoted $X \rightarrow Y$. 
The \textbf{pseudo-complement} of any element $X$, noted with the negation $\neg X$, is defined as $\neg X= X \rightarrow 0 $ (this definition of negation implements the fundamental principle of non-contradiction). A pseudo-complement $\neg X$ is a complement $\bar{X}$ if $X \wedge \neg X = 0$ and $X \vee \neg X  = 1$. A Boolean algebra is a Heyting algebra in which for all elements $Y$ we have the equality $(\neg Y \vee Z)=(Y\Rightarrow Z)$. The lattice of an open set of a topological space is a typical example \cite{Diaconescu1975}. \\

\textbf{Probability multivalued logic.} Doering and Isham  \cite{Doering2008} proposed to provide a foundation of physics based on topos theory, and further developed a framework to interpret both quantum and classical probabilities with a multivalued logic \cite{Doering2012}. Independently and with a different construction, Baudot, Bennequin and Vigneaux proposed information topos on the (pre)-sheaf (site) of probabilities where conditioning provides the arrows \cite{Baudot2015,Baudot2015a,Vigneaux2017} (the two constructions were respectively introduced to each other and presented at the conferences "Categories and Physics 2011" in Paris).  
It is possible to illustrate the multiple truth values logic of probability in some simple elementary examples which further underline that where set theory could be considered a deterministic theory, topos theory may be conceived as a non-deterministic extension of it. Probability values are taken as valuations or truth-values. Simpson developed the general case where every measure is considered as a $\sigma$-continuous valuation \cite{Simpson2012}. It means that the usual boolean tables for meet and join are replaced by their homolog in probability which is continuous (for real-valued probability), a long table of a continuum of truths, instead of binary. To obtain some understandable elementary examples, we need to introduce integer partitions and consider a  rational field of probability, such that probabilities take values in the rational numbers $\mathbb{Q}$ and are given by the basic and usual empirical ratio $n/m$, as described by Kolmogorov (cf. Tapia and colleagues \cite{TapiaPacheco2017} and Baudot and colleagues \cite{Baudot2018}). First we recall the usual Boolean operator tables, of the operators joint and meet, for example: \\
\begin{tabular}{|c|c|c|}
	\hline 
	$X$  & $Y$  & $X\wedge Y$ \tabularnewline
	\hline 
	\hline 
	$\top$  & $\top$  & $\top$ \tabularnewline
	\hline 
	$\top$  & $\bot$ & $\bot$\tabularnewline
	\hline 
	$\bot$ & $\top$ & $\bot$\tabularnewline
	\hline 
	$\bot$ & $\bot$ & $\bot$\tabularnewline
	\hline 
\end{tabular}
\begin{tabular}{|c|c|c|}
	\hline 
	$X$  & $Y$  & $X\vee Y$ \tabularnewline
	\hline 
	\hline 
	$\top$  & $\top$  & $\top$ \tabularnewline
	\hline 
	$\top$  & $\bot$ & $\top$\tabularnewline
	\hline 
	$\bot$ & $\top$ & $\top$\tabularnewline
	\hline 
	$\bot$ & $\bot$ & $\bot$\tabularnewline
	\hline 
\end{tabular}\\
We rewrite those tables with 0 replacing $\bot$ (contradiction, "false") and 1 replacing $\top$ (tautology, "true") in a matricial form, giving us:\\ 
\begin{tabular}{|c|c|c|}
	\hline 
	$P(X\vee Y)$  & $P(X)=0$  & $P(X)=1$ \tabularnewline
	\hline  
	$P(Y)=0$  & $0$  & $1$ \tabularnewline
	\hline 
	$P(Y)=1$  & $1$ & $1$  \tabularnewline
	\hline 	
\end{tabular}
\begin{tabular}{|c|c|c|}
	\hline 
	$P(X\wedge Y)$  & $P(X)=0$  & $P(X)=1$ \tabularnewline
	\hline  
	$P(Y)=0$  & $0$  & $0$ \tabularnewline
	\hline 
	$P(Y)=1$  & $0$ & $1$  \tabularnewline
	\hline 	
\end{tabular}\\

Such logic is a finite deterministic case of logic, which in terms of probability follows a finite 0-1 law, the smallest probability field with two elements $E$ and $\emptyset$ described by Kolmogorov which corresponds in what follows to the case $m=2$ with its singleton partition $\{2\}$. For non-deterministic probability logic, a truth table is given for each integer partition of $m$, the integer number of observations (also called repetitions, trials, sample size). In the following example of  $\vee$ and $\wedge$ operator tables, we consider $m=2$ and $m=3$ and the integer partition of $2$ $\{1,1\}$ and the integer partition of $3$ $\{1,2\}$, such that we have $1/2+1/2=1$ and $1/3+2/3=1$ respectively. For $m=2=1+1$, we have: \\
\begin{tabular}{|c|c|c|c|c|}
	\hline 
	$P(X\vee Y)$  & $P(X)=0$  & $P(X)=1/2$ & $P(X)=1/2$  & $P(X)=1$ \tabularnewline
	\hline  
	$P(Y)=0$  & $0$  & $1/2$ & $1/2$ & $1$ \tabularnewline
	\hline 
	$P(Y)=1/2$  & $1/2$  & $1/2$ & $1/2$ & $1$ \tabularnewline
	\hline 
	$P(Y)=1/2$  & $1/2$  & $1/2$ & $1/2$ & $1$ \tabularnewline
	\hline 
	$P(Y)=1$  & $1$ & $1$ & $1$ & $1$  \tabularnewline
	\hline 	
\end{tabular}\\
\begin{tabular}{|c|c|c|c|c|}
	\hline 
	$P(X\wedge Y)$  & $P(X)=0$ & $P(X)=1/2$ & $P(X)=1/2$  & $P(X)=1$ \tabularnewline
	\hline  
	$P(Y)=0$  & $0$ & $0$ & $0$ & $0$ \tabularnewline
	\hline 
	$P(Y)=1/2$  & $0$ & $1/2$ & $1/2$ & $1/2$  \tabularnewline
	\hline 
	$P(Y)=1/2$  & $0$ & $1/2$ & $1/2$ & $1/2$  \tabularnewline
	\hline 
	$P(Y)=1$  & $0$ & $1/2$ & $1/2$ & $1$  \tabularnewline
	\hline 	
\end{tabular}\\

\begin{figure} [!h]
	\centering
	\includegraphics[height=2.3cm]{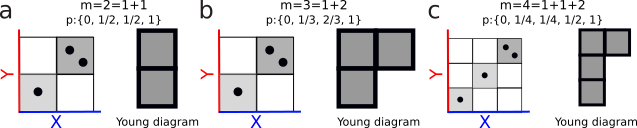}
	\caption{\textbf{Examples of probability integer partition. a,} the case $m=2=1+1$, representation in data space of two variables $X,Y$ (left), and the associated Young diagram (right). \textbf{b,} the case $m=3=1+2$. \textbf{c,} the case $m=4=1+1+2$.  }
	\label{figurrigthe_Supp_partition lattice}
\end{figure}

For $m=3=1+2$, we have: \\
\begin{tabular}{|c|c|c|c|c|}
	\hline 
	$P(X\vee Y)$  & $P(X)=0$  & $P(X)=1/3$ & $P(X)=2/3$  & $P(X)=1$ \tabularnewline
	\hline  
	$P(Y)=0$  & $0$  & $1/3$ & $2/3$ & $1$ \tabularnewline
	\hline 
	$P(Y)=1/3$  & $1/3$  & $1/3$ & $2/3$ & $1$ \tabularnewline
	\hline 
	$P(Y)=2/3$  & $2/3$  & $2/3$ & $2/3$ & $1$ \tabularnewline
	\hline 
	$P(Y)=1$  & $1$ & $1$ & $1$ & $1$  \tabularnewline
	\hline 	
\end{tabular}\\
\begin{tabular}{|c|c|c|c|c|}
	\hline 
	$P(X\wedge Y)$  & $P(X)=0$ & $P(X)=1/3$ & $P(X)=2/3$  & $P(X)=1$ \tabularnewline
	\hline  
	$P(Y)=0$  & $0$ & $0$ & $0$ & $0$ \tabularnewline
	\hline 
	$P(Y)=1/3$  & $0$ & $1/3$ & $1/3$ & $1/3$  \tabularnewline
	\hline 
	$P(Y)=2/3$  & $0$ & $1/3$ & $2/3$ & $2/3$  \tabularnewline
	\hline 
	$P(Y)=1$  & $0$ & $1/3$ & $2/3$ & $1$  \tabularnewline
	\hline 	
\end{tabular}\\
More rigorous and extended notations should be $P(X=a_i)=1/3$ instead of the  abbreviate $P(X)=1/3$, underlining the necessary introduction of random variables, also called observables, in the theory (see \cite{Baudot2015,Baudot2015a,Vigneaux2017,Baudot2015a,TapiaPacheco2017}; philosophically this assumes that there is no probability without an observer). An introduction to the world of partitions can be found in the work of Stanley \cite{Stanley2011} and Andrews \cite{Andrews,Andrews1998} and MacDonnald's book \cite{Macdonald1995}. These tables correspond to the usual joint and meet for events; notably, they obey the inclusion-exclusion theorem $P(X\vee Y)=P(X)+P(Y)-P(X\wedge Y)$. From a logical point of view they correspond to usual multivalued $G_m$  logic as introduced by G\"{o}del for which $P(X\vee Y)=\operatorname{max}(\{P(X),P(Y)\}$ and $P(X\wedge Y)=\operatorname{min}(\{P(X),P(Y)\}$ \cite{Godel1932} (see Gottwald for a review of many-valued logic \cite{Gottwald2004} and other operators). The generalization to more than 2 random variable multivariate cases can be achieved via tensorial logical tables. 
In general, we have the following theorem:
\begin{theorem}
	\textbf{(correspondence between integer partition and logical tables).} Let $(\Omega,\mathcal{F},P)$ be a finite probability space where $P$ is a finite (empirical) probability measure with sample size $m$, then the set of logical tables is in one to one correspondence with the set of integer partitions of $m$.      
\end{theorem}
This multiplicity of logic tables in the finite context reflects the multiplicity of logics exposed in the work of  Sorensen and Urzyczyn, which established that there is no single finite Heyting algebra that satisfies Soundness and Completeness \cite{Sorensen2006} (but they are however sufficient to preserve the semantics stated in theorem 2.4.8 \cite{Sorensen2006}). Unfortunately, the asymptotic limit of such logic is quite unknown, particularly hard, and is being investigated by Hardy and Ramanujan. Considering the construction of random reals with an inaccessible cardinal by Solovay, it appears natural to call these probability values finite/accessible rational random rationals. However, what follows suggests that there exist several ways to complete such discrete random field to the continuous field, namely euclidean and p-adic completion, following  Ostrowski's theorem. 
Regardless, we hence leave off here from a trail of an elementary probabilistic logic proposed to be relevant for biological structures, cognition, consciousness and physics.  

\subsubsection{Information functions and set theory}\label{info_functions}

Firstly we need to restate the usual functions of information established by Shannon \cite{Shannon1948} and Hu kuo Ting \cite{Hu1962}, specifically those used in this review: 
\begin{itemize}
	\item Shannon-Gibbs entropy of a single variable $X_j$ is defined by \cite{Shannon1948}: 
	\begin{equation}\label{singleentropy}
	H_1=H(X_{j};P_{X_j})=k\sum_{x \in [N_j] }p(x)\ln p(x)=k\sum_{i=1}^{N_j}p_i\ln p_i
	\end{equation}
	where $[N_j]=\{1,...,N_j\}$ denotes the  alphabet of $X_j$.
	
	\item Joint entropy is defined for any joint-product of $k$ random variables  $(X_1,...,X_k)$ and for a probability joint-distribution $\mathbb{P}_{(X_1,...,X_k)}$  by \cite{Shannon1948}:
	\begin{equation} \label{jointentropy multiple}	
	H_k  = H(X_{1},...,X_{k};P_{X_{1},...,X_{k}})  =  k\sum_{x_1,...,x_k\in [N_1\times...\times N_k]}^{N_1\times...\times N_k}p(x_1.....x_k)\ln p(x_1.....x_k) 
	\end{equation}
	where $[N_1\times...\times N_k]=\{1,...,N_j\times...\times N_k\}$ denotes the  alphabet of $(X_1,...,X_k)$.
	
	\item The mutual information of two variables $X_{1},X_{2}$ is defined as \cite{Shannon1948}: 
	\begin{equation}\label{mutual information}
	I(X_{1};X_{2};P_{X_{1},X_{2}})=k\sum_{x_1,x_2\in[N_1\times N_2]}^{N_1\times N_2}p(x_1.x_2)\ln \frac{p(x_1)p(x_2)}{p(x_1.x_2)}
	\end{equation}
	It can be generalized to k-mutual-information (also called co-information) using the alternated sums given by equation \ref{Alternated sums of information}, as originally defined by McGill \cite{McGill1954} and Hu Kuo Ting \cite{Hu1962}, giving:
	\begin{equation}\label{n-mutual information}
	I_k=I(X_{1};...;X_{k};P)=k\sum_{x_1,...,x_k\in [N_1\times...\times N_k]}^{N_1\times...\times N_k}p(x_1.....x_k)\ln \frac{\prod_{I\subset [k];card(I)=i;i \ \text{odd}} p_I}{\prod_{I\subset [k];card(I)=i;i \ \text{even}} p_I} 
	\end{equation}
	For example, 3-mutual information is the function: 
	\begin{equation}\label{3-mutual information}
	I_3=k\sum_{x_1,x_2,x_3\in[N_1\times N_2\times N_3]}^{N_1 \times N_2\times N_3}p(x_1.x_2.x_3)\ln \frac{p(x_1)p(x_2)p(x_3)p(x_1.x_2.x_3)}{p(x_1.x_2)p(x_1.x_3)p(x_2.x_3)}
	\end{equation}
	For $k\geq3$, $I_k$ can be negative \cite{Hu1962}. 	
	
	\item  The  total correlation introduced by Watanabe \cite{Watanabe1960}, called integration by Tononi and Edelman \cite{Tononi1998} or  multi-information by Studen\'{y} and  Vejnarova \cite{Studeny1999} and which we note $C_k(X_1;...X_k;P)$, is defined by:
	\begin{equation}\label{total correlation}
	\begin{split}
	C_k &= C_k(X_1;...X_k;P)=\sum_{i=1}^k H(X_i) - H(X_1;...X_k)=\sum_{i=2}^{k}(-1)^{i}\sum_{I\subset [n];card(I)=i}I_i(X_I;P)\\
	& =k\sum_{x_1,...,x_k\in[N_1\times...\times N_k]}^{N_1 \times ...\times N_k}p(x_1....x_k)\ln \frac{p(x_1...x_k)}{p(x_1)...p(x_k)}
	\end{split}
	\end{equation}
	For two variables the total correlation is equal to mutual-information ($C_2=I_2$). The total correlation has the pleasant property of being a relative entropy between marginal and joint variables and hence of always being non-negative.  
	
	\item The conditional entropy of $X_{1}$ knowing (or given) $X_{2}$ is defined as \cite{Shannon1948}:
	\begin{multline}\label{conditionalentropy}
	X_{2}.H_{1}= H(X_{1}|X_{2};P)=k\sum_{x_1,x_2\in[N_1\times N_2]}^{N_1*N_2}p(x_1.x_2)\ln p_{x_2}(x_1) \\
	=  k \sum_{x_2\in\mathscr{X}_2}^{N_2}p(x_2). \left( \sum_{x_1\in\mathscr{X}_1}^{N_1}  p_{x_2}x_1  \ln p_{x_2}x_1 \right)
	\end{multline}
	Conditional joint-entropy, $X_3.H(X_1,X_2)$ or $(X_1,X_2).H(X_3)$, is defined analogously by replacing the marginal probabilities with the joint probabilities.  
	
	\item The conditional mutual information of two variables $X_{1},X_{2}$ knowing a third $X_3$ is defined as \cite{Shannon1948}: 
	\begin{equation}\label{conditional mutual information}
	X_3.I_2=I(X_{1};X_{2}|X_3;P)=k\sum_{x_1,x_2,x_3\in [N_1\times N_2\times N_3]}^{N_1\times N_2\times N_3}p(x_1.x_2.x_3)\ln \frac{p_{x_3}(x_1)p_{x_3}(x_2)}{p_{x_3}(x_1,x_2)} 
	\end{equation}
\end{itemize}
The chain rules of information are (where the hat denotes the omission of the variable):
\begin{equation}\label{chain rule gener}
H(X_1;...;\widehat{X_i} ;...;X_{k+1};P)  = H(X_1;...;X_{k+1};P) - (X_1;...;\widehat{X_i} ;...;X_{k+1}).H(X_i;P) 
\end{equation}
That can be written in short as $H_{k+1} - H_k = (X_1,...X_k).H(X_{k+1})$
\begin{equation}\label{chain rule gene infomut}
I(X_1;...;\widehat{X_i} ;...;X_{k+1};P) = I(X_1;...;X_{k+1};P) + X_i.I(X_1;...;\widehat{X_i} ;...;X_{k+1};P) 
\end{equation}
That can be written in short as $I_{k-1} - I_k = X_k.I_{k-1}$, generating the chain rule \ref{chain rule gener} as a special case.
We have $I_1=H_1$. We have the alternated sums or inclusion-exclusion rules \cite{Hu1962,Matsuda2001,Baudot2015a}: 
\begin{equation}\label{Alternated sums of entropy}
H_n(X_1,...,X_n;P)=\sum_{i=1}^{n}(-1)^{i-1}\sum_{I\subset [n];card(I)=i}I_i(X_I;P)
\end{equation}
\begin{equation}\label{Alternated sums of information}
I_n(X_1;...;X_n;P)=\sum_{i=1}^{n}(-1)^{i-1}\sum_{I\subset [n];card(I)=i}H_i(X_I;P)
\end{equation}
For example: $H_3(X_1,X_2,X_3)=I_1(X_1)+I_1(X_2)+I_1(X_3)-I_2(X_1;X_2)-I_2(X_1;X_3)-I_2(X_2;X_3)+I_3(X_1;X_2;X_3)$

\begin{figure}
	\centering
	\includegraphics[height=5cm]{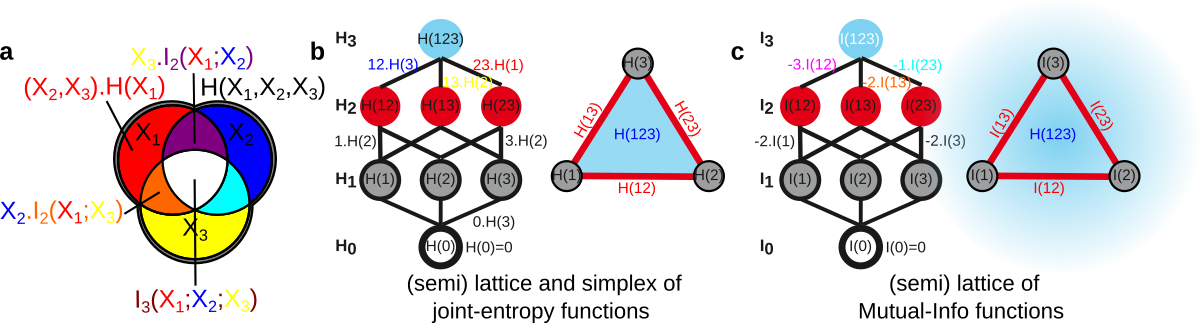}
	\caption{\textbf{Naive set-theoretic and lattice representation of information function. a,} Venn diagram representation of the various information functions justified by the theorem of Hu Kuo Ting \cite{Hu1962} (see text). \textbf{b,} semilattice of joint-entropy with conditional entropies (coface map in simplicial sets \cite{Weibel1995}) implementing the chain rule, for example using the abbreviated notations $H(12)=H(1)+1.H(2)$.A corresponding simplicial representation of these joint entropies which can be easily realized by the non-negativity properties of conditional entropies. \textbf{c,} semilattice of mutual informations with conditional mutual informations implementing the chain rule, for example using the abbreviated notations $I(123)=I(12)-3.I(12)$. Since the $I_k$ can be negative for $k\geq3$, there is no obvious corresponding simplicial representation; see Yeung for more details \cite{Yeung2003,Yeung2007,Yeung1997}. }
	\label{venn_diag}
\end{figure}

\textbf{Hu Kuo Ting and Yeung theorem:} The theorem of Hu Kuo Ting and Yeung \cite{Hu1962}\cite{Yeung2007} establishes a bijection between information functions and finite additive (measurable) functions, for which the set theoretic operators $\cup,\cap,/$ correspond to Joint $(;)$, Mutual $(,)$ and conditional $(/)$ information operation respectively. This important theorem has been neglected in information theory for some time and rediscovered independently in a more superficial form many times within the community. \\
\textbf{Hu Kuo Ting - Yeung theorem:} For a given sequence of variables $X_{1},X_{2}...$ and their distribution $P$ there exist a corresponding sequence of sets $A_{1},A_{2}...$ and an additive function $\varphi$ on the ring $\mathbb{U}$ generated by the sequence $A_{i},\;i=1,2,...$, such that: $H(Q(X_{i_{1}},X_{i_{2}},...,X_{i_{n}}))=\varphi(Q(A_{i_{1}},A_{i_{2}},...,A_{i_{n}})$
for all collections of variables, $X_{i_{1}},X_{i_{2}},...,X_{i_{n}}$,
and all operations, $Q$ denoting a symbol generated by a finite number of operations $\cup,\cap,/$.\\

Csisz\'{a}r and K\"{o}rner have proposed an alternative "geometric" proof of Hu \cite{Csiszar2011} and also suggested the converse correspondence of additive functions with information functions by way of symmetric argument.
Hu's theorem and its converse establish\textbf{a bijection between additive functions and information functions}, which is a deep equivalence between set and measure theory and information theory; any additive function can be written in term of information and vice versa. Figure \ref{venn_diag} illustrates the consequence of this theorem, allowing a naive handling of information functions with Venn diagrams and supporting the simplicial-boolean lattice decompositions studied by Shannon \cite{Shannon1953} and Han \cite{Han1975}.  
One can estimate the importance of such a theorem with regard to integration and measure theory. Considering the axiomatic setting of Solovay, that is, a set theory without the axiom of choice in which all sets are measurable, the universality of information functions appears justified. Considering Solovay's axiomatic system, any function is measurable and information functions are hence in bijection with all functions. The universality of function has already appeared in the context of Riemann zeta functions \cite{Karatsuba1992,Voronin1975}, which are related to polylogarithms. \\
From the algebraic point of view, Hu Kuo Ting's first theorem of his 1962 paper on "information quantities" establishes explicitly that \textbf{information functions are the set of finitely additive functions on the ring of random variables}. This result justifies the consideration of information functional modules on random-variable modules and the information cohomology constructed in \cite{Baudot2015a,Vigneaux2017} can be understood as a generalization of this result. His first theorem therefore supersedes and condenses many of the results on information that were found a posteriori.    

\subsubsection{The information of a formula/thought}

In this section, we investigate whether a mathematical formula has an information content that could be quantified.
As previously discussed, Denef and Loeser proposed a formalism based on motivic measures that give a field independent measure of mathematical formula \cite{Hales2005}. Here, we propose an informational version and the possibility, based on the probabilistic logic formulation we have presented, of considering any thought as a mathematical formulation; the mathematical nature of thoughts, an idea which still has life in it. In this section, we revisit G\"{o}del's arithmetic coding procedure in a finite probabilistic context and show that Shannon's entropy decodes and assigns a (real-valued) information measure to finite mathematical formulae. Unlike in the deterministic case, for which the information (complexity) of a string is in general not computable, the entropy of a probabilistic object can be computed. Kolmogorov defined the Algorithmic information or complexity, $K(X)$, of an object (a string $X\in{0,1}^*$) to be the length of the shortest program that produces (print) the object and halts on a given universal Turing machine $T$:
$K(X)=\min \{ |p| :  C_T ( p )= X\}$,
where $|p|$ denotes the length of the program in bits, $C_T: {0,1}^*\rightarrow{0,1}^*$ is a partial recursive function (that is computed by the Turing machine $T$) and $C_T ( p )$  is the result of running the program $p\in{0,1}^*$ on the Turing machine $T$. Zvonkin and Levin \cite{Zvontin1970a} showed that Shannon entropy of binary iid variables equals the averaged randomness-complexity $K$ in the limit of infinitely long strings (see Th. 5.18 for a precise statement on this \cite{Zvontin1970a}). 
The fundamental theorem of arithmetic (the unique-prime-factorization theorem of Euclid) states that any integer greater than 1 can be written as a unique product (depending on the ordering of its factors) of prime numbers (see Hardy and Wright \cite{Hardy1979}). We write any integer $n$ as its prime decomposition, called its standard form:
\begin{equation}
n=p_1^{\alpha_{1}}p_2^{\alpha_{2}}...p_k^{\alpha_{k}},~(\alpha_{1}>0,\alpha_{2}>0,...,\alpha_{k}>0,p_1<p_2<...<p_k)
\end{equation}
\begin{equation}
\forall n\in\mathbb{N},~n>1,\;n=\prod_{p~\text{prime}}^{\infty}p^{\alpha_{p}}
\end{equation}
where $\alpha_{p}\in\mathbb{N}$ is a natural integer coefficient depending on the prime $p$. Including 1 implies the loss of uniqueness, since the prime factorization of 1 contains 0 exponents ($1=2^0.3^0.5^0...=3^0$), and if we allow zero exponents, the factorization ceases to be unique.   
A standard method of extending the fundamental theorem of arithmetic to rational numbers is to use the p-adic valuation of $n$, noted $v_p(n)$, to ascribe the exponent $v_p(n)=\alpha_{p}$ to all prime numbers in the product and to then give an exponent $v_p(n)=0$ to those that do not divide $n$. The decomposition into prime factors of rational numbers requires considerations of the possibly negative exponents $\alpha_{p}\in \mathbb{Z}$ and $v_p(\frac{n'}{m})=v_p(n')-v_p(m)$, the so-called p-adic norm (see Khrennikov and Nilson \cite{Khrennikov2004} for definitions and non-deterministic dynamic applications), giving this representation of a rational number 
\begin{equation}
n=\frac{n'}{m}=2^{v_2(n)}3^{v_3(n)}...p_k^{v_k(n)},~(v_p(n)\in \mathbb{Z}, ~ p_1<p_2<...<p_k)
\end{equation}
\begin{equation}
\forall n\in\mathbb{Q},\;n=\frac{n'}{m}=\prod_{p~\text{prime}}^{\infty}p^{v_p(n)}
\end{equation}
, and every rational number as a unique prime factorization. \\

\textbf{G\"{o}del code :} We will firstly introduce G\"{o}del's logic and methods. The relation between indeterminism, uncertainty and logical undecidability has been a leitmotiv of many works. G\"{o}del's approach was called the arithmetisation program of logic. The basic hypothesis of G\"{o}del is based on the fact that the formula of a formal system can be viewed as finite sequences of basic symbols (variables, logical constants, and parentheses or separators), allowing one to define which sequences of these basic symbols are syntactically correct formula (or not) and, from this, which finite sequences of formula provide correct proofs (or not). To do so he designed a numbering-code that assigns bijectively a natural integer to each sequence \cite{Goedel1931}. Hence, a formula is a finite sequence of natural numbers, and a proof schema is a finite sequence of finite sequences of natural numbers. G\"{o}del could then prove that, under this hypothesis, there exist some theorems that are independent and which can neither be proved or disproved. To do so he defined a map from logical statements, that is, any sequence of mathematical symbols, to natural numbers, which further allows deciding whether logical statements can be constructed or not. Given any statement, the number it is converted to is called its G\"{o}del number, defined by:   
\begin{equation}
\alpha(x_{1},x_{2},x_{3},\dots,x_{n})=2^{x_{1}}.3^{x_{2}}.5^{x_{3}}...p_{n}^{x_{n}}
\end{equation} 
In the original work, the first $12$ "powers" are occupied by basic symbols, and the numerical variable occupies the powers
$p\geq13$ \cite{Nagel1959}:\\
\begin{tabular}{|c|c|c|}
	\hline 
	G\"{o}del Number & Symbol & meaning\tabularnewline
	\hline 
	\hline 
	1 & $\neg$  & not\tabularnewline
	\hline 
	2 & $\vee$  & or\tabularnewline
	\hline 
	3 & $\supseteq$  & if ... then (implication)\tabularnewline
	\hline 
	4 & $\exists$  & There exist\tabularnewline
	\hline 
	5 & $=$  & equals\tabularnewline
	\hline 
	6 & $0$  & zero\tabularnewline
	\hline 
	7 & $s$  & the immediate successor of\tabularnewline
	\hline 
	8 & $($  & left parenthesis\tabularnewline
	\hline 
	9 & $)$  & right parenthesis\tabularnewline
	\hline 
	10 & $,$  & comma\tabularnewline
	\hline 
	11 & $+$  & plus\tabularnewline
	\hline 
	12 & $\times$  & times\tabularnewline
	\hline 
\end{tabular} \\
For example, the formula $x_{1}=x_{1}$ is coded by the G\"{o}del number
$\alpha(13,5,13,0,0,\dots,0)=2^{13}.3^{5}.5^{13}7^{0}...p_{n}^{0}$,
and $(\exists x_{1})(x_{1}=sx_{2})$ is coded by the G\"{o}del number 
$\alpha(8,4,13,9,8,13,5,7,17,9,0,0,\dots,0)$. This "function" sends every formula (or statement that can be formulated in a theory) into a unique number, in such a way that it is always possible to retrieve the formulas from the G\"{o}del numbers, but also to say whether an arbitrary number corresponds to a G\"{o}del number. The G\"{o}del number of the sequence $(x_{1},x_{2},x_{3},\dots,x_{n})$ is more generally called a pairing function, noted $f(x,i)=x_{i}.$. $i$ is always in the range of $1,\ldots,n$ (and in the previous case  the indices correspond to the labels of the primes).\\

Now that we have introduced G\"{o}del's arithmetic coding, we can apply his method to rational (empirical) probabilities fields and show that the Shannon entropy function is a "decoding function" that sends any number back to its formula with a one to one correspondence.
We first define an extended G\"{o}del number as the p-adic norm $v_p(\frac{n'}{m})$ and identify its value as G\"{o}del did and as summarised in the table above, the only difference being that we now dispose of negative integers in order to facilitate the code.  
\begin{theorem}
	\textbf{(Fundamental theorem of arithmetic Information)} Let $H(X,P_{\mathbb{Q}})$
	be the information function over a rational probability field $P_{\mathbb{Q}}$. Then:
	\begin{equation}
	H(X;P_{\mathbb{Q}})=-\sum_{p ~ \text{prime}}v_p(n)\log p
	\end{equation}  
	where $v_p(n)\in\mathbb{\mathbb{Z}}$ is a relative integer coefficient depending on the prime $p$ $v_p(n)=v_p(\prod_{i=1}^n p(x_{i})^{p(x_{i})})$.
\end{theorem}
Proof: the probabilities over the rational field $P_{\mathbb{Q}}$, and an elementary probability $p_{j}$, which can be written according to the fundamental theorem of arithmetic (for readability noting a prime with $q$ symbol):
\begin{equation}
p_{j}=\frac{n'}{m}=\prod_{q~\text{prime}}q^{v_q(p_{j})}
\end{equation}
where  $0<p_{j}\leq1$ and $\sum_{j=1}^{n}p_{j}=1$, and $v_q(p_{j})\in \mathbb{Z}$ are relative integer coefficients depending on the prime $q$. Entropy function $H(X;P_{\mathbb{Q}})$ is, according to Shannon's axiom, a continuous function of the $p_i$  and can be written in the form: 
\begin{equation}
H(X;P_{\mathbb{Q}})=k\sum_{i=1}^n p(x_{i})\log p(x_{i})=-\log\prod_{i=1}^n p(x_{i})^{p(x_{i})}
\end{equation}
It follows from elementary algebra that $p(x_{i})^{p(x_{i})}$ has a prime decomposition with relative integer exponents, and hence the theorem. $\Box$

This theorem applies to any information function $H(Q(X_{1},X_{2},...,X_{n}),\mathbb{P_{\mathbb{Q}}})$ as defined by Hu Kuo Ting \cite{Hu1962}, namely, joint-entropies, mutual informations, conditional entropies and conditional mutual informations (see also \cite{Baudot2018}), as they are linear combinations of entropies. Notably, considering all information functions, since mutual information can be negative for k variables, $k>2$, the set of information values is in one to one correspondence with $\mathbb{Z}$. Such bijection can only be achieved by considering a variable space of at least 3 dimensions. We hence have established that the following corollary:\\
\textbf{Corollary - Information-G\"{o}del code : } $H(Q(X_{1},X_{2},...,X_{n}),\mathbb{P_{\mathbb{Q}}})=h(v_{q}(n),q)$
is a G\"{o}del code.

\begin{figure} [!h]
	\centering
	\includegraphics[height=12cm]{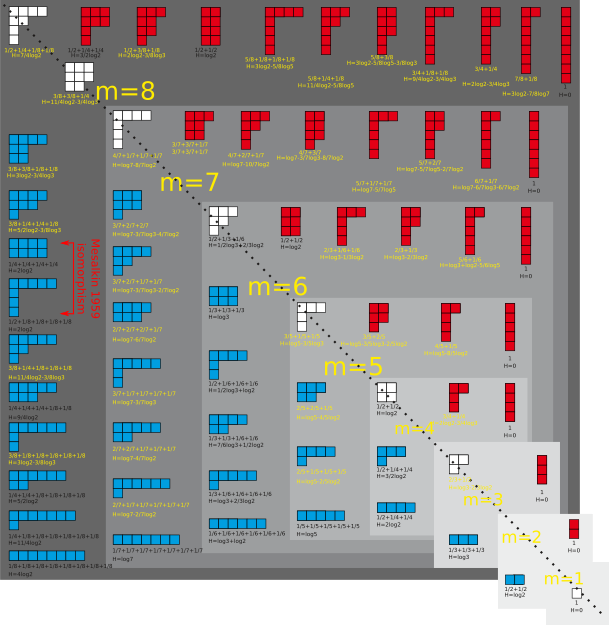}
	\caption{\textbf{The smallest empirical probability fields for a number of total observations $m=2,3,4,5,7,8$} represented using Young's diagram of the associated partitions. The associated entropy $H$ is written below each partition in its prime decomposition form. Note that entropy is an increasing function from right to left and top to down, and that the two equal entropies for $m=8$ correspond to isomorphic partitions according to a theorem of Mesalkin (1959 \cite{Ornstein1970}), which is a special case of Kolmogorov-Ornstein Isomorphism theorem  $(1/4,1/4,1/4,1/4)\approx (1/2,1/8,1/8,1/8,1/8)$. See section \ref{Geometries-Homeostasis}. (Figure adapted and modified from the work of R. A. Nonenmacher (CC)) }	
	\label{integer_partition}
\end{figure}
Figure 	\ref{integer_partition} gives the Young diagram for all integer partitions of $m$ with $m=2,3,4,5,7,8$ as previously associated with the associated logical tables. 
The two partitions of 8 which have the same entropy $(1/4,1/4,1/4,1/4)\approx (1/2,1/8,1/8,1/8,1/8)$ are isomorphic according to the theorem of Mesalkin (1959 \cite{Ornstein1970}, a special case of Kolmogorov-Ornstein Isomorphism theorem (see section \ref{Geometries-Homeostasis}), and their associated tables and logic shall be considered as equivalent.

This is just a preliminary result on the elementary logic of information; the characterization of this logic lies beyond what has been so far been achieved. Notably, more work needs to be done involving the consideration of the elementary context of integer partition probabilities, introduced in the previous section, and the extension to negative coding values that offers a possibly richer logic. The results so far also provide some original geometrical insight into logic and probability, allowing the future study of mathematical formula with Euclidean and p-adic metrics \cite{Khrennikov2004}. 
In constructive logic, the implication $\Rightarrow$ is a power-like operator and provides a "direction" to logical flow; it would be interesting to investigate such directional flow from the irreversible and causal point of view of thermodynamics. \\
There is another question as regards statistical independence and independence of axioms. The undecidability of a proposition $X$ in a theory $\Omega$, suggesting that $X$ is independent of the other proposition $Y$ in $\Omega$, could  correspond to independence in probability such that it would be possible to say that $X$ being independent in $\Omega$ is "equivalent" to $P_Y(X)=P(X)$, or that the joint theories associated with $X$ and $Y$ factorize $P(Y.X)=P(X)P(Y)$ in $\Omega$. In such a case the additivity or subadditivity of the information decoding function quantifies the independence or dependence of the propositions in the theory $\Omega$ (in a topological sense). In more up-to-date terms, Cohen's technique of forcing may have a simple probabilistic analog. In a very pragmatic empirical sense, a mathematical theory is also a human belief, stated and written by humans (or machines) implementing human beliefs. If one  considers probabilistic-information theory as the relevant model for cognition, then there exists a probabilistic theory of mathematics that encompasses mathematicians' entire mathematical product. 

\subsection{Homology, the shapes and the language of perception}

\subsubsection{Homological history and cognition}

Topology is the science that characterizes objects by the relation interactions of their components. It is the domain of mathematics dedicated to shapes or patterns which classifies them by identifying their invariants. Here we give a little historical taste of what topology is or could be, highlighting its cognitive aspects and motivations, already made explicit by its original founders. Some historical reviews of the different aspects of topology, i.e. algebraic and differential, of topology can be found in the work of Milnor \cite{Milnor2011}, of Weibel \cite{Weibel1999} and of Dieudonn\'{e} \cite{Dieudonne1989}. Topology was first born under the name of \textbf{Analysis Situs}, notably in the writings of Leibniz \cite{Leibniz1679}. Analysis Situs is inseparable from all his further work, his quest for a universal characteristic that first took form in differential calculus, on a \textbf{qualitative geometry}, consisting in a language allowing one to "algebraically-logically" manipulate geometrical figures. Leibniz's model was a universal cognitive and consciousness model; he developed the concept of the \textbf{monad}, which is at the same time a physical substance and a semantic qualia unit element; monads are exact, irreducible, real and perfect \cite{Leibniz1714}. Monads, according to Liebniz, can compose hierarchically forming new monads inheriting properties from the originals, and the structure and algebra ruling them can be conceived of as the analysis situs. They are physical atoms of knowledge and of sensations, a monist view contrasting with usual mind-body dualism. Hence, in Leibniz's view, the whole world is perfect, optimal. One can still recognize Leibniz's view in modern studies of monads, also called triples by Barr and Wells \cite{Barr1985}. Leibniz's view is indeed still at work in what is presented here, notably his mind-body model, \textbf{monist and pluralist}, physical and mathematical, and his idea of perfection, which when re-expressed in probabilistic modern terms, although optimistic, sounds much more generally like a basic hope or expectancy. Leibniz's opus is also at the roots of information theory in the form of binary calculus, and Uchii recently proposed an extended work on monadology, information and physics \cite{Uchii2015,Uchii2014,Uchii2014a}.
After Liebniz, Euler made a contribution by solving the 7 bridges problem and defining his "characteristic" $\chi$ , the main topological invariant that Leibniz was unable to find \cite{Euler1759}. Betti and Riemann, following Abel, developed the foundations of homology by classifying surfaces \cite{Riemann1857}, then Poincar\'{e} introduced with his analysis situs most of the basic theorems and concepts in the discipline \cite{Poincare1895a}. Topology was born as geometry cleaned of its "unnecessary" undecidable axioms, a geometry without metric assumptions. It was notably conceived in Poincar\'{e} work and theorems (such as uniformization theorem and his conjecture) to maintain geometrical unity in mathematics as a consequence of the discovery of the existence of geometrical diversity, i.e. legitimate non-Euclidian geometries, the diversity of geometry. Poincar\'{e} directly related analysis situs to the cognitive process of constructing a continuous space from discrete experience, and even proposed to explain the Weber-Fechner law of sensory adaptation on this basis as reported in appendix \ref{topology of psychophysic}. This obviously constitutes the first mathematical model of perceptual adaptation, explicitly topological, more than a century ago. Since then many homology theories have appeared \cite{Dieudonne1989}, each characterizing different, more or less general mathematical structures. However, these theories appeared to have analog structures and the work of unifying them began in the second half of the 20th century. The working principle of homology theory followed by Eilenberg, Maclane and Grothendieck has been to "dig deeper and find out", such that homology theory has continued to define new, more general and enormous homologies, generalizing simplicial homology by (the equivalent) singular homology, then by homological algebras, then by topos, and then by conjectural motives, introducing categories and functors as central working tools (see Cartier's review \cite{Cartier2001} and Eilenberg's biographical memoir \cite{Bass2000}). The result generalizes them to differential Lie, associative algebra, arithmetic, etc. The simple consideration of the swathes of mathematics which are concentrated under the little names of functor Ext and Tor is sufficient to illustrate the principle of cognitive process proposed here, namely that "understanding is compressing". 
In the original view of Grothendieck, the ascension towards unified-general cohomology theory followed 3 steps: schemes, topos, and finally motive theory \cite{Cartier2001}. The aim of motivic cohomology is to nevertheless handle geometry and algebra equivalently, but also number theory: notably, one aim was to solve an algebraic subcase of Riemann's conjecture, the Weil conjecture. The structure of this general cohomology became progressively more clear, notably thanks to the work of Beilinson, Bloch and Goncharov. Voevodsky, following an original approach, proposed a formalisation of motivic cohomology based on triangulated categories.

\subsubsection{Groups and action: ambiguity and uncertainty according to Galois}

\begin{figure} [!h]  	
	\centering
	\includegraphics[height=10cm]{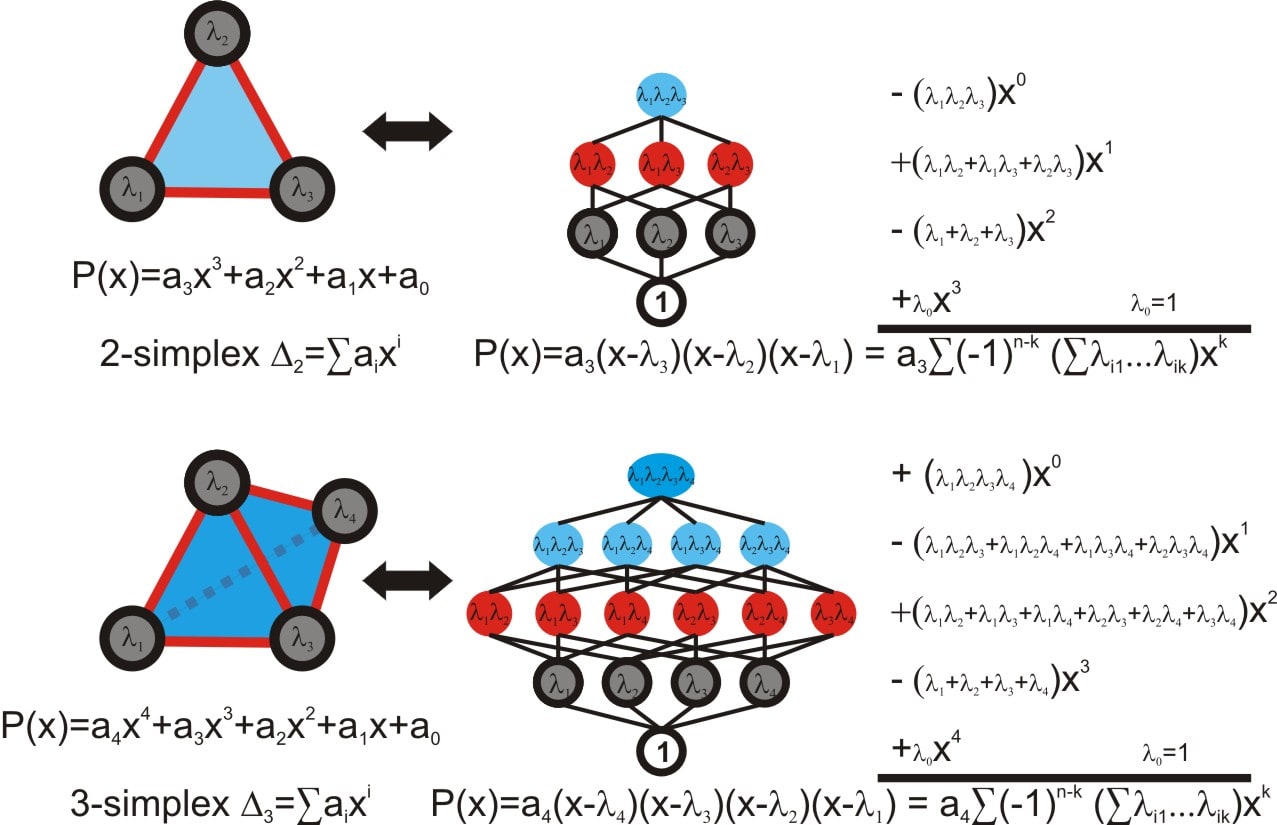}
	\caption{Example of degree 3 (top) and 4 (bottom) polynomial univariates (in one variable) in their additive (left) and multiplicative-factorized forms (right), together with their corresponding simplex.}
	\label{figure_roots}
\end{figure}

Following Poincar\'{e} but also Grothendieck, and exaggerating a little as they did, one could say that topology is the story of group, a "Long March through Galois theory" \cite{Grothendieck1997}. Group theory originates in the study by Galois of the permutations of solutions, called roots, to the polynomial equation $P(x)=0$; what he called \textbf{ambiguity}. It transpires that this notion of ambiguity captured by groups is related to the notion of uncertainty captured by information, and that the cohomology of information and random variable structure has the structure of Galois group cohomology, a guiding idea of Bennequin's \cite{Bennequin2014,Baudot2015a}; see section \ref{infotopo_synthesis}.\\
Galois theory conveys in itself a cognition theory, as summarised by Deleuze: \textit{"the group of the equation characterizes at one point, not what we know of the roots, but the objectivity of what we do not know about them. Conversely, this non-knowledge is no longer a negative, a deficiency, but a rule, \textbf{a learning which corresponds to a fundamental dimension of the object}."} \cite{Deleuze2000}. This idea was further developed by Andr\'{e} \cite{Andre2008,Andre2007a,Andre2007b}. Bennequin applied this Galoisian epistemological principle of knowledge to propose a Galoisian physics \cite{Bennequin1994}. In what follows, consciousness is defined in terms of group and the actions of a group; hence we need a brief definition of and introduction to those concepts. 
Permutations are central objects in (discrete) group theory and combinatorics, and provide a definition of \textbf{symmetry} in finite sets. The fundamental theorem of algebra states that any general polynomial of degree $n$, $P(x)=a_{n}x^{n}+a_{n-1}x^{n-1}+ \cdots +a_{1}x+a_{0}$, where the coefficients $a_i$ are real or complex numbers and $a_i \neq 0$ (or any integral domain of a ring) have $n$ complex roots $\lambda_1, \lambda_2, \cdots , \lambda_n$. The roots are not necessarily distinct, and if they are indistinct they are called degenerate, and they hence encode the multiplicity of indistinct solutions \label{degenerateroot}. We can therefore also write the polynomial as a product $P(x) = a_{n}(x-\lambda_1)(x-\lambda_2)\cdots (x-\lambda_n)$.
Expanding the product on the right hand side of the equation provides a symmetric polynomial in the roots $\lambda_i$ that exhibit a Newton Binomial powerset structure (cf. figure \ref{figure_roots} for examples with $n=3,4$.) Newton's binomial method \label{binomial coefficient}) is as follows:

\begin{equation} \label{Vieta}
P(x)= a_{n}(x-\lambda_1)(x-\lambda_2)\cdots (x-\lambda_n) = a_{n} [\sum_{k=0}^n (-1)^{n-k}(\sum_{1\leq i_1 \leq i_2 \leq \cdots \leq i_k \leq n }\lambda_{i_1}.\lambda_{i_2}...\lambda_{i_k})x^k]
\end{equation}

Or in the notations used for information structures: 

\begin{equation} \label{Vieta2}
P(x)= a_{n}(x-\lambda_1)(x-\lambda_2)\cdots (x-\lambda_n) = a_{n} \left[\sum_{k=0}^n (-1)^{n-k}\left(\sum_{I\subseteq[n],|I|=k} \prod_{j=1}^{n}\lambda_{j}^{\mathbf{1}_{I}(i_j)}\right)x^k\right]
\end{equation}
, where $\mathbf{1}_{I}(i_j)$ is the indicator function of the subset $I=\{i_1,...,i_k\}$ of $[n]=\{1,...,n\}$. 

\begin{figure} [!h]  	
	\centering
	\includegraphics[height=6cm]{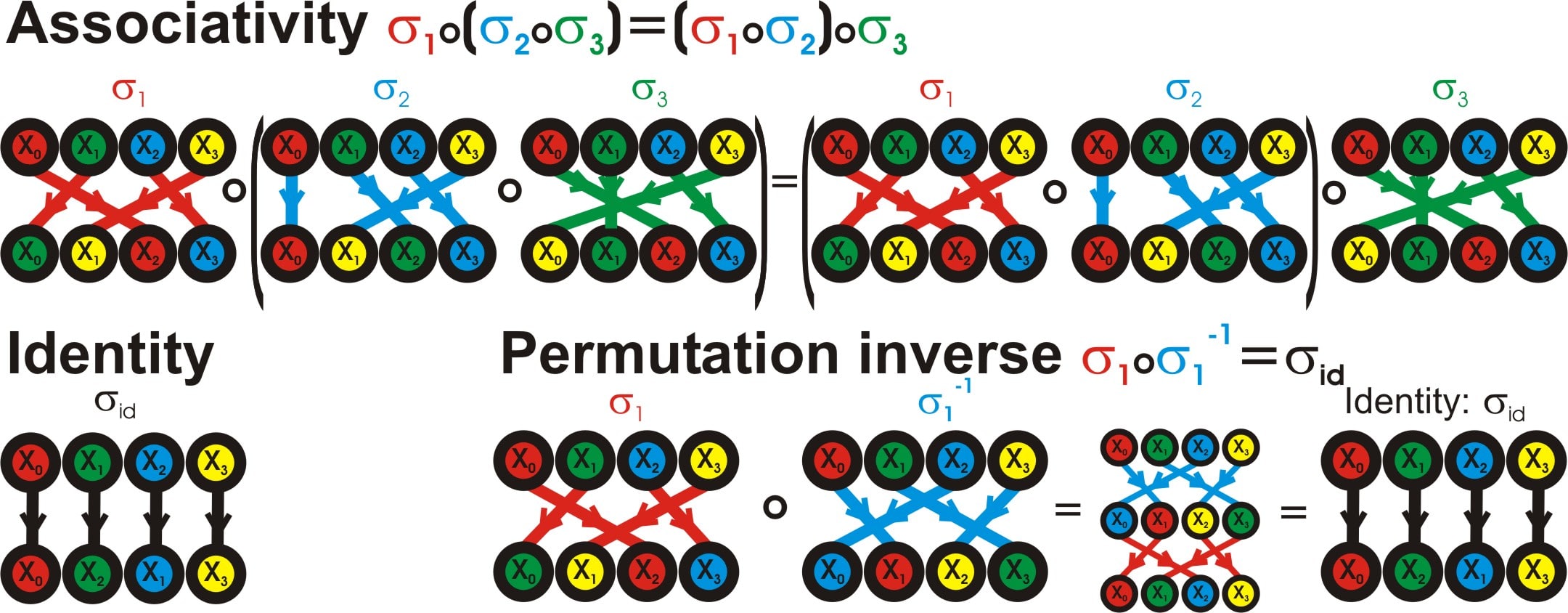}
	\caption{Illustration of the main three properties imposed by the axioms of a group on a given example of permutations. Permutations and their graphical representation are introduced in what follows.}
	\label{figure_axioms}
\end{figure}

It provides an elementary example of group structure and of the relationship between addition and product which has been generalized in numerous ways. It also provides the most elementary appearance in algebra of topological alternated sums. Identifying the coefficient $a_i$ with the coefficients of the equation \ref{Vieta} gives Vieta's formulas. If the leading coefficient $a_n=1$, then it is called a monic polynomial and the set of univariate monic polynomials with coefficient in a ring is closed under the operation of multiplication (the product of the leading terms of two monic polynomials is the leading term of their product), and forms a monoid (with the operation of multiplication, the identity element is the constant polynomial 1). \\
Thanks notably to the work of Bourbaki, a group is now well defined according to few axioms:\\
\textbf{Group :} A group is a set, $G$, together with an operation  $\star$  that combines any two elements $x$ and $y$ to form another element $x\star y$. To be a group, the set and operation, $(G,\star)$, must satisfy four axioms:
\begin{itemize}
	\item \itshape{Closure:} for all $x,y\in G$, the result of the operation, $x\star y$ is also in $G$.  
	\item \itshape{Associativity:} for all $x,y,z\in G$ , $(x\star y)\star z=x\star(y\star z)$. 
	\item \itshape{Identity element:} there exists an element $e$ in $G$, such that for all $x\in G$ , $e\star x=x\star e=x$. 
	\item \itshape{Inverse element:} for all $x\in G$ , there exists an element $y$ in $G$ such that $x\star y=y\star x=e$. 
\end{itemize}
Figure \ref{figure_axioms} gives an illustration of those axioms. $\mathbb{Z}$, the set of relative integers, forms a group with the operation of addition; this is a countably infinite cyclic group. One should be aware that the simplicity of these axioms hides the rich structures groups may exhibit, as stated by Borcherds, for example \cite{Cook2009}. The richness of these group structures is captured and encoded by homology, prefiguring the following sections of our paper. \\
\textbf{Symmetric group \cite{Berger2009}:} The symmetric group on a finite set of $n$ symbols, noted $S_{n}$, is the group whose elements are all permutations (bijection) of the $n$ elements of a finite set $\Omega$ and whose group operation is the composition of such permutations. The identity element is the identity permutation. More generally, if $\Omega$ is a non-empty set, we denote by $S_{\Omega}$ the group of permutations (that is, bijections) of $\Omega$ under the composition of maps $\sigma_i.\sigma_j=\sigma\circ\eta$. This latter includes the case of group of infinite order. 
It is easy to verify the permutations on a set forms a group under composition according to the axioms of a group (closure, associativity:
function composition is associative: $\sigma\circ\left(\eta\circ\rho\right)=\left(\sigma\circ\eta\right)\circ\rho$, identity, inverse); this group is termed \textbf{the symmetric group} of $\Omega$, noted $S_{\Omega}$ in general, and $S_{n}$ in the case where $\Omega=\left\{ 1,2,...,n\right\} $.\\
The common way to relate the elements of a group to a particular transformation of an object is to consider the action of the group. The Cayley's theorem states that every group $G$ is isomorphic to a subgroup of the symmetric group acting on $G$. The usual definition of an action is the following: \\ 
\textbf{Group action:} Let $G$ be a group and $X$ be a set. Then a left-group action $f^{\ast}$ of $G$ on $X$ is a function $f^{\ast} : G \times X \rightarrow X : (g,x)\rightarrow f^{\ast}(g,x)$, usually noted $g.x$ and called a left action (or translation) of $G$ on $X$,  that satisfies the two axioms:
\begin{itemize}
	\item Identity:   $\forall x \in X$ and $e$ the identity element of $G$ , we have  $e.x = x$.
	\item associativity:   $\forall (g,g') \in G^2$ and $\forall x \in X$, we have $(g'g).x = g'.(g.x)$, where $g'g$ denotes the result of applying the group operation of $G$ to the elements $g$ and $g'$, and  $g'g \in G$ and $g.x \in X$. 
\end{itemize}
If we define the morphism $\phi^{\ast}$ associated to the action $\forall g \in G$,  $\forall x \in X$, such that $g.x = (\phi^{\ast}(g))(x)$, then these axioms are equivalent to saying that the group $G$ acts on $X$ (on the left) if we have the morphism of the group $\phi ^{\ast}: G \to  S_X$, from $G$ into the symmetric Group $S_X$ of $X$. Such a morphism is called a representation of the group $G$.\\
The dual action called the right action is defined by inverting the order in $g$ and $g'$:
$f_{\ast} : G \times X \rightarrow X : (g,x)\rightarrow f_{\ast}(g,x)$, usually noted $x.g$ and called a right action (or translation) of $G$ on $X$. This satisfies the two axioms:
\begin{itemize}
	\item Identity:   $\forall x \in X$ and $e$ the identity element of $G$ , we have  $e.x = x$.
	\item associativity:   $\forall (g,g') \in G^2$ and $\forall x \in X$, we have $(gg').x = g'.(g.x)$, where $gg'$ denotes the result of applying the group operation of $G$ to the elements $g'$ and then $g$, and  $gg' \in G$ and $g.x \in X$. 
\end{itemize}
Dually, if we define the morphism $\phi_{\ast}$ associated to the action $\forall g \in G$,  $\forall x \in X$, such that $x.g = (\phi_{\ast}(g))(x)$, then these axioms are equivalent to saying that the group $G$ acts on $X$ (on the right) if we have the morphism of the group $\phi _{\ast}: G \to  S^{opp}_X$, from $G$ into the opposite symmetric Group $S^{opp}_X$ of $X$. Such a morphism is a representation of the group $G$ dual to left one. The opposite group $S^{opp}_X$ of the symmetric group $S_X$ is the set of permutations of $X$ with the law of composition $ (f, g) \mapsto f \star g = g \circ f$.  We go from left to right dual using the fact that $(gg')^{−1} = g'^{−1}g−1^{−1}$ and composing with the inverse operation of the group.

After Galois, Lie’s work and motivations for studying Lie groups were intended to extend Galois theory to differential equations by studying the symmetry groups of differential equations. The  resulting differential Galois theory was studied by Picard-Vessiot, and Chevalley and Eilenberg later formalized the cohomology of Lie algebra \cite{Chevalley1948}. One of the motivations of homological algebra was then to unify Galois discrete cohomology with Lie continuous cohomology.  

\subsubsection{What is topology?}\label{What is topology}

\begin{figure}
	\centering
	\includegraphics[height=12cm]{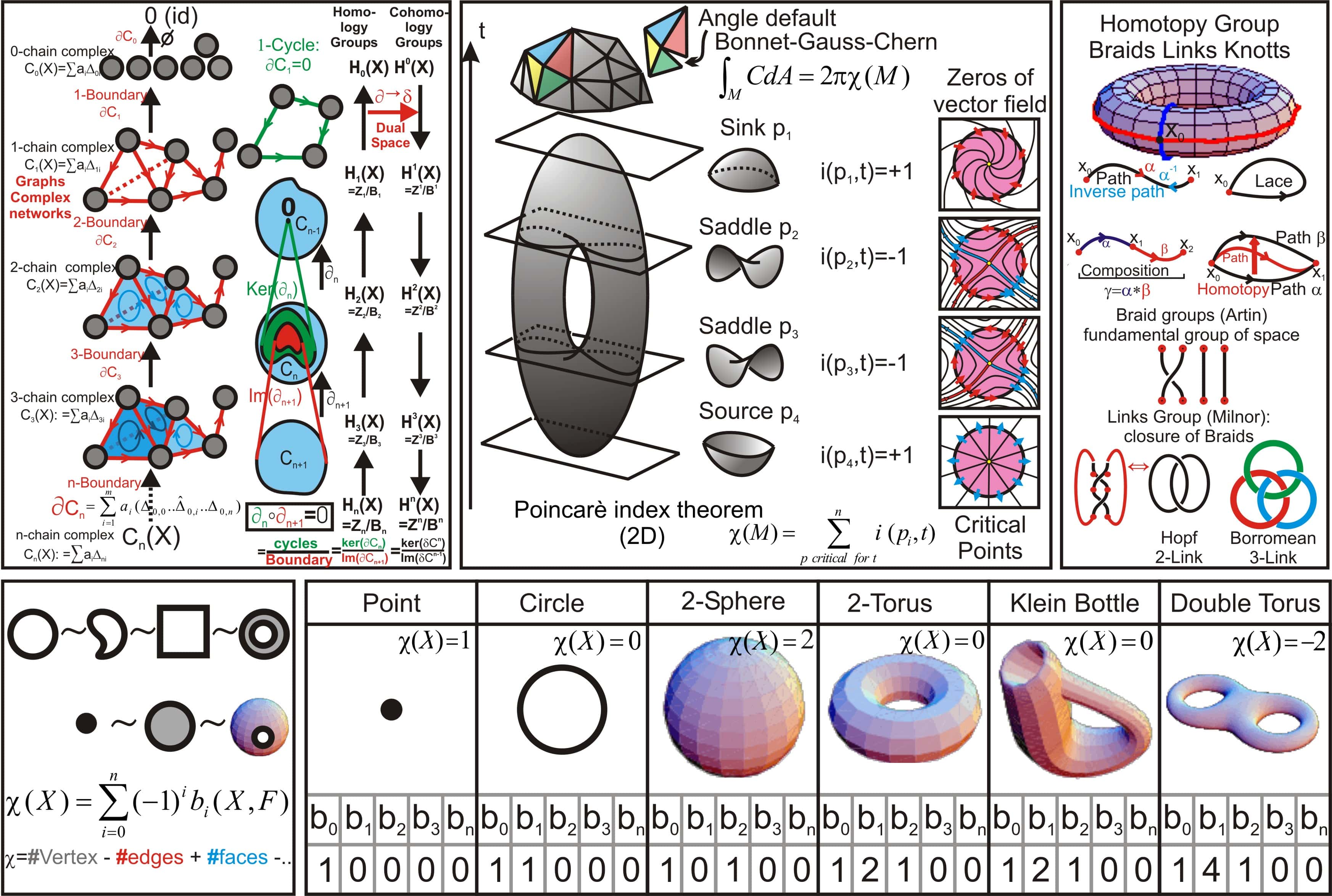}
	\caption{A snapshot of algebraic topology (see text).Top: (left) A simplicial chain complex, boundary operator, cycle. (Middle) (Morse) Homology counts the number of critical points, source and sink counts +1 and saddles -1. (Adapted and modified with permission from Banchoff \cite{Banchoff1970}, and from Ghrist \cite{Ghrist2014}). (Right) Homotopy: an equivalence relation between paths. Bottom: (left) homology: equivalence "up to continuous deformation". (Right) Homology counts the number of holes, i.e. algebraically independent components, in each dimension.}
	\label{Homology}
\end{figure}

We will now give a short and informal snapshot of algebraic topology (see Figure \ref{Homology}). Classical expositions can be found in Alexandroff \cite{Alexandroff1932a}, Hatcher \cite{Hatcher2002} and Ghrist 's book \cite{Ghrist2014}. The researchers working in complex systems or neural networks are familiar with the idea of topology: a complex network where a graph is a 1-chain complex, a 1-complex. In the study of complex systems, the seminal works of Erdos and Renyi \cite{Erdoes1959} and then Watts and Strogatz \cite{Watts1998} showed that combinatorial features and statistics of connections of networks affect their dynamical, statistical and critical behavior (reference reviews can be found in \cite{Dorogovtsev2008} and \cite{Newman2010}). The considerations of these studies relies on a tiny 1D window of the n-dimensional (or degree) landscape of topology. The meaning of this high dimensional generalisation by the topology has a simple and intuitive interpretations; whereas a network is an assembly of pairs of elements (of neurons for example), homology investigates assemblies with arbitrary numerous elements, a good mathematical start to formalizing neural assemblies (or other "groups"). We indeed believe that information theory may provide appropriate tools to ultimately render some of the difficult frameworks of algebraic topology as intuitive as they should be.    
Simplicial homology (commutative, with addition as an operation, for example) writes a complex network as a group here commutative, e.g as an algebraic sum of its $m$ edges, of its elementary constituents in one dimension, weighted by coefficients with value in a group or a field (or in modules): $C_1(S,\mathbb{F})=\sum_{i=0}^{m-1} a_i \Delta_{1,i}  $. The orientation and the weighting of the network are implemented by the coefficients $ a_i $. Homology provides an alternative and a generalization of adjacency and incidence matrices. For example, the coefficients {(0,1)} of the simplest adjacency matrix are assimilated to the field with two elements $\mathbb{F}_2$  etc. Homology can then be considered as providing a generalization to n-dimensions of complex networks. \\
\textbf{Homology:} an homology is defined by two things: an n-degree (or n-dimensional in special cases like the simplicial) complex $C_n$ and a boundary operator $\partial_{n}$. What follows is illustrated in the top left panel of Figure \ref{Homology}. By the defining condition $\partial\circ\partial=0$, the application of the differential operator $\partial$ to the complex generates a sequence of n-complex $C_k$ or k-chains, as follows:
\begin{equation}
\xrightarrow{\partial_{n+1}} C_n \xrightarrow{\partial_{n}} C_{n-1}\xrightarrow{\partial_{n-1}}... C_1 \xrightarrow{\partial_{1}} C_0 \xrightarrow{\partial_{0}}0
\end{equation}  \\
This is the basic principle; now let us investigate what a complex is.
\textbf{n-complex:} a simplicial n-complex $C_n$ is written as a weighted sum of its elementary n-simplices with a composition operation symbolized by the of addition $C_n(S,\mathbb{F})=\sum_{i=0}^{m-1} a_i \Delta_{n,i}$. The building blocks, n-simplex $\Delta_n$ (in the simplest case of simplicial homology) are triangles generalized to all dimensions; a point-vertex is a 0-simplex, an edge a 1-simplex, a triangle a 2-simplex, a tetrahedron a 3-simplex and so on; they are also called the k-faces of the complex. The most basic definition of an abstract complex is a family $C$ consisting of finite subsets of a given set of vertices $V={x_1,...,x_n}$, such that the 2 following conditions hold: i) $\{x_i\}\in C$ for all $\{x_i\}\in V$  ii) If $X \in C$ and $Y \subseteq X$, then $Y \in C$. In simple but ambiguous words, a complex contains all its subfaces. n-complexes are organized in a sequence of decreasing degree-dimensions which are also inclusive (or projective) sequences. An edge is included in a triangle which is included in a tetrahedron and so on.\\ 
\textbf{n-boundary:} We go from one dimension $n$ to another $n-1$ by a boundary operator $\partial_n: C_n \rightarrow C_{n-1}$, a homomorphism. It literately makes a dimension reduction, just as we saw conditioning do in probability. The simplest expression of a boundary operator in simplicial homology consists of an alternating sum of the complexes obtained by deleting one of the vertices each time. By definition, the boundary of a boundary must be zero ($\partial_n\circ\partial_{n-1}=0$ where $0$ denotes the mapping to the identity in $C_{n-1}$); this implies that the sequence is inclusive and that the image of the $n+1$ boundary is included in the kernel of the $n$ boundary ($Im(\partial_n)\subseteq Ker(\partial_{n-1})$). This defining condition $\partial\circ\partial=0$ or $\partial ^2=0$, that is, the boundary of a boundary is 0, is fundamental and implements the non-contradiction principle (considering $Y$ to be the boundary of $X$, that is $Y = \operatorname{Cl}(X)\cap \operatorname{Cl}(\Omega-X)$ and then considering the boundary of $Y$. Since $\operatorname{Cl}(X)$ and $\operatorname{Cl}(\Omega-X)$ are both closed, their intersection is also closed, and hence $\operatorname{Cl}(Y)=Y$, and $\operatorname{Cl}(\Omega-Y)= \operatorname{Cl}(\Omega)-\operatorname{Cl}(Y)$, and moreover considering that $\Omega$ is the whole space, we also have $\operatorname{Cl}(\Omega)=\Omega$. Hence the boundary of the boundary of $X$ is $Y \cap (\Omega-Y)$, that is the intersection of any set with its complement, that is the empty set and hence the consistency-non-contradiction axiom). \\

\textbf{n-cycles:} It allows us to define the n-cycles as null n-boundaries, that is, n-boundaries that equal zero ($\partial_n=0$), literally an n-chain-complex without a boundary (or with an identity boundary, that is, a closed chain). \textbf{Homology groups:} Homology groups are defined as the quotient group of the kernel of the n-boundary by the image of the n+1-boundary ($H_n(S,\mathbb{F})= Ker(\partial_{n})/ Im(\partial_n+1)=Z_n(S,\mathbb{F})/B_n(S,\mathbb{F})$. They hence quantify holes, empty cycles. Betti numbers are the nth rank of the simplicial homology group, its number of generators. Cohomology is the dual of homology and uncovers more information with invariants, including torsion and the change of the group of the coefficients. For oriented complexes, we go from homology to cohomology via Poincare duality, in general via the universal coefficient theorem which states that $H^n(S,\mathbb{F})\approx Hom(H_n(S),{F})\oplus Ext^1(H_{n-1}(S),{F})$, where $Ext$ and $Hom$ are functors. It was one of the first motivations of cohomology to account for both finite groups and Lie groups. The chains become co-chains ($C^n$), boundaries $\partial_n$ coboundaries ($\delta^n$), cycles cocycles ($\delta^n=0$); in other words, the sequence is reversed. 
It was the algebraic aspect, the beauty of topology relies upon that it is expressed equivalently the geometrical aspects. Topology is the science that categorizes and distinguishes the shapes of spaces. We have already seen the geometrical realization of simplexes with the probability simplex shown in Figure \ref{figure_probability_simplex_complex}. Simplicial n-complexes are discretization or n-triangulation of continuous n-dimensional manifold $M$ (piecewise-linear manifolds). Homology is an equivalence relation on the manifolds up to continuous deformation (cf. Figure \ref{figure_probability_simplex_complex} bottom). For example, the circle is topologically equivalent (homeomorphic) to the square, the point to the disc, etc. It thus appears that the homology of two objects is different if they differ in their number of holes, and homology accounts for the holes, which are algebraically independent components in each dimension. Betti numbers quantify the number of holes in each dimension; their alternating sum equals the Euler characteristic, which is the main topological invariant $\chi(S)= \sum_{i=0}^{m-1}(-1)^i b_i (S,\mathbb{F})$. 
If one has a height function $h$ as in Morse theory, homology counts the critical points of the manifold. A saddle point is a hyperbolic unstable critical point and counts for -1. The sources and sinks count for + 1. The sum of the critical points equals the Euler characteristic, $\chi(M)= \sum_{p critical for h}^{m-1} i (p_i,h)$, which also equals the integral of the curvature $C$ for 2D compact without boundary, according to the Bonnet-Gauss-Chern theorem ($\int_MC\mathrm{d}A=2\pi\chi(M)$). A reference work on Morse theory is Milnor's book \cite{Milnor1964} and Forman's review of the discrete version \cite{Forman2002}.\\
\textbf{Homotopy:} Homotopy is an equivalence relation between paths or geodesics in manifolds with inverses, and an associative operation of composition (equivalence classes form a group called homotopy groups, noted $\pi_n(S,x_0)$ where $x_0$ is the based point. $\pi_1$ is called the fundamental group). In dimension 1, the holes can be detected by obstruction to retract a loop (closed curve) into a point, and if two loops can be deformed into one another they are homotopically equivalent and define the same hole. A hole in a manifold implies the existence of non-homotopic laces (e.g. if there are two non-homotopic laces on the torus). These definitions can be generalized to higher dimensions, and an obstruction to deform-retract a closed oriented 2D surface into a point can detect a two-dimensional hole and so on. $S$ is said to be n-connected if and only if its first n-homotopy groups are 0 ($\pi_i(X) \equiv 0~, \quad -1\leq i\leq n$, notably if it is non-empty $\pi_{-1}(S) \equiv 0~$ and path-connected $\pi_{0}(S) \equiv 0~$). Postnikov provided a generic method to determine the homology groups of a space by means of homotopy invariants \cite{Postnikov1951}. Links (see also knots), such as the Hopf 2-link or the Borromean 3-link, form homotopy groups \cite{Milnor1954} that can be formalized as the closure of compositions of braids (signed permutations that form Artin's group). It is obvious in the case of n-links that the first i-linking numbers ($i<n$) vanish: the rings of a Borromean link are unlinked in pairs, which is a purely emergent/collective property. \\
Concerning neuroscience and cognition, as already mentioned in the cell assembly section, following the development of topological data analysis (which is mainly based on persistence homology \cite{Carlsson2009}), several studies have been dedicated to the application of topological data analysis to visual activity by Singh and colleagues \cite{Singh2008}, to neural networks by Petri and colleagues \cite{Petri2014} and to neural activity patterns and assemblies by Curto and Itskov \cite{Curto2008}. Linguistic structure has also been studied using algebraic topology methods. Port and colleagues used persistent homology to detect the presence of structures and relations between syntactic parameters globally across all languages and within specific language families \cite{Port2018}. Topology also provides the mathematical ground for the electromagnetic view of cognition proposed in the first chapter. Even without going into the complex details of Topological Quantum Field Theories, the basic of Kirchhoff's voltage and current conservation laws which state that the algebraic sum of currents at every node  of an electrical circuit (formalized as a simplicial 1-complex) is equal to 0, is a direct consequence of the first homology group, i.e., a chain I is a 1-cycle $\partial I=0$. The formalization of electrical circuits as a homology theory was developed by Lefschetz, and the electromagnetic generalization is treated in the work of Zeidler (\cite{Zeidler2011} chap 22 and 23). Wheeler founded his \textbf{"austerity principle" of physics} on the definition of a boundary \cite{Wheeler1982}, and the chapter 15 of his heavy gravitation book presents why the homological view is fundamental for general relativity \cite{Misner1973}. This convergence of quantum field and gravitation on different homological formalisms has provided the basis for the main gravity quantization investigations  \cite{Rovelli2008}. Wheeler has been a major actor of the physical theory of information, the "it from the bit" , notably sustaining that \textit{"all things physical are information-theoretic in origin and that this is a participatory universe"}\cite{Wheeler1990}. 
We now discuss the formalism of information topology underlying the numerous studies that have applied information theory to studies of perception or consciousness, and formalize the way in which machine learning principles and algorithms are topological by nature.

\subsubsection{Information topology synthesis: consciousness's complexes and thermodynamic} \label{infotopo_synthesis}

The characterization of entropy in the context of topology started with a surprising coincidence in the work of Cathelineau \cite{Cathelineau1988} on the scissor congruences introduced in section \ref{measure}. As briefly introduced in \cite{Baudot2018}, the appearance of the functional equation of entropy, and hence entropy itself, pertains to motives, the unfortunately yet importantly conjectural side of topology which gathers very different approaches, starting with the investigations of Kontsevitch, Gangl and Elbaz-Vincent \cite{Kontsevitch1995,Elbaz-Vincent2002,Connes2009,Marcolli2011,Baez2011,Baez2014}. The formalism of information topology developed by Baudot, Bennequin and Vigneaux in \cite{Baudot2015a,Vigneaux2017} is based on probability, namely on information structures formalized as follows.\\  

\begin{figure}[! h]
	\centering
	\includegraphics[height=4.5cm]{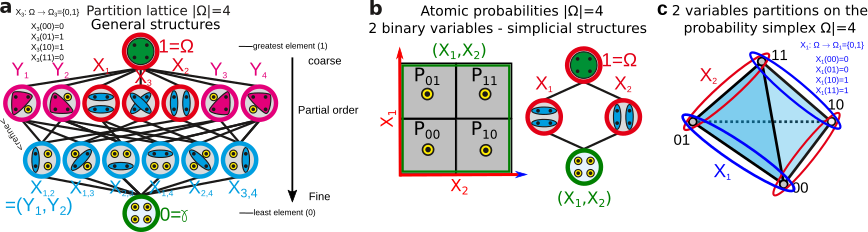}
	\caption{\textbf{Example of general and simplicial information structures. a,} Example of a lattice of random variables (partitions): the lattice of partitions of atomic events for a universe of 4 events $|\Omega|=4 $ (for example two coin tossing $\Omega = \{00,01,10,11\}$). Each event is depicted by a black dot in circles  representing the variables. The operation of the joint-variable noted $(X,Y)$ or $X\otimes Y$, of two partitions, is the less fine partition $Z$ which is finer than $X$ and $Y$ ($Z$ divides $Y$ and $X$, or $Z$ is the greatest common divisor of $Y$ and $X$). The joint operation has an identity element noted $1 = \Omega$ (noted 0 in what follows), with $X,1 = X,\Omega = X$ and is idempotent $ (X,X) = X^2 = X $ (the non-contradiction principle stated by Boole, cf. section \ref{logic of thoughts}, giving a codegeneracy map in cohomology or simplicial sets). The structure is partially ordered set (poset) and endowed with a refinement relation. \textbf{b,} Illustration of the simplicial structure (sublattice) used for data analysis ($|\Omega| = 4 $). \textbf{c,} Illustration of the random variable partitioning of the probability simplex in the same example as in b. (Adapted and modified with permission from Baudot, Tapia and Goaillard \cite{Baudot2018,TapiaPacheco2017})}
	\label{partition_lattice}
\end{figure}

\begin{figure} [!h]
	\centering
	\includegraphics[height=14cm]{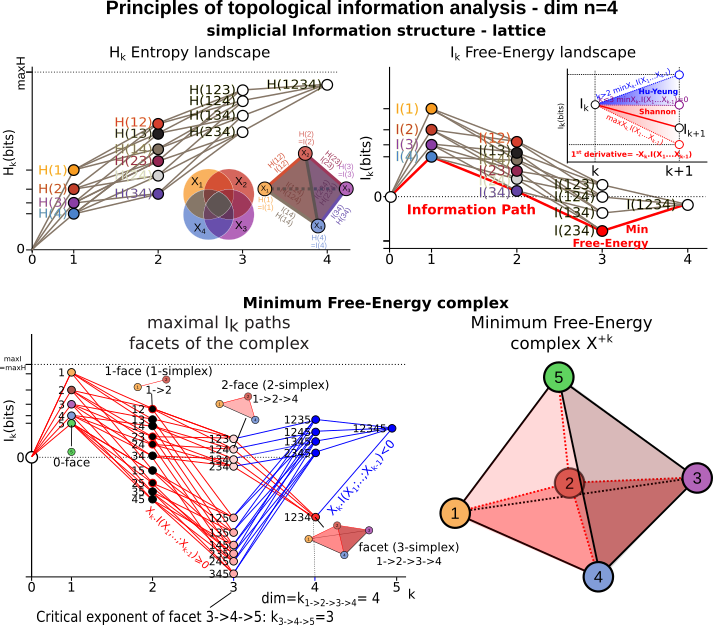}
	\caption{\textbf{Principles of the analysis of dimension 4}. \textbf{Top:} an example of entropy landscapes $H_k$ and mutual information $I_k$ (free energy) for 4 variables (semi-lattice). In red is represented an information path (piecewise linear function $IP(k)$) and the first free energy minima of critical dimension 3. The cartoon illustrate the Shannon's and Yeung's topological cone arising from standard and non-Shannonian information inequalities and that bounds the paths \cite{Yeung2003,Baudot2018}. \textbf{Bottom:} An example of a complex of minima of free energy and its $I_k$ landscape, for which the facets are represented in red (positive conditional information path of maximum length) (adapted and modified with permission from Baudot, Tapia and Goaillard \cite{Baudot2018}).}	
	\label{Principles_TIDA}
\end{figure}

\textbf{Information structures:} Random variables are partitions of the atomic probabilities of a finite probability space $(\Omega,\mathcal {B},P)$. The operation of joint-variable $(X_1, X_2)$ is the less fine partition, which is finer than $ X_1 $ and $ X_2 $; the whole lattice of partitions $\Pi$ \cite{Andrews1998} hence corresponds to the lattice of joint-variables \cite{Fresse2004,Baudot2015a}. A general \textbf{information structure} is defined in usual set-theoretic terms as a triplet $(\Omega,\Pi,P)$, and hence covers all the possible equivalence classes on atomic probabilities. A more general and modern expression is given in category and topos theory, in \cite{Baudot2015a,Vigneaux2017}. The image law of the probability $P$ by the measurable function of the joint-variables $(X_1, ..., X_k)$ is noted $(X_1, ..., X_k; P)$. Figure \ref{partition_lattice} gives a simple example of the lattice of partition for a universe of 4 atomic probabilities, a sub-simplicial lattice which is combinatorially computable on data.
The fact that the lattice is a partially ordered set (poset) endowed with a refinement relation is central; it means that there is an intrinsic hierarchy of informational structure, just as in the general model of physical cognition of Schoeller, Perlovsky, and Arseniev \cite{Schoeller2018}. Concerning classification-recognition tasks of machine learning, information structures can be considered as universal: as a partition is equivalent to an equivalence class all possible classification are represented in an information structure. For example, this lattice can be understood as an algebraic formalization of deep networks, that is, networks with hidden layers of neurons for which the rank (dimension given in what follows) in the lattice gives the rank of a hidden layer and the connections correspond to coface maps (roughly, elementary projections or dimension reduction or increase). The random variables formalize neurons that are intrinsically probabilistic and possibly multivalued, generalizing binary and deterministic neurons such as McCulloch and Pitts' formal neurons. As discussed in the section on electrodynamic and digital coding \ref{Neural coding}, such a generalization is biologically relevant and even necessary.
The other common interpretation of this poset hierarchical structure, probably equivalent to the previous one (at least in ergodic systems), is that the ordering of the lattice provides a \textbf{multi-scale, coarse to fine analysis} (cf. figure \ref{partition_lattice}a), and each rank of the lattice provides an information analysis at the corresponding organizational level, as already formalized and applied by Costa et al \cite{Costa2002,Costa2005}, who called it multiscale entropy in the context of time series. Hence, such formalism can be applied in the context of multiscale systems such as the one illustrated in Figure \ref{scale_function} (in theory), and the entropy necessarily increases as more and more variables join, e.g. while progressing in organizational scales (cf. Figure \ref{Principles_TIDA}a).  

\textbf{Action:} In this general information structure, we consider the real module of all measurable functions $F(X_1, ..., X_k; P) $. We consider the conditional expectation-mean (corresponding to informational conditioning) the action of a variable $ Y $ on the module, noted:
\begin{equation}
Y.F (X_1, ..., X_k; P)=  k \sum_{y\in\mathscr{Y}}^{N_y}p(y).F(X_1, ..., X_k; P/Y=y)
\end{equation}
where $P/Y=y$ denotes the conditioning of the probability by the event $Y=y$, such that the action corresponds to the usual definition of conditional entropy given in section \ref{conditionalentropy}. Centrally, the action of conditioning is associative  \cite{Baudot2015a,Vigneaux2017}.
This action is also extremely important with regard to the theory of cognition; we used it in the section on homeostasis \ref{Homeostatic plasticity} to define invariance, and we dedicate a more mathematically rooted presentation in the next section \ref{Geometries-Homeostasis}. Notably, Vigneaux was albe to generalize all the formalisms presented here to Tsallis entropies by considering a deformed action (integer powers of probability in the expectation) \cite{Vigneaux2017}, also giving a straightforward extension to quantized information. 

The complexes of measurable functions of random variables  $X^k=F(X_1, ..., X_k;P)$ and the cochain complex $ (X^k,\partial^k) $ are noted as:
\begin{equation*}
0 \xrightarrow{} X^0 \xrightarrow{\partial^0} X^1 \xrightarrow{\partial^1} X^{2} \xrightarrow{\partial^2} ... X^{k-1}\xrightarrow{\partial^{k-1}} X^{k}
\end{equation*}
, where $ \partial^k $ is the coboundary with a left action proposed by Hochschild for associative structures and rings \cite{Hochschild1945}, for Galois cohomology (see Tate's work \cite{Tate1991}), and for homological algebra (see Cartan et Eilenberg's work  \cite{Cartan1956} and for non-homogenous bar complex (see Maclane \cite{MacLane1975})  is noted as:
\begin{equation}
\begin{split}
(\partial^k)F(X_1;X_2;...;X_{k+1};P)={} & X_1.F(X_2;...;X_{k+1};P)\\
& +\sum_{i=1}^{k}(-1)^{i} F(X_1; X_2;...;(X_i,X_{i+1});...;X_{k+1};P)\\
& + (-1)^{k+1} F(X_1;...;X_{k};P)
\end{split}
\end{equation} 

For the first degree $k = 1$, the 1-coboundary is $(\partial^1)F(X_1;X_2) = X_1.F(X_2)-F(X_1,X_2) + F(X_1) $ and the 1-cocycle condition  $(\partial^1)F(X_1;X_2)=0 $ gives $F(X_1,X_2) =F(X_1) + X_1.F(X_2)$, which is the fundamental chain law of information (cf. equation \ref{chain rule gener}). Following Kendall \cite{Kendall1964} and Lee \cite{Lee1964}, it is possible to deduce from this chain law the functional equation of information and to uniquely characterize Shannon entropy as the first class of cohomology, up to the arbitrary multiplicative constant $k$ \cite{Baudot2015a,Vigneaux2017}. It constitutes the main theorem that founded information topology. It appears by direct computation in this cohomology that mutual informations with an odd number of variables are minus the coboundary of even degrees $\partial^{2k}=-I_{2k+1}$. Obtaining even mutual informations is achieved by reproducing the Hodge decomposition of Hochschild cohomology constructed by Gerstenhaber and Shack \cite{Gerstenhaber1987,Weibel1995,Kassel2004}. We construct for this a double complex $(X^{\bullet,\bullet},\partial,\partial_\ast)=(X^{k',k''},\partial^{k',k''},\partial_\ast^{k',k''}), ~ (k',k'')\in \mathbb{N}\times\mathbb{N}$ endowed with the preceding coboundary $\partial$ and the same coboundary with a symmetric action $\partial_\ast$ (left and right, commutative) \cite{Gerstenhaber1987,Weibel1995,Kassel2004}. 
As a result, the odd mutual informations are minus the even coboundary $\partial^{2k}=-I_{2k+1}$, the even mutual-informations are minus the odd symmetric coboundaries $\partial_{\ast}^{2k-1}=-I_{2k}$, and the mutual informations are the coboundaries of the total complex with an alternated sign $\partial_{tot}^k=(-1)^{k+1}I_{k+1}$.\\
The independence of two variables ($ I_2 = 0 $) is then directly generalized to k-variables and gives the cocycles $I_k = 0$.\\ As a conclusion concerning the probabilist interpretation of cohomology, information cohomology quantifies statistical dependencies and the obstruction to factorization.\\
What is the interest of these mathematical tools for cognition? The uniqueness of the obtained functions implies, in the case classical finite probabilistic application to empirical data, that the information functions are not only "good" but also the only ones to quantify statistical dependences and independences in the multivariate case. The finite-discrete symmetries of permutation groups, which are the structural ambiguity and the (co)differentials arising from Galois's theory, are equivalent to uncertainties and shared information arising from the "mathematical theory of communication". To comment on such a surprising and important fact,  mutual informations are indeed (co)differential operators, a purely continuous operation arising from a finite and discrete context. Hilbert noted in his work on infinity, "the first unreasonable impression given to us by natural phenomena and matter is that of continuity" \cite{Hilbert1924}: while physics repeatedly proved that objectively the input of our senses is finite and discrete, our consciousness construct the impression of continuity \cite{Hilbert1924}. As expressed by Poincar\'{e}, the construction of our continuous perception from discrete data can be proposed to be a cohomological operation by nature (even explaining Weber-Fechner's law) that mutual informations naturally fulfill. This is an important contribution of Galois's theory, further pursued by Lie, Picard-Vessiot and others, that allows us to conceive of the notion of continuity and of derivation yet holding in the discrete world, extending the classical Newtonian view.   
The second point of interest is that cohomology is the science of the forms (patterns) of spaces. Information topology hence provides a preliminary theory of the shapes of probabilistic structures on which it is possible to develop methods of pattern recognition-characterization for machine learning and the quantification of epigenetic landscapes for biological adaptive dynamics, following Waddington and Thom  \cite{Baudot2018,TapiaPacheco2017}. \\
The third point of interest lies in the fact that this cohomology can be expressed as a \textbf{Topos} on a probability site, which  allows the establishing of the multivalued constructive logic described in an elementary finite context in section \ref{topos}. Such logic can provide a basis for a probabilistic, biological and cognitive logic.\\

Regarding data analysis and physics, information topology allows us to quantify the structure of statistical interactions within a set of empirical data and to express these interactions in terms of statistical physics, machine learning and epigenetic dynamics \cite{Baudot2018}. The combinatorics of general variable complexes being governed by Bell's numbers, their effective computation on classical computers is illusory. To circumvent those computational hardness, we define the sub-case of the simplicial cohomology of information, with an algorithmic complexity that can be implemented, but that neglects some of the possible dependencies. The computational hardness of consciousness in discussed in section \ref{Computational mind2} in the perspective of Artificial Intelligence and classical Turing definitions of computation. The exhaustive computation of the simplicial-binomial combinatoric of $I_k$ and $H_k$ (see Figure \ref{Principles_TIDA}, the set of subsets of $n$ variables) is thus reduced to a complexity in $2^n$, computable in practice up to $n = 21$ with basic resources. The set of the entropy values $H_k$ and mutual information $I_k$ for all subsets of $n$ are represented by the entropy landscapes $H_k$ and information landscapes $I_k$ as a function of the dimension $k$, as illustrated in Figure \ref{Principles_TIDA}. The entropies $H_k$ quantify uncertainty on variables, and the mutual informations $I_k$ quantify statistical dependencies. An information path $IP(k)$ is defined as a piecewise linear function, as illustrated in Figure \ref{Principles_TIDA}. Its first derivative is equal to minus the conditional mutual information, which allows the characterization of the first minima of the paths based on the negativity of the conditional information and the non-Shannonian cone and information inequalities studied by Yeung \cite{Yeung2003} (cf. Figure \ref{Principles_TIDA}).

\begin{figure} [!h]
	\centering
	\includegraphics[height=9cm]{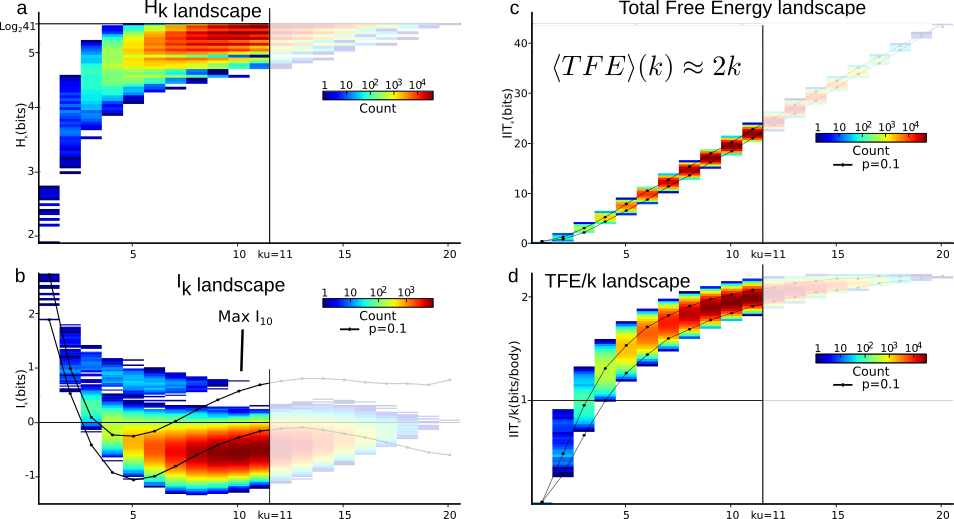}
	\caption{\textbf{Information Topology Analysis of neural genetic expression: $I_k$, $H_k$ and Total Free Energy (TFE) landscapes}. The figure presents the Information Topology Analysis of the expression in single cell qPCR of 41 genes in 10 dopaminergic (DA, Subtancia Nigra pars compacta) and 10 non Dopaminergic (nDa, neighboring Ventral Tegmental Area) neurons from adult TH-GFP mice pre-identified by fluorescence \cite{Baudot2018,TapiaPacheco2017}. \textbf{a:} entropy  $H_k$ and \textbf{b:} mutual information $I_k$ (free energy) landscapes. The vertical bars indicates the dimension above which the estimations of information become too biased due to the finite sample size, a phenomenon known as the curse of dimensionality (the undersampling dimension is $k_u = 11$, p value 0.05). The  significance value obtained from shuffled distributions for $p = 0.1$ are depicted by the black lines and the doted lines. This test is based on the random shuffles of the data points that leaves the marginal distributions unchanged, as proposed by \cite{Pethel2014}. It estimates if a given $I_k$ significantly differs from a randomly generated $I_k$, a test of the specificity of the k-dependence. (adapted and modified with permission from Baudot, Tapia and Goaillard \cite{TapiaPacheco2017,Baudot2018}). \textbf{c:} The total free energy (TFE) or Integrated Information landscape quantifying consciousness according to Tononi and Edelman and \textbf{d:} the landscape of the TFE per body. }	
	\label{applied_TIDA}
\end{figure}

The main theorems, definitions and data analysis in Baudot, Tapia and Goaillard \cite{TapiaPacheco2017,Baudot2018} establish the following results, here included with comments about their relevance regarding consciousness and neural processing theories:
\begin{itemize}
	\item The marginal information $I_1$ are generalized internal energies and the $I_k$ are the free energy contributions of the k-body interaction. Figure \ref{applied_TIDA}b illustrates the $I_k$ landscape quantifying the free energy contributions as the function of the dimension (number of bodies) for the genetic expression of two neural populations. The maximum of $I_{10}$ identifies the 10 Dopaminergic neurons. The total correlation proposed by Watanabe and Studeny \cite{Watanabe1960,Studeny1999} to quantify dependences, or the  Integrated Information proposed by Tononi and Edelman to quantify consciousness \cite{Tononi1998},  $G_k =\sum_{i=2}^{k}(-1)^{i}\sum_{I\subset [n];card(I)=i}I_i(X_I;P)$, is the total free energy (TFE). Figure \ref{applied_TIDA}c illustrates the TFE landscape in the same context as previously. TFE hardly distinguishes the two population, but instead presents a quite homogeneous linear behavior on average, $\langle TFE \rangle\approx 2k$, meaning that the total free energy adds linearly with adding new bodies. This is illustrated by TFE per body (or TFE rate) in Figure \ref{applied_TIDA}d. In agreement with IIT theory that assigns consciouness according to those measure \cite{Paulson2017,Tononi2016}, the conclusion is that genetic expression participate to consciousness, to its slow component as discussed in section \ref{jacob} on epigenetic regulation timescales. Although it remains to be achieved effectively, we propose that the same principles and methods apply to electrical of neural imaging recordings.   
	We rediscover the (semi-)classical definitions of internal energy as a special case for phase space independent identically distributed variables (Gibbs distributions) and the usual relation of thermodynamics. See Adami and Cerf \cite{Adami1999} and Kapranov \cite{Kapranov2011} for an exposition of this classical result:
	\begin{equation}\label{thermodynamical relation}
	H(X_1,...,X_n)=\langle E\rangle - G=U-G
	\end{equation}
	The marginal component, the internal energy, corresponds to a self-interaction, a reflexive component of consciousness that completes the model of Tononi and Edelman. Such a formalism could hence account for both the reflexive and the qualitative aspects of consciousness consistently introduced in our first chapter \ref{neuronal_postulate}, in agreement with the Leibniz's monadic hypothesis.     
	\item Information paths are in bijection with symmetric group and stochastic processes. These paths correspond to the automorphisms of the partition lattice. We directly obtain a topological generalization of the second principle of thermodynamics. This theorem generalizes Cover's theorem for Markov chains \cite{Cover1994} and allows one to conjecture the existence of a Noether theorem for stochastic processes and discrete symmetries, notably following Baez and Fong \cite{Baez2013}. Such a theorem should be considered as the topological version of the first principle of thermodynamics. On the $H_k$ landscape illustrated in Figure \ref{applied_TIDA}a, this theorem imposes that any path can only "go up". Information paths and landscape directly account  for standard causal criteria, like Granger causality and Transfer entropy, that generalize the later to the non-gaussian case \cite{Barnett2009} and defined by Schreiber as a pairwise conditional mutual information \cite{Schreiber2000}. Of course the generalization to the multivariate case together with the consideration of positivity and negativity is a major interest and shall be investigated further. In \cite{Baudot2018} (p.29), we give a very preliminary view of how space-time could emerge from purely topological considerations (without metric), and we consider that a formalism of the space-time shape of k interacting bodies should provide the ultimate expression of what consciousness is.    
	These paths allows the formulation of sums over paths appearing in statistical field theory, but in a discrete finite classical and informational context. The remarkable difference compared to the usual path integrals relies on that no infinite energy divergence can occur. The hope is that such an information formalism will give a discrete finite expression of electrodynamics and of renormalization groups (however without artificial renormalization \cite{Dirac1929,Feynman1985}). This would complete the electrodynamic theory of consciousness given an exposition of in the first chapter with the statistical physical informational view presented here. 
	\item The longest paths to the minima (equilibrium points) form the complex of minima of free energy. This complex formalizes the principle of minimal free energy in topology in complicated cases where multiple local minima co-exist, the central paradigm of frustrated systems in statistical physics \cite{Vannimenus1977,Mezard2009a}.
	
	\item This simplicial complex provides a generic and consensual definition of a complex system, thus generalizing complex (1-complex) networks to larger dimensions. The multiplicity of these minima (facets) defines and quantifies diversity. 
	This complex is proposed to provide a thermodynamical and mathematical formalization of the complexes developed in integrated information theory \cite{Tononi2016,Oizumi2014,Tononi1998}. The possible coexistence of several facets that define the complex may explain the apparently paradoxical unity and diversity of consciousness: a conscious experience, corresponding to one facet, does not forbid the existence of some other conscious experience possibly less or more complex (of a different dimension), and that may be estimated as an unconscious process by the first one. Cognitively, a facet shall be understood as a composite memory process, a classical analog of what Griffiths,  Omnes, and Gell-Mann and Hartle, called the consistent histories \cite{Griffiths1984,Omnes1988,Gell-Mann1990}. The quantification of consciousness proposed by Tononi and Edelman corresponds, for phase space variables, to free energy, and appears to be in agreement with the free energy principle proposed by Friston as an explanation for embodied perception \cite{Friston2006}. Indeed, the complex of minima of free energy can be understood as a topologically discrete and finite version of the free energy principle of Friston that can be applied in the multivariate case with heterogeneous variables. Information topology also agrees in principles with the model of "projective consciousness" of Rudrauff and colleagues \cite{Rudrauf2017}. This model proposes that the passage to a conscious perception relies on a change of geometry by fixing and changing of frames, from the affine or 3D-Euclidean to the 3D-projective, and is related to information since the action of the change of frame acts on the internal variable of the probability organized by a partial free energy. In this framework, it is also a mechanism of minimization of free energy which guides the changes of frames.We moreover propose to replace the "self-evident" axioms proposed in the work of Tononi and colleagues \cite{Oizumi2014} by the axioms of measure and probability theory, ultimately in the constructive logic framework that is sketched in the section dedicated to information topos \ref{topos}, and developed in the cited references. Such axiomatization may allow to pursue the "artificial" consciousness opus of Turing and Wiener in some more refined, modern and hopefully computationally efficient formalism (cf. section on the computational mind \ref{Computational mind2}).   
	The concept of "networks of networks" \cite{DAgostino2014a} corresponds topologically to the hypercohomology provided by the double complex of Hodge decomposition (complexes of complexes in a homological sense, or a derived functor). It hence may also account for the Dehaene-Changeux model, which involves global neuronal workspaces and which is a "meta neural network", a network of neural networks constructed with neural integrate-and-fire neurons, thalamo-cortical columns and long-range cortical area networks \cite{Dehaene2005,Dehaene2011,Dehaene2006}. Moreover, the minima of the complex corresponds to critical points which can be considered to correspond to the consciousness transition of their model. 
	\item The application to data and simple theoretical examples shows that the positive maxima of $I_k$ identify the variables that co-vary the most, which could be called covariant assemblies or modules in the neuronal context. Figure \ref{homeostasis}d (top) shows the kind of dependences identified by the maxima for 3 variables ($I_3$). We hence propose that such positive modules provide a statistically rooted definition of neural assemblies, generalizing correlation measures to the  nonlinear cases \cite{Reshef2011}. For example, the maximal $I_{10}$ module in Figure \ref{applied_TIDA}b could be legitimately called the DA cell assembly. The negative minima of $I_k$, commonly called synergistic interactions \cite{Brenner2000} or negentropy following Schr\"{o}dinger \cite{Schroedinger1944}, identify the variables that most segregate the population, and hence detect clusters corresponding to exclusive differential activity in subpopulations. This negativity of Free Energy component is discussed in \cite{Baudot2018} in the perspective of physic, and provides a topological signature of condensation phenomenon corresponding to the clustering of data point. It refines the negentropy principle of Schr\"{o}dinger, stating that living systems feed upon negentropy or free-energy, by showing that even free-energy can have some negative components. 
	It is remarkable that the pattern identified by positive and negative information corresponds to the two fundamental dual tasks of psychophysics, e.g. binding and segmentation, respectively. Moreover, minima of mutual information correspond in examples, and conjecturally in general to links, like the Borromean link (cf. section \ref{What is topology}). For example, the minima of $I_3$ for three Bernoulli variables is -1 bit' the variables are independent in pairs but linked at 3 by a purely 3-dimensional effect, a purely emergent collective interaction. 
\end{itemize}

These methods establish a topological version of the Boltzmann and Helmholtz machines in machine learning \cite{Ackley1985, Dayan1995}, named the Poincar\'{e}-Shannon machine. They also give a topological and algebraic answer, already present in essence in the work of Hu \cite{Hu1962}, to the questions of information decomposition that have been the subject of numerous publications and data applications, for instance the proposal of a non-negative composition by Williams and Beer \cite{Williams2010}, the "unique information" of Bertschinger and his colleagues \cite{Olbrich2015, Bertschinger2014}, Griffith and Koch \cite{Griffith2014} and the applications of the resulting information decomposition to the development of the neural network \cite{Wibral2017}, and neuromodulation \cite{Kay2017}.\\
In conclusion, those topological tools allow us to conciliate at least five important theories of consciousness, namely the global neuronal workspace model, the integrated information (IIT), the free energy principle, the projective model, and the dynamic logic, and confer on them an interesting topological foundation, allowing those theories to evolve and be improved with further discoveries in mathematics. Notably, it answers to the critics and requests concerning IIT further stated by Seth,  Izhikevich, Reeke, and  Edelman, \textit{"that characterizing the relevant complexity of such a system will require a multidimensional analysis[...] qualia space is a high-dimensional space in which the axes reflect dimensions on which phenomenally experienced conscious scenes are discriminated"} \cite{Seth2006}.
The original contribution of this model and of the topological view compared to those 5 theories, underlines the fact that the essential properties of consciousness rely on structure and shape, not a single function, a single number or scalar. Moreover, the formalism highlights the fact that conscious experience, and also biological structures in general, correspond to discrete symmetries, to local energy minima, and to dynamical stochastic process. Considering the fact that symmetry could be a mathematical definition of aesthetics, which is historically a canonical definition, the formalism also further joins the model of physical cognition and that of dynamic logic by Schoeller, Perlovsky and Arseniev \cite{Schoeller2018}: a Galoisian theory of e-motivs or e-motions, an ambiguous theory, \textit{"between crystal and smoke"} \cite{Atlan1979}, order and disorder, uncertainty and certainties (shared uncertainties) of the self and its environmental constitutive interactions.  In simple words, it justifies the subjective view that the world is beautiful, including you: the nicest conclusion we could find concerning a mathematical and physical theory of qualitative cognition.

\subsection{Dynamics, Geometries, action invariance and Homeostasis of consciousness}\label{Geometries-Homeostasis}

\subsubsection{The invariances to group actions of perception }\label{invariances to group actions}

\begin{figure} [!h]
	\centering
	\includegraphics[height=10cm]{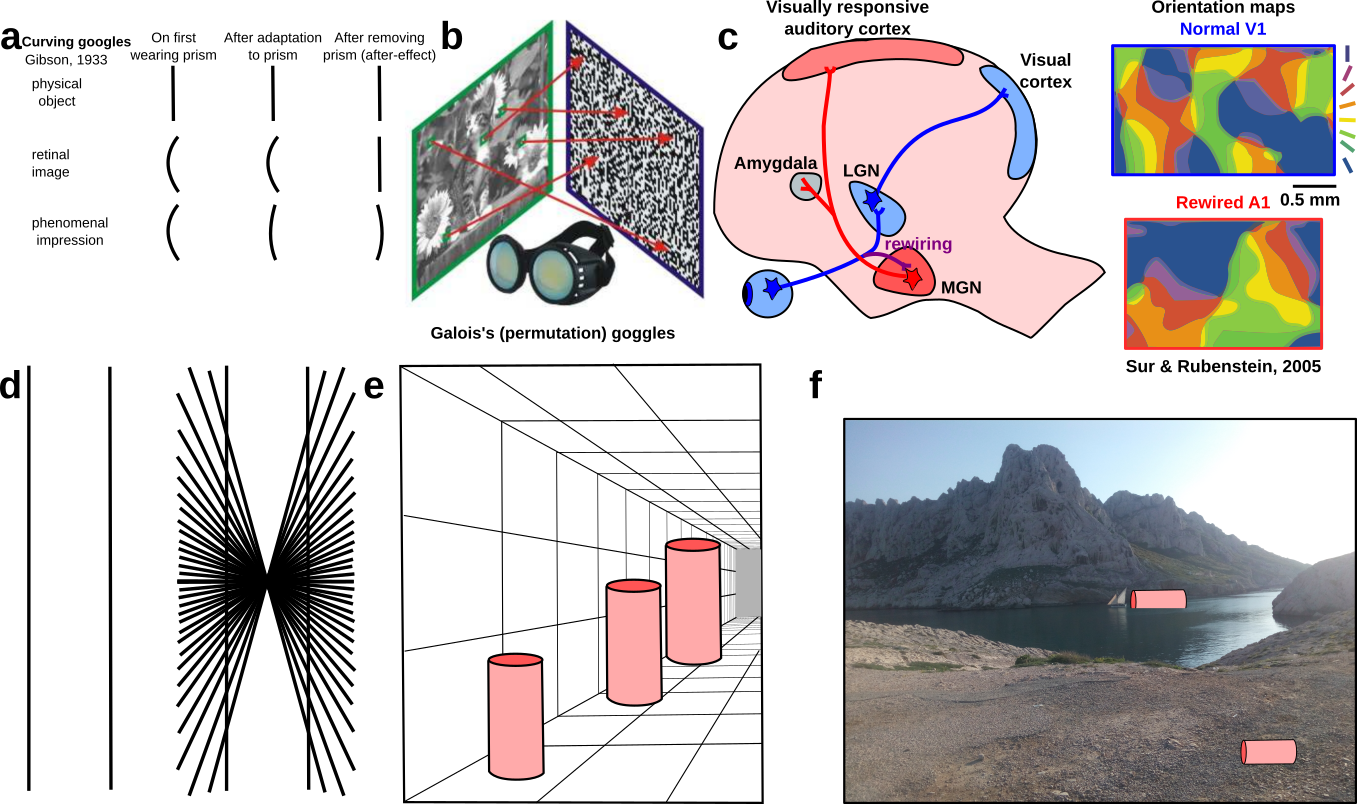}
	\caption{\textbf{Invariance to transformation and perceptual adaptation. The projective geometry aspects of perceptual space. a,} the adaptation to wearing goggles that impose curvature to visual retinal input, as reported by Gibson \cite{Gibson1933}. After 3 to 4 days the subject recovers an almost normal phenomenological perception of straight lines while removing the glasses induces an after-effect of curvature in the opposite direction (adapted and modified from Gibson \cite{Gibson1933}). \textbf{b,} A "gedankenexperiment" of permutation goggles that imposes a fixed given permutation of the photoreceptor input to the visual system, implementing the action of the symmetric group and testing the invariance to permutation of perception and consciousness. \textbf{c,} the experiment of rewiring visual input onto the auditory cortex realized by the team of Sur \cite{Sharma2000,Sur2001,Sur2005,Roe1990}. The ablation of the inferior colliculus, which normally provides auditory input, induces a rewiring of the optic nerve to the medial geniculate nucleus (MGN), which in turn induces an activity-dependent differentiation of the primary auditory cortex (A1) into a functional architecture of an almost-normal primary visual cortex (V1) exhibiting spatial retinotopy and orientation selectivity. The drawing on the left roughly reproduces orientation selectivity maps exhibiting typical organization of pinwheels on the cortical surface of a normal V1 neuron and a rewired A1 obtained using an optical imaging technique (adapted and modified with permission from  Sharma and colleagues \cite{Sharma2000}, see text). \textbf{d,e,f,} 3 different classical optical illusions induced by projective geometry (adding a point at infinity), implemented here by adding a contextual perspective. In \textbf{d,} the vertical lines appear curved whereas their retinal image is "straight" and parallel, and in \textbf{e and f,} the size of the barrels appears to depend on the perspective cues whereas their retinal image has the same size.}	
	\label{invariance}
\end{figure}

\textit{\textbf{"The research of invariant is the fundamental fact of perception"}}; this is in essence the fundamental principle proposed by Gibson, which gave rise to his ecological theory of perception \cite{Gibson1979} after Cassirer had introduced groups into the theory of perception \cite{Cassirer1938}, but it is also the central principle in Piaget's structuralist formalization of cognitive development \cite{Piaget1970}. A more mathematically rooted exposition of such principle, supporting modern neurophysiological results, can be found in Bennequin \cite{Bennequin2014}. The principle of considering an invariance to transformation as a kind of adaptive process was first highlighted by the "transforming Goggle experiments" of Straton \cite{Stratton1896,Stratton1897} and Erismann and Kohler \cite{Kohler1962}, which consisted in the study of visual and visuo-motor adaptation and the after-effects of long-term wearing of reversing mirror, prismatic or colored goggles. For example, after starting to wear goggles that invert left and right or flip the individual's vision upside-down, their vision progressively (i.e. within few days) goes back to their "usual" perception, demonstrating an adaptive visuomotor invariance to mirror-symmetry-transformation of the perceived world. As illustrated in Figure \ref{invariance}a, Gibson studied adaptation to deforming goggles that imposed curvature on the retinal image and discovered an invariance to a curving transformation that can be considered as diffeomormism or homeomorphism \cite{Gibson1933}. Figure \ref{invariance}a also presents the after-effects just after removing the curving goggles, manifested by a phenomenal perception curved in the opposite direction. It is possible to imagine other goggles associated with discrete transformation, such as the Galois or permutation goggles illustrated in  Figure \ref{invariance}b which permutes the light flux arriving on all photoreceptors with a given fixed permutation. According to the known architecture of visual processing, it is likely that adults would be barely able adapt to such a transformation, that would destroy the usual spatial retinotopic relations; or, adaptation would take time. However, from what is known of the development of the visual system, as exemplified by the rewiring experiment of Sur and colleagues \cite{Sharma2000,Sur2001,Sur2005,Roe1990}, we can infer that nonetheless, a newborn wearing such goggles would develop "normal" vision, but that the normal development and the fine wiring of the visual system is naturally endowed with an invariance to the action of permutation as a result of being predominantly ruled by activity-dependent plasticity. Sur et al's experiment consists of an ablation of the inferior colliculus (which provides the normal auditory input), which induces retinal afferents to innervate the medial geniculate nucleus (MGN), which is the normal relay of the auditory system. Such rewiring, which can be considered as a kind of permutation, induces a plastic differentiation of the primary auditory cortex (A1) that reproduces (with deformations) the usual main functional characteristics of the primary visual cortex (V1), complete with retinotopic and orientation-direction selectivity \cite{Sharma2000,Sur2001,Sur2005,Roe1990}. One could further reasonably infer from the "meta-plasticity" of the time-dependent plastic rules previously stated, that the developmental process would lead to an invariance in space-time permutation of the visual input, given that the permutations concern a time window of reasonably small duration. 

Invariance to transformation, formalized within an adequate group formalism, is a major stream in theoretical psychology; a review can be found in the work of Boring \cite{Boring1952} and Curtis \cite{Cutting1983}. Since those seminal works, psychophysics research has provided more precise formalization and empirical verification of Gibson's fundamental statement that identifies invariance to transformation with perceptual adaptation, and the problem since then has been to characterize the \textbf{"geometry of perceptual space"}, what could be called the \textbf{"shape of consciousness"}. Koenderink et al were able to reveal the fact that visual perceptive spaces partially present a "distorted" intrinsic affine \cite{Todd2001} and projective structure \cite{Koenderink2002}, as they verify Varignon's and Pappus's theorem using bisection and collinearity judgments tasks respectively. The effect of such projective geometry can be illustrated by classical optical illusions induced by perspective cues or contexts as illustrated in Figure \ref{invariance}d, e and f. These groups of transformations are organized into a Lie subgroup hierarchy, a Euclidian transformation being a special case of affine transformation, which is a special case of projective transformation, which is a special case of isomorphism. 
However, many experiments have revealed that perceptual space is more complex, and departs from homogeneous (or constant curvature), affine, projective or flat cases. Several experiments have demonstrated that the intrinsic curvature of perceptual space is non-Euclidean \cite{Battro1976,Indow1991,Koenderink2000} and that the curvature of perceptual space varies with position \cite{Indow1991,Koenderink2000}. With individual observers, 60\% of them display a negative curvature, while the other 40\% display a positive curvature \cite{Battro1976}. Koenderink and colleagues propose that these large variations in the metric structure of perceptual space reveal that the underlying "geometry of observers" depends on contextual factors (see Suppes \cite{Suppes1977}) such as objects in the visual field or the observer's contextual attention \cite{Todd2001}. To conclude, there appear to be no preferred intrinsic and fixed stable Riemannian metrics of perceptual space, indicating a possibly weaker and more generic topological invariance. This is indeed a central proposition of this review: perceptual geometry changes with experience, and these changes are called adaptation or learning - a topological change - whereas homeostasis is the signature of the resulting acquired stable geometry.
With regard to visual cortex functional architecture, Ermentrout and Cowan, and then Bressloff and colleagues, account for spontaneous activity pattern giving rise to typical visual hallucination with a planar model of V1 under an action of the Euclidean group $E(2)$ (translations, rotations, and a reflection) on the plane $\mathbb{R}^2$ \cite{Ermentrout1979} or on the plane with a hypercolumn pinwheel orientation structure $\mathbb{R}^2 \times S^1$ which preserves the structure of lateral cortical connections presented in the association field section - the shift-twist action \cite{Bressloff2001}. The previous thought experiment on permutation and plastic rewiring of Sur and colleagues indicates that the group of transformation should be much more generic than the Euclidean group $E(2)$, which is 2 dimensional and a very specialized-differentiated group of transformation. Indeed, as further developed in a generic groupoid formalism by Golubitsky et al \cite{Golubitsky2006a} and reviewed in \cite{Golubitsky2006b}, the group action approach can generally exploit symmetries in the connectivity of neural networks, with example of application in the primary visual cortex, but also in locomotor or vestibular system, giving rise to biologically relevant activity patterns.
As a conclusion on these studies, learning or adapting is acquiring a specialized geometry and the associated peculiar invariance structure, and obeys a topological hierarchical sequence of subgroups. Such a principle is, in essence, the basic definition of structural stability originally proposed by Thom and Petitot, discussed in the next paragraph, on which they based a general morphodynamic and semiotic theory \cite{Thom2002,Thom1977,Petitot1983}.

\subsubsection{Geometrical and Topological invariance, isomorphism }\label{Geometrical and Topological invariance}

\textbf{Invariance, stability and shape in mathematics:} After Riemann formalized his multiply extended manifoldness (measure) theory of space \cite{Riemann1854} and the discovery of non-Euclidian cases, finding an algebraically sound definition of a geometry was a central quest of mathematical research at the beginning of the 20th century. A consensual expression was given by Klein in the formulation of his Erlangen Program: a geometry is the action of a group on a space. "Given a manifoldness, and a group of transformations of the same; to develop the theory of invariants relating to that group" \cite{Klein1872, Birkhoff1988}. Klein's view of geometry generalized Riemann's original theory by replacing metrics with a group. Klein proposed the study of the homogeneous manifold: a structure $(M,G)$ consisting of a manifold $M$ and a group $G$ acting transitively on $M$, replacing Riemann's concept of a structure $(M,d)$ consisting of a manifold on which a metric $d(p,q)$ is defined by a local distance differential $ds^{2}=\sum g_{ij}dx_{i}dx_{j}$ \cite{Birkhoff1988}. The concept of invariance was then pursued in the general context of topology, defining topological invariance under the name of structural stability in the work of Thom \cite{Thom1977}, in the work of Smale \cite{Smale1967} in the context of differentiable dynamical systems, and of Mumford \cite{Mumford1994} in the context of algebraic varieties. Topological invariance is a weak invariance; topological invariants are the properties that are conserved under arbitrary deformations (homeomorphism) that preserve neighborhoods (local), sustaining the classical view of a rubber sheet geometry. Such invariance to deformation defines equivalence classes called isomorphisms. Thom defines structural stability as follow: \textit{“In every case where arbitrary small perturbation of initial conditions can conduct to very important variations in the subsequent evolution [...], it is possible to postulate that the phenomenon is deterministic; but it relies on a pure metaphysical statement inaccessible to any empirical verification. If one wonders controllable experimental properties, we will have to replace the unverifiable hypothesis of determinism by the empirically verifiable property of “structural stability”: "A process (P) is structurally stable, if a small variation of initial condition lead to a process (P') Isomorphic to (P) (in this sense that a small transformation in space-time, a $\epsilon$-homeomorphism, in geometry brings back the process (P') on the process (P))"} \cite{Thom1983}. \\ 

\subsubsection{Dynamical aspects of information, isomorphism, stability, and homeostasis}\label{Dynamical}
\begin{figure} [!h]
	\centering
	\includegraphics[height=10cm]{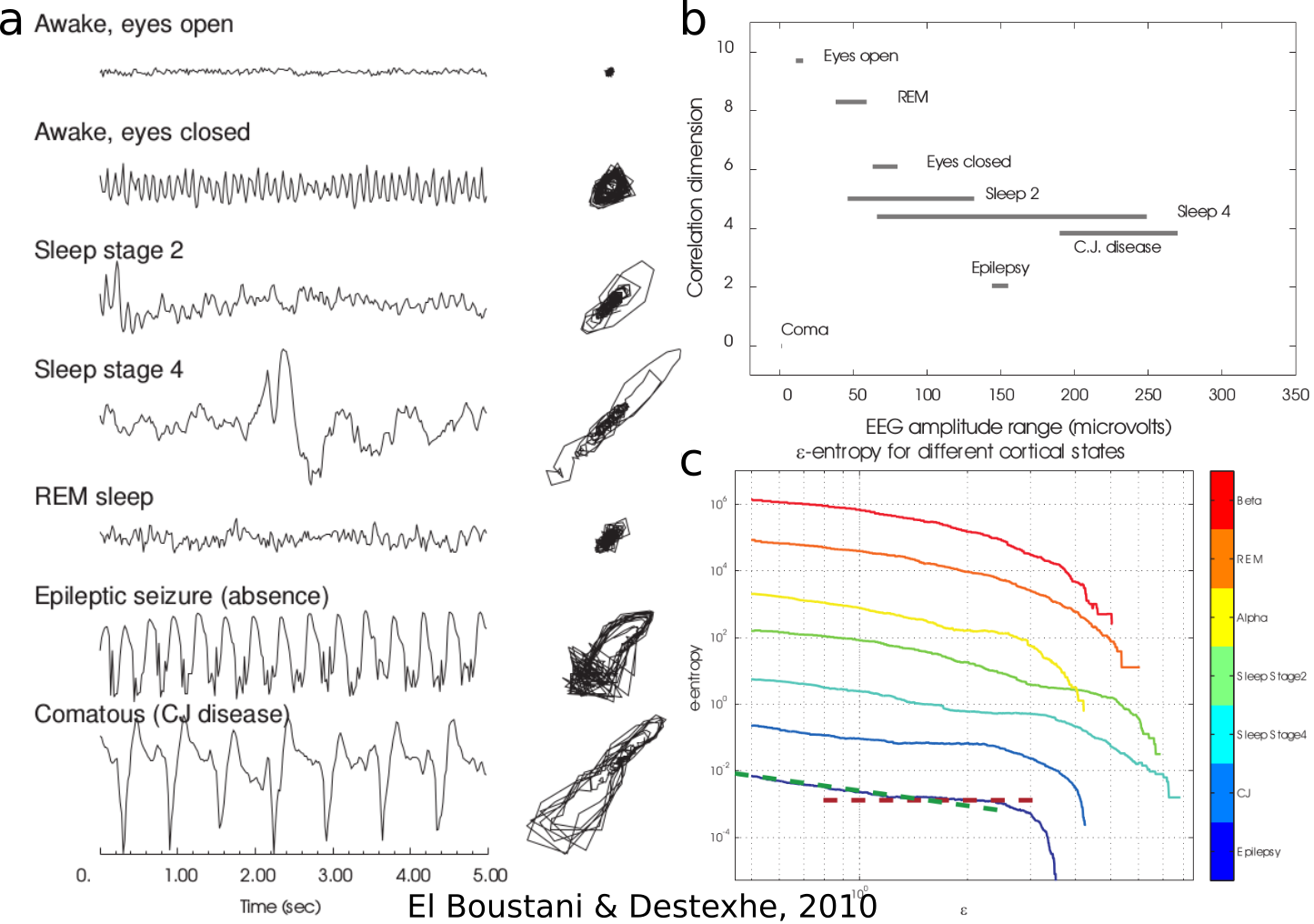}
	\caption{\textbf{Dimension, stability and $\epsilon$-entropy analysis of various arousal states. a,} 5 seconds of EEG recordings with the same amplitude scale (left) and their associated phase portraits during different brain states in humans. \textbf{b,} The correlation dimension is plotted as a function of the amplitude range of the EEG for different states. \textbf{c,} Scale-dependent $\epsilon$-entropy for
		different brain states' EEG recordings. The plateau in slow wave sleep and pathological states indicates the existence of a low-dimensional attractor on the corresponding scales (adapted and modified with permission from Destexhe and El-Boustani \cite{Destexhe1992} and \cite{El-Boustani2010})}	
	\label{eeg_entropy}
\end{figure}

This section asks what is the relation between dynamical system approaches of consciousness and information topology. At all scales of organization of nervous system, dynamical systems provided a guiding framework to model and predict the activity. For example at the cellular level, the electrical dynamic is investigated at length by the mean of dynamical system in the book of Izhikevich \cite{Izhikevich2007}, and the dynamical system study of neural network was pioneered by Sompolinsky and colleagues \cite{Sompolinsky1988}. What follows investigate dynamical systems from information topology point of view, in their simplest discrete and finite case, leaving the continuous cases, conjecture to be handled by the Lie algebra cohomology, for further studies. It hence provides only some preliminary directions and results upon the unification of those two fields that will be the subject of a more complete and formal work. 
\textbf{Invariance to conditioning (conditional expectation):} \label{dynamical_system} Information topology relies fundamentally on the action of random variables on information function, known as conditioning in usual information terms, a peculiar expectation integration summation with respect to a given variable. The invariance condition in information is explicitly given by $Y.H(X;P)=H(X;P)$, the definition given in the section on homeostasis \ref{Invariancedef}, which is equivalent to the statistical independence of $X$ and $Y$. Hence, a geometry in the sense of Klein, in the context of probability and of random variables and processes, can be defined by the preceding condition of invariance of information functions under the action of random variable, or in more biological terms, under the perturbation of the variable $Y$, and can be called an informational geometry. 
Such a stability or invariance condition is directly related to information negativity, since we have the following theorem: if $X$ is invariant to $Y$, then for any variable $Z$ we have $I(X;Y;Z)\leq 0$. In the current theory of information, which is commutative, the invariance is "symmetric", namely, if $X$ is invariant to $Y$, then $Y$ is invariant to $X$.
We saw that with regard to mutual informations, the $I_{k+1}$- or ($k+1$)-dimensional dependencies quantify the default of invariance for the conditioning of a k-dimensional system
$I(X_1;.;X_k;..;X_n)=I(X_1;.;\hat{X_k};..;X_n)-X_k.I(X_1;.;\hat{X_k};..;X_n)$, where the hat denotes the omission of the variable.\\
\textbf{Dynamical systems and information isomorphism:}
Following Thom's isomorphism and the structural stability framework, we can propose an isomorphism theorem for information structures and relate the stability condition for a stochastic process to classical results in dynamical systems, in which Shannon entropy plays an important role. The cohomology of dynamical systems proposed by Bergelson, Tao and Ziegler \cite{Bergelson2010} indeed appears to be the usual Hochschild cohomology with left action on which information cohomology is based, as notably discussed and shown in Tao's blog \cite{Tao2008a}:\\
\textbf{Information structure isomorphism:} let $X^n$ and $Y^n$ be two complexes of random variables; $X^n$ and $Y^n$ are information isomorphic if for whatever subset $X^k$ of $X^n$ and whatever subset $Y^k$ of $Y^n$ the information $I(X^k;Y^k)=I(X^k)=I(Y^k)$ (or equivalently $H(X^k;Y^k)=H(X^k)=H(Y^k)$).\\ 
Proof:  $H(X^k,Y^k)=H(X^k)$ is equivalent to $X^k.H(Y^k)=0$ and thus to the fact that $Y^k$ is a deterministic function of $X^k$. Reciprocally if $H(X^k;Y^k)=H(Y^k)$, then $X^k$ is a deterministic function of $Y^k$. Hence $Y^k$ and $X^k$ are isomorphic. If it is true for whatever subset $[k]$ of $[n]$, it is true for $[n]$ $\Box$. \\
This theorem includes as a special case of Bernoulli shifts, a part of the Ornstein-Kolmogorov isomorphism theorem which states: \\
\textbf{Ornstein-Kolmogorov Isomorphism theorem \cite{Ornstein1971,Ornstein1970}}:
All Bernoulli shifts with the same entropy are isomorphic.\\
Proof: let us note the two Bernoulli shifts $X^n$ and $Y^n$; since they are Bernoulli shifts they are independent processes \cite{Ornstein1971,Ornstein1970}) and hence $I(X^k)=I(Y^k)=0$ for all subsets of $k\geq2$ elements of $X^n$ and $Y^n$. Moreover, the variables $X_1,...,X_n$ are by definition identically distributed, and we hence have $H(X_1)=...=H(X_n)$, which is also the case for $Y_1,...,Y_n$ and $H(Y_1)=...=H(Y_n)$. In such a case, the preceding informational isomorphism condition $H(X^k;Y^k)=H(X^k)=H(Y^k)$ is reduced to the condition $H(X_1)=H(Y_1)$, the Kolmogorov-Ornstein theorem.$\Box$\\
Figure \ref{integer_partition} provides an illustration of Bernoulli shifts that are isomorphic, discovered by Mesalkin \cite{Ornstein1970}. \textbf{From the cognitive point of view, we propose that two informationally isomorphic processes have the same qualitative experience.}
Entropy, since the work of Kolmogorov, Sina\"{\i} and Ornstein, has been one of the main invariants of ergodic theory and has driven the study of dynamical systems, as discussed at length in Katok's review \cite{Katok2007}. The quantification of dynamical systems by entropy relies on attaching a number to an action of a countable group $G$ that preserves the probability measure in Borel space $X$. The Kolmogorov-Sinai entropy is Shannon entropy on this basis, and a short review of the development of the theory to deal with non-amenable groups is provided by Gaboriau \cite{Gaboriau2016}. The Ornstein-Kolmogorov theorem works for group action when the group is $\mathbb{Z}$, and Ornstein and Weiss could showed that it holds for any countable amenable groups including commutative groups \cite{Ornstein1987}. The introduction of amenable groups allows making the bridge with the constructive logic that avoids the Axiom of infinite Choice presented in section \ref{measure}. Von Neumann defined amenable groups as groups with an invariant mean, which includes all finite and all solvable groups, in order to isolate the groups that are subject to the Banach-Tarski paradox \cite{Neumann1929}. The following theorem credited to Tarski is more explicit: $G$  is non-amenable if and only if $G$ is paradoxical. Hence in constructive mathematics, or in Solovay's theory, all groups are amenable, and Ornstein-Kolmogorov isomorphism holds without restriction. So considering the constructive theory of cognition and consciousness, information theory provides a generic quantification of consciousness structure. These studies led Ornstein to conclude that some deterministic dynamical systems and Newtonian dynamics cannot be distinguished from probabilistic systems (and are described by the same informational invariants) \cite{Ornstein2004}.
Concerning stability and instability quantification and entropy, these developments notably led Pesin to develop a theory for which the entropy of a measure is given exactly by the total expansion in the system, the sum of all positive Lyapunov (expansive/unstable) exponents \cite{Pesin1977}:
\begin{equation}
H(X^n;P)=\sum_{i=1}^n \lambda^+_i \operatorname{dim}E_i
\end{equation} 
where $P$ is a Riemann measure of the Riemannian manifold $M$, and holds if and only if $P$ is a Sinai-Ruelle-Bowen (SRB) measure. Although we are conscious that the context of information cohomology is different from Pesin's theory, conditional mutual information and its sign play an analog role of Lyapunov exponents  Lyapunov exponents, whose sign indicates stability or instability, while the complex of free energy summing over information paths with positive conditional information appears analog to Pesin's formula. The context is different, however' instead of $n$ Lyapunov exponents for ergodic theory, in the simplest simplicial case we have $n.2^n$ conditional informations. Lyapunov exponents, correlation dimensions and entropy have been used to characterize arousal states, commonly considered as levels of consciousness, further supporting the view that "fully" conscious awake states are high-dimensional chaotic dynamics, usually called complex states. Such a dynamical system characterization and quantification of consciousness could be termed a Newtonian theory of consciousness. EEG recordings, because of their macroscopic resolution, impose an important underestimation of the dimensions and complexity of arousal states. Figure \ref{eeg_entropy} presents the results of the study of El Boustani and Destexhe into EEG recordings of various arousal states, ranging from coma to awake, their associated correlation dimensions and their $\epsilon$-entropy (related to the finite-size Lyapunov exponent). $\epsilon$-entropy is
a generalization of the Kolmogorov-Sinai entropy rate proposed by Gaspard and Wang, \cite{Gaspard1993} which is defined for a finite scale $\epsilon$ and time delay $\tau$ by $h(\epsilon,\tau)=\frac{1}{\tau} \lim_{m \rightarrow \infty} \frac{1}{m} H_m(\epsilon,\tau)$, where $ H_m(\epsilon,\tau)$ is the entropy estimated with a box partition of the phase space for box size given by $\epsilon$ on the attractor, reconstructed with a time delay $\tau$ and an embedding dimension $m$.

\textbf{Homeostasis and multi-equilibrium:} The definition of homeostasis given in  section \ref{homeostasis}  and its associated figure corresponds to the equilibrium condition of vanishing conditional mutual information, conditional independence, and, in the context of information paths and topology, to the minima of free energy, hence corresponding to the usual definition of equilibrium in an organism. Given the definition of a minimum free energy complex, all complex systems are in a homeostatic state or are the result of a homeostatic process, while adaptation and learning correspond to changes in the geometry, changes of the minima and hence of the complex. From an empirical point of view, this definition of homeostasis corresponds exactly to the concept and measures of homeostasis in biology; a process is homeostatic when the perturbation X or the removing of X (like a K.O.) changes nothing in the observed system (the information structure). The signature of such invariance to $X$ is hence a null slope in the corresponding information path segment. \textbf{Homeostasis hence corresponds to the maintenance of the shape, of the structure-function}.

\subsection{Computational mind - from Cybernetic and AI to biological beliefs} \label{Computational mind2}

\textit{"Every organic body of a living being is a species of divine machine, or a natural automaton, which infinitely surpasses all artificial automata."}  Leibniz \cite{Leibniz1714}. 

\subsubsection{Computation, machines and consciousness}
In his editorial to "Special issue: Challenges in neuroscience" \cite{Stern2017}, Stern asked the insidious question "Why do computers presently lack consciousness, and when might they acquire it?", which implicitly assumes the truth of the statement "computer presently lack consciousness" and avoids any discussion about the fundamental problem that has motivated some key research and developments since Turing. What has been presented here supposes, on basis of the principle of observability, that current computers have a particular form of consciousness, but also offers such a status to a coffee machine, that basically implements thermodynamic principles. The Nagel's question quoted in the first chapter "What is it like to be a bat?", has hence become the question "What is it like to be a coffee machine?" \cite{Nagel1974}, which, at first glance, appears much easier to answer: probably quite boring, except maybe when they add milk. Before laughing, beware that this is not so far from the definition of a mathematician by Erd\"{o}s: \textit{"A mathematician is a device for turning coffee into theorems"}. Indeed, the review of logic, Turing machines, artificial intelligence and machine learning made here has shown that they should be considered some of the first efficient synthetic biology results, or at least synthetic cognitive theories, with historical and scientific results supporting such a thesis. Just as Turing creating his famous test, it is only possible to judge the consciousness or intelligence of a system from its output and input, its actions in the real world - all the observables that can be measured from external point of view. Wiener, one of the fathers of cybernetics, also suppported such a conclusion from the early days of the field. The question of computer or robots rights, or those of other synthetic biology constructions, is difficult, and will be probably asked in the future. An extension and improvement of human rights as occurred during the 18th century will probably have to be considered. We think that the principle of invariance to an action, which highlights a diversity of possible equivalence relations, possibly contextual, may provide a much richer and softer fundamental relation than the rigid equality stated in human rights (which is not biologically rigorous), while still respecting the fundamental equivalence of humans' right actions. Basically current computers and current humans, are not equal; they are in some cases equivalent but in most cases clearly different with respect to some tasks. Computers notably surpass usual human abilities in the reliability and precision of their computations, a fact that has allowed computing machines to acquire the trust of mathematicians. Mathematicians consider computing machine as deriving rigorous mathematical proofs and theorems (such as in the case of the proof assistant software coq \cite{Bertot2015}) despite the possible errors induced by thermal noise (which are pragmatically considered as less likely than human errors). In other words, mathematicians consider computers as their equivalent with respect to the mathematical task of producing theorems (which is their very basic purpose). This is one reason, in our very subjective opinion, to respect them as intelligent beings. With regard to AI algorithms, in 1997, Deep Blue beat Kasparov at chess, and today, AIs have beat human players in a variety of different games, from Go \cite{Silver2017} to the card game Poker and many Atari video games with the same algorithm and set up \cite{Mnih2015}. Leibniz's view, summarised by the citation at the beginning of this section, turned to be partially wrong; artificial automata now beat humans in specific tasks when the sizes of possible states or actions remain relatively small. Such games tests can be considered as restricted task-specific Turing tests which are not linguistically directed. The sets of inputs and tasks humans treat and achieve represent a significantly bigger space of possible states and actions, including motion in 3 dimensions, linguistic abilities etc., although this is difficult to quantify. The improvements of computers' performance has been possible, notably if not mainly, thanks to the computational power increase that occurred within the few last decades and algorithm improvements, together with the decrease in the cognitive pretension of the tasks. Retrospectively, inaugurating AI in the 1950's with a test like the Turing test was a complete underestimation and misunderstanding of the computational task and of the underlying cognitive resources, and has been the source of many failures, notably of the subsequent renaming of the fields AI, cybernetics, cognitive science, machine learning etc.. In the numerical world of the web, the CAPTCHA security system designed to prevent robots from visiting websites is nothing but a reversed Turing test. Humans are now effectively faced with new forms of intelligence and of consciousness not so far from his own, and the predicted increase of computational capacity will aim to complete the panel of human tasks that machines can effectively achieve. 
The main question now is whether it is possible to significantly increase computational power, notably by effectively taking advantage of the non-deterministic or quantum nature of computation, which would bring the consciousness structure and level (dimension) of our machines close to our. 

\subsubsection{Computational hardness of consciousness} \label{subsection 2}

A general conclusion that comes out of information topology analysis regarding consciousness and biological structures concerns their computational hardiness with respect to the usual computational definitions based around Turing machines and boolean classical logic. Their application to data reveals that the dependences and independences sustaining consciousness and participating in the free energy functional in high dimensions exist in extremely large numbers; there is combinatorial explosion of interactions analogous to the effect that occurs in Van der Walls interactions \cite{Baudot2018}. Such a problem is well known in physics, which has dedicated many-body interactions to it, notably in Density Functional Theory, and Kohn in his Nobel lecture called this computational problem the exponential wall \cite{Kohn1999}. 
In the case of general classical information structures, not even considering quantum, computational complexity follows Bell's combinatoric in  $\mathcal{O}(exp(exp(N^n))$) for $n$ $N$-ary variables; for example, considering 16 variables that can take 8 values each, we have $8^{16}=2^{48} \approx 3.10^{14}$ atomic probabilities and the partition lattice exhibits around $e^{e^{2^{48}}-1}\geq2^{200} $ elements to compute, and hence requires a significantly new form of computing resources - our classical Turing machines, clusters or hypercomputers cannot face such complexity \cite{Baudot2018,TapiaPacheco2017}. Considering a restricted  simplicial structure, complexity falls to $2^n$, which can be explored in few hours with $n=21$ variables using a simple personal computer. Yet 21 is a small figure with respect to an Avogadro number of particles, a mole of matter, multiplied by 6 (for each position and momenta), and even with such restrictions, computing the information structure and energies would require other methods. Many studies pointed out the fascinating capacity of even "simple" biological systems to solve computationally hard tasks efficiently \cite{Aloupis2012}\cite{Nakagaki2000} (see the field of research on swarm intelligence), and the present results emphasize this view of the unreasonable effectiveness of natural sciences in mathematics and computation \cite{Wigner1960} (a trivial observation since mathematics is produced by natural humans). Non-deterministic Turing machines, whose time complexity overcomes deterministic Turing ones, appear pertinent to computationally formalize such a biological calculus \cite{Blazewicz2012}. As we outlined previously the constructive probabilistic logic that goes hand in hand with information topology, it would be reasonable to ask what computational resource would be adequate to effectively compute it. Analog computing, culminating with quantum computing, appears as an obvious possibility. With a small step beyond this reflection, it appears that human should indeed be a reasonable computable resource for informational and probabilistic logic calculus, and one can reasonably ask the motivation for the idea of replacing or outperforming human cognition. Alternatively, it is also possible to consider in the future a co-evolution of human and machine cognition and consciousnesses, the pitfall being the possibility of the creation of a new slavery status version 10.4 (considering the equivalent output, equivalent freedom and equivalent rights of humans and machines in this hypothetical situation).        

\section{Conclusion - the global ecological synthesis}

The most obvious conclusion of this work is that consciousness is a natural and physical phenomenon, in principle ubiquitous, revealing itself in many different forms, that our human, highly specialized consciousness can hardly understand, imagine or even conceive. As a biologist or naturalist considering observed interdependencies and interactions, the almost trivial conclusion is that respect is a necessary condition for the stable and normal development of the self. This synthesis was proposed by the ecological theory of mind and biology inaugurated by Gibson \cite{Gibson1979} and later formulated clearly by Atick \cite{Atick1992} in information and adaptive terms. On the mathematical side it is currently promoted and developed by Baez, Fritz and Leinster and all the collaborators of the azimuth project and the Complex System community represented by the CS-DC following Bourgine. These are the most useful aspects we could find about the qualitative aspects of consciousness theory; the rest is just for "the honor" (... or the beauty...) "of the human spirit", whatever that spirit may be, following Hilbert and Dieudonn\'{e} \cite{Dieudonne1987a,Hilbert1924}. 
Information topology should be conceived as an environmentally embedded theory of cognition, a global ecology, providing a basic preliminary formalization and quantification of ecology (the modern name of Analysis Situs). The usual definition of ecology is the science that studies relations among living beings (animals, plants, micro-organisms, etc.) with their habitat and environment as well as with other living beings. Information topology formalizes an ecosystem as a complex system, i.e. a complex of free energy minima, and these methods provide rigorous means of quantification:
\begin{itemize}
	\item statistical, collective interactions in ecosystems including synergistic interactions.
	\item  diversity (biodiversity).
\end{itemize}
These methods include tools relevant to the issues of sustainable development:
\begin{itemize}
	\item  Risk identification: entropies quantify uncertainty.
	\item  Resource identification: Mutual information quantifies available (free) energies. 
\end{itemize}
We hope that the quantitative aspects of informations will be of help in the ecological and social fields. 
However, from this exposition it appears clearly that the quantification of the precise information in a real system, such as a protein, a neuron, a cell or a "network", is far from being achieved, and that we have access to a very tiny window on what information structures really are in biological systems due to limited computational and experimental resources. Moreover, the quantification and monitoring, a la Laplace or the Human Brain Project, of all this information, of a given precise experimental model and form of cognition, is probably not that interesting or even useful, beyond the answering of certain precise, physiologically-motivated questions.  Beyond the question of the mathematical and physical nature of consciousness, we believe that the interesting problematics rely on the methods and tools that are used or constructed to gain in precision upon such a positive answer, and researchers in cognitive science and neuroscience gave a large panel of refined methods to answer yes to this question. From a theoretical point of view, the challenges are based more around principles and machine learning development: the fields to explore are immense, from the fundamental to more applied computational problems. Probability theory is essentially unknown \cite{Gromov2018} and has to be rewritten in a broader, more empirically rooted, and modern mathematical context; a good attempt in the field of category theory is currently being worked on by Fritz and Perrone \cite{Fritz2018}. The empirical view tends to think that integer partitions will ultimately play a foundational role, underlining the important role of the sample size  divisor "m", as outlined here (although this was more a question than an answer). As regards information theory, the work to be achieved is just as large, firstly because probability theory and information are now fully indistinguishable, as Kolmogorov suspected they would become \cite{Kolmogorov1983}, although they are more precisely probably interrelated by a dual Homology-cohomology relation which is yet beyond our current knowledge. Secondly, because most of the conditions that are classically imposed on information theory and statistical mechanics are unnecessary, the essence of information theory still works without ergodic, Markov, iid or asymptotic hypotheses; some of the urgent questions regarding those aspects are listed in \cite{Baudot2018}. It just points out that the mathematical theory of information and communication is not yet written: we only have elementary cues of what it will be. Notably, we hope that the Asymptotic Equi-Partition theorem will occur as a special case of a multivalued constructive logic. We saw that renormalization techniques still provide some of the most promising tools concerning data analysis; however, these methods are also essentially not understood. The fundamental objection of Feynman and Dirac is that neglecting infinite energy quantities is a mathematical weakness of the theory that still holds. Here again, constructive logic (bearing in mind that "all functions are continuous", a view that was originally considered as the default, may be helpful) and information topology provide a possible backbone for such further development. Moreover, we have only tackled very indirectly here the fundamental problem of consciousness and space-time and the question of how we acquire "distinct" representations of space and time following Piaget, since the answer from the topological point of view is beyond the current methods.\\
It should be clear that a precise and unified description and account of complex phenomenon such as the consciousness we experience unavoidably requires the use of the big machinery of algebraic topology and category, and even challenges it. The most basic reason for this is that it contains in its very constitutive foundation the germs of diversity, which are lost when one adds very few supplementary axioms or considers more specialized theories.

\appendix
\section{The topology of psychophysic according to Poincar\'{e}} \label{topology of psychophysic}

"\textbf{The Physical Continuum \cite{Poincare1902}.} We are next led to ask if the idea of the mathematical continuum is not simply drawn from experiment. If that be so, the rough data of experiment, which are our sensations, could be measured. We might, indeed, be tempted to believe that this is so, for in recent times there has been an attempt to measure them, and a law has even been formulated, known as \textbf{Fechner's law}, according to which sensation is proportional to the logarithm of the stimulus. But if we examine the experiments by which the endeavour has been made to establish this law, we shall be led to a diametrically opposite conclusion. It has, for instance, been observed that a weight A of 10 grammes and a weight B of 11 grammes
produced identical sensations, that the weight B could no longer be distinguished from a weight C of 12 grammes, but that the weight A was readily distinguished from the weight C. Thus the rough results of the experiments may be expressed by the following relations: $A=B,\;B=C,\;A<C$, which may be regarded as the formula of the physical continuum. But
here is an intolerable disagreement with the law of contradiction, and the necessity of banishing this disagreement has compelled us to invent the mathematical continuum. We are therefore forced to conclude that this notion has been created entirely by the mind, but it is
experiment that has provided the opportunity."\\

\textbf{Physical continuum of several Dimension }\cite{Poincare1905} "I have explained in 'Science Hypothesis' whence we derive the notion of physical continuity and how that of mathematical continuity has arisen from it. It happens that we are capable of distinguishing two impressions one from the other, while each is indistinguishable from a third. Thus we can readily distinguish a weight of 12 grams from a weight of 10 grams, while a weight of 11 grams could neither be distinguished from the one nor the other. Such a statement, translated into symbols, may be written: $A=B,\;B=C,\;A<C$. This would be the formula of the physical continuum, as crude experience gives it to us, whence arises an intolerable contradiction that has been obviated by the introduction of the mathematical continuum. This is a scale of which the steps (commensurable or incommensurable numbers) are infinite in number, but are exterior to one another instead of encroaching
on one another as do the elements of the physical continuum, in conformity with the preceding formula. The physical continuum is, so to speak, a nebula not resolved; the most perfect instruments could not attain to its resolution. Doubtless if we measured the weights with a good balance instead of judging them by the hand, we could distinguish
the weight of 11 grams from those of 10 and 12 grams, and our formula would become: $A<B,\;B<C,\;A<C$. But we should always find between A and B and between B and C new elements D and E, such that $A=D,\;D=B,\;A<B,\;B=E,\;E=C,\;B<C$, and the difficulty would only have receded and the nebula would always
remain unresolved; the mind alone can resolve it and the mathematical continuum it is which is the nebula resolved into stars. Yet up to this point we have not introduced the notion of the number of dimensions. What is meant when we say that a mathematical continuum or that a physical continuum has two or three dimensions? First we must introduce the notion of cut, studying first physical continua. We have seen what characterizes the physical continuum. Each of the elements of
this continuum consists of a\textbf{ manifold of impressions}; and it may happen either that an element can not be discriminated from another element of the same continuum, if this new element corresponds to a manifold of impressions not sufficiently different, or, on the contrary, that the discrimination is possible; finally it may happen that two elements indistinguishable from a third, may, nevertheless, be distinguished one from the other. That postulated, if A and B are two distinguishable elements of a continuum C, a series of elements may be found, $E_{1},E_{2},...,E_{n}$ all belonging to this same continuum C and such that each of them is indistinguishable from the preceding, that $E_{1}$ is indistinguishable from A and $E_{n}$ indistinguishable from B. Therefore we can go from A to B by a continuous route and without quitting C. If this condition is fulfilled for any
two elements A and B of the continuum C, we may say that this continuum C is all in one piece. Now let us distinguish certain of the elements of C which may either be all distinguishable from one another, or themselves form one or several continua. The assemblage of the elements thus chosen arbitrarily among all those of C will form what I shall call the cut or the cuts. Take on C any two elements A and B. Either we can also find a series of elements $E_{1},E_{2},...,E_{n}$, such: \\
(1) that they all belong to C;\\
(2) that each of them is indistinguishable from the following, $E_{1}$ is indistinguishable from A and $E_{n}$ indistinguishable from B;\\
(3) and beside that none of the elements E is indistinguishable from any element of the cut. Or else, on the contrary, in each of the series, $E_{1},E_{2},...,E_{n}$ satisfying the first two conditions, there will be an element $E$ indistinguishable from one of the elements of the cut.\\

In the first case we can go from A to B by a continuous route without quitting C and without meeting the cuts; in the second case that is impossible. If then for any two elements A and B of the continuum C, it is always the first case which presents itself, we shall say that C remains all in one piece despite the cuts. Thus, if we choose the cuts in a certain way, otherwise arbitrary, it may happen either that the continuum remains all in one piece or that it does not remain all in one piece; in this latter hypothesis we shall then say that it is divided by the cuts. It will be noticed that all these definitions are constructed in setting out solely from this very simple fact, that two manifolds of impressions sometimes can be discriminated, sometimes can not be. That postulated
if to divide a continuum, it suffices to consider as cuts a certain number of elements all distinguishable from one another, we say that this continuum is of one dimension; if, on the contrary, to divide a continuum, it is necessary to consider as cuts a system of elements themselves forming one or several continua, we shall say that this continuum is of several dimension. If to divide a continuum C, cuts forming one or several continua of one dimension suffice, we shall
say that C is a continuum of two dimension; if cuts suffice which form one or several continua of two dimensions at most, we shall say that C is a continuum of three dimensions; and so on. To justify this definition it is proper to see whether it is in this way that geometers introduce the notion of three dimensions at the beginning of their works. Now, what do we see? Usually they begin by defining surfaces as the boundaries of solids or pieces of space, lines as the boundaries of surfaces, points as the boundaries of lines, and they affirm that the same procedure can not be pushed further.
This is just the idea given above: to divide space, cuts that are called surfaces are necessary; to divide surfaces, cuts that are called lines are necessary; to divide lines, cuts that are called points are necessary; we can go no further, the point can not be divided, so the point is not a continuum. Then lines which can be divided by cuts which are not continua will be continua of one dimension; surfaces which can be divided by continuous cuts of one dimension will be continua of two dimensions; finally space which can be divided by continuous cuts of two dimensions will be a continuum of three dimensions."...\\
"The formula $A>C,\;A=B,\;B=C$, which summed up the data of crude experience, implied an intolerable contradiction. To get free from it it was necessary to introduce a new notion while still respecting the essential characteristics of the physical continuum of several dimensions. The mathematical continuum of one dimension admitted of a scale whose divisions, infinite
in number, corresponded to the different values, commensurable or not, of one same magnitude. To have the mathematical continuum of n dimensions, it will suffice to take n like scales whose divisions correspond to different values of n independent magnitudes called coordinates. We thus shall have an image of the physical continuum of n dimensions, and this image will be as faithful as it can be after the determination not to allow the contradiction of which I spoke above."

\section{The objective poetry} \label{The objective poetry}

"Who's there?", What is life?... Those questions pertain to anybody, as do their answers. Academic science provides many trails of an answer to those questions, but art, and notably poets, have worked out nice answers. What was presented here was a vulgar and laborious version of what Rimbaud called the “objective poetry”, a quest of “a universal language”, a universal living algebra.
\textit{"Algebra is nothing but a written geometry; geometry is nothing but a depicted algebra"} (Sophie Germain). It is this common essence of geometry and algebra that topology aims to catch. As originally defined by Leibniz with analysis situ or qualitative geometry, topology also aims to put in correspondence two dual worlds, quantities-numbers and qualitative-forms, which is indeed the original idea of Harmonia in mathematics and science developed by Pythagorean school. Scientific research is also a quest, a spiritual and aesthetic quest, as expressed by Schoeller and colleagues \cite{Schoeller2018}. In this sense, \textbf{Mathematics is pure Poetry}, and science an objective poetry, the intimate language of nature and of our sensations.
Indeed, the essence of the ideas proposed here was stated much more nicely by de Nerval: 

\textit{“ Well then - all things feel! ”} Pythagoras\\
\textit{Golden Verses\\
	Man ! Free thinker - do you believe that you alone can think\\
	In this world, where life bursts forth in everything :\\
	Forces you hold your freedom dispose,\\
	But from all your advices the universe is absent.\\
	Respect in the beast an acting spirit : ...\\
	Each flower is a soul of the bloomed Nature ;\\
	A mystery of love in the metal repose :\\
	"All things feel !" - And everything on your being is powerfull !\\
	Fears in the blind wall a glance watching you\\
	Even to the matter a verb is attached ...\\
	Do not make it serve to some impuous use !\\
	Often in the obscure being lives a hidden god ;\\
	And like a nascent eye covered by its lids,\\
	A pure spirit grows under the bark of stones !}\\

Gerard de Nerval, 1853.  

\section*{Supplementary material}
The software Infotopo that computes all basic information functions and the Information Topological Analysis is available at https://github.com/pierrebaudot/INFOTOPO \label{supplementary material} 

\section*{Acknowledgement}
This work was funded by the European Research Council (ERC consolidator grant 616827 \textit{CanaloHmics}), developed at UNIS Inserm 1072 - Universit\'{e} Aix-Marseille , and thanks previously to supports and hostings since 2007 of Max Planck Institute for Mathematic in the Sciences (MPI-MIS) and Complex System Instititute Paris-Ile-de-France (ISC-PIF) and Institut de Math\'{e}matiques de Jussieu - Paris Rive Gauche (IMJ-PRG). This work is dedicated to Daniel Bennequin and addresses a deep and warm acknowledgement to the researchers who participated to its development: i) for the electrophysiological and psychophysical part: Fr\'{e}d\'{e}ric Chavane, Yves Fr\'{e}gnac, Sebastien Georges, Manuel Levy, Jean Lorenceau, Olivier Marre, Cyril Monier, Marc Pananceau, Peggy Series, ii) for the gene expression and homeostasis part: Jean-Marc Goaillard, Monica Tapia  iii) for the topological part: Daniel Bennequin, Juan-Pablo Vigneaux iii) for their encouragement, support and help: Henri Atlan, Fr\'{e}d\'{e}ric Barbaresco, Habib B\'{e}nali, Paul Bourgine, Andrea Brovelli, J\"{u}rgen Jost, Guillaume Marrelec, Ali Mohammad-Djafari, Jean-Pierre Nadal, Jean Petitot, Alessandro Sarti, Jonathan Touboul.\\
The author declare no competing financial interests.

\bibliographystyle{unsrt}
\bibliography{bibtopo}

\begin{thebibliography}{100}

\bibitem{Tononi1998}
G.~Tononi and G.M. Edelman.
\newblock Consciousness and complexity.
\newblock {\em Science}, 282:1846--1851, 1998.

\bibitem{Baudot2015a}
P.~Baudot and D.~Bennequin.
\newblock The homological nature of entropy.
\newblock {\em Entropy}, 17(5):3253--3318, 2015.

\bibitem{Bennequin2014}
D.~Bennequin.
\newblock {\em Remarks on Invariance in the Primary Visual Systems of Mammals},
  pages 243--333.
\newblock Neuromathematics of Vision Part of the series Lecture Notes in
  Morphogenesis Springer, 2014.

\bibitem{Riemann1876}
B.~Riemann.
\newblock Philosophical opuscule.
\newblock {\em in Bernhard Riemann's Gesammelte mathematische Werke und
  Wissenschaftlicher Nachlass, University of California Libraries}, 1876.

\bibitem{Poincare1905}
H.~Poincare.
\newblock {\em The Value of Science}.
\newblock Paris: Flammarion. http://www3.nd.edu/~powers/ame.60611/poincare.pdf,
  1905.

\bibitem{Poincare1902}
H.~Poincare.
\newblock {\em Science and Hypothesis}.
\newblock London W. Scott.
  https://archive.org/details/scienceandhypoth00poinuoft, 1902.

\bibitem{Thom1983}
R.~Thom.
\newblock {\em Mathematical models of morphogenesis}.
\newblock Ellis Horwood series, 1983.

\bibitem{Yartsev2017}
M.M. Yartsev.
\newblock The emperor’s new wardrobe: Rebalancing diversity of animal models
  in neuroscience research.
\newblock {\em Science}, 358(6362):466--469, 2017.

\bibitem{Paulson2017}
S.~Paulson.
\newblock The spiritual, reductionist consciousness of christof koch: What the
  neuroscientist is discovering is both humbling and frightening him.
\newblock {\em Nautilus}, 2017.

\bibitem{Nagel1974}
T.~Nagel.
\newblock What is it like to be a bat?
\newblock {\em The Philosophical Review}, 4:435--450., 1974.

\bibitem{Montesquieu1748}
Montesquieu.
\newblock {\em De l'Esprit des Lois. chapitre XV: De l'esclavage des nègres.}
\newblock édition de Robert Derathé, 1748.

\bibitem{Hegel1807}
G.W.F. Hegel.
\newblock {\em Phenomenology of Spirit}.
\newblock Oxford: Clarendon Press, 1977, translation by A.V. Miller, 1807.

\bibitem{Changeux1983}
J.P. Changeux.
\newblock {\em L'homme neuronal}.
\newblock Fayard Collection Pluriel, 1983.

\bibitem{Edelman2000}
G.M. Edelman and G.~Tononi.
\newblock {\em A Universe of Consciousness. How Matter becomes Imagination}.
\newblock Basic Books, New York,, 2000.

\bibitem{Dehaene2000}
S.~Dehaene and J.P. Changeux.
\newblock Reward-dependent learning in neuronal networks for planning and
  decision making.
\newblock {\em Prog Brain Res}, 2000.

\bibitem{Dehaene2011}
S.~Dehaene and J.~Changeux.
\newblock Experimental and theoretical approaches to conscious processing.
\newblock {\em Neuron}, 70(2):200--227, 2011.

\bibitem{Dehaene2006}
S.~Dehaene, J.P. Changeux, L.~Naccache, J.~Sackur, and C.~Sergent.
\newblock Conscious, preconscious, and subliminal processing: a testable
  taxonomy.
\newblock {\em Trends Cogn Sci}, 10(5):204--211, 2006.

\bibitem{Sperry1961}
R.W. Sperry.
\newblock Cerebral organization and behavior: The split brain behaves in many
  respects like two separate brains, providing new research possibilities.
\newblock {\em Science}, 133:1749–1757, 1961.

\bibitem{Gazzaniga1967}
M.~Gazzaniga.
\newblock The split brain in man.
\newblock {\em Scientific American}, 217(2):24--29, 1967.

\bibitem{Chalmers1995}
D.~Chalmers.
\newblock Facing up to the problem of consciousness.
\newblock {\em Journal of Consciousness Studies}, 2(3):200--219, 1995.

\bibitem{Jackson1982}
F.~Jackson.
\newblock Epiphenomenal qualia.
\newblock {\em Philosophical Quarterly}, 32(127):127–136, 1982.

\bibitem{Ramachandran2003}
V.S. Ramachandran and E.M. Hubbard.
\newblock More common questions about synesthesia.
\newblock {\em Scientific American}, 2003.

\bibitem{Leibniz1686}
G.~Leibniz.
\newblock {\em Discours de metaphysique VI.}
\newblock 1686.

\bibitem{Leibniz1714}
G.W. Leibniz.
\newblock {\em The Monadology}.
\newblock Opera Philos. (Oxford) English translation by Robert Latta 1898,
  1714.

\bibitem{Galvani1791}
L.~Galvani.
\newblock De viribus electricitatis in motu musculari commentarius (commentaire
  sur l'effet de l'électricité sur le mouvement musculaire).
\newblock {\em Bononiae - Institutus Scientiarium
  http://sciences.amisbnf.org/node/1102}, 1791.

\bibitem{Gazzaniga1998}
M.S. Gazzaniga.
\newblock {\em The New Cognitive Neurosciences (2nd edition).}
\newblock MIT Press, 1998.

\bibitem{Sarraf2016}
S.~Sarraf and J.~Sun.
\newblock Functional brain imaging: A comprehensive survey.
\newblock {\em arXiv:1602.02225v4}, 2016.

\bibitem{MacKinnon2003}
R.~MacKinnon.
\newblock Potassium channels.
\newblock {\em FEBS Letters}, 555:62--65, 2003.

\bibitem{Salari2017}
V.~Salari, H.~Naeij, and A.~Shafiee.
\newblock Quantum interference and selectivity through biological ion channels.
\newblock {\em Sci Rep.}, 7(41625), 2017.

\bibitem{Neher1976}
E.~Neher and B.~Sakmann.
\newblock Single-channel currents recorded from membrane of denervated frog
  muscle fibres.
\newblock {\em Nature}, 260:799--802, 1976.

\bibitem{Neher1991}
E.~Neher.
\newblock Ion channels for communication beetween and wiwith cells.
\newblock {\em Nobel Lecture, December 9, 1991}, 1991.

\bibitem{Sakmann1991}
B.~Sakmann.
\newblock Elementary steps in synaptic transmission revealed by currents
  through single ion channels.
\newblock {\em Nobel Lecture, December 9, 1991}, 1991.

\bibitem{Jiang2002}
Y.~Jiang, A.~Lee, J.~Chen, M.~Cadene, B.T. Chait, and R.~MacKinnon.
\newblock The open pore conformation of potassium channels.
\newblock {\em Nature}, 417:523–526, 2002.

\bibitem{Grosman2000}
C.~Grosman, M.~Zhou, and A.~Auerbach.
\newblock Mapping the conformational wave of acetylcholine receptor channel
  gating.
\newblock {\em Nature}, 403(6771):773--776, 2000.

\bibitem{Millonas1998}
M.M. Millonas and D.A. Hanck.
\newblock Nonequilibrium response spectroscopy of voltage-sensitive ion channel
  gating.
\newblock {\em Biophysical Journal}, 74:210–229, 1998.

\bibitem{Simmons2005}
P.J. Simmons and R.~de~Ruyter~van Steveninck.
\newblock Reliability of signal transfer at a tonically transmitting, graded
  potential synapse of the locust ocellar pathway.
\newblock {\em J. Neurosci.}, 25(33):7529–7537, 2005.

\bibitem{Bi1998}
G.Q. Bi and M.M. Poo.
\newblock Synaptic modifications in cultured hippocampal neurons: dependence on
  spike timing, synaptic strength, and postsynaptic cell type.
\newblock {\em J Neurosci.}, 18(24):10464--10472, 1998.

\bibitem{Mainen1995}
Z.F. Mainen and T.J. Sejnowski.
\newblock Reliability of spike timing in neocortical neurons.
\newblock {\em Science}, 268(5216):1503--1506, 1995.

\bibitem{Hodgkin1952}
A.L. Hodgkin and A.F. Huxley.
\newblock A quantitative description of membrane current and its application to
  conduction and excitation in nerve.
\newblock {\em The Journal of Physiology}, 117(4):500–544, 1952.

\bibitem{Bryant1976}
H.L. Bryant and J.P. Segundo.
\newblock Spike initiation by transmembrane current: a white-noise analysis.
\newblock {\em J Physiol.}, 260(2):279--314, 1976.

\bibitem{Baudot2006}
P.~Baudot.
\newblock Natural computation: much ado about nothing? an intracellular study
  of visual coding in natural condition.
\newblock Master's thesis, Paris 6 university, 2006.

\bibitem{Baudot2013}
P.~Baudot, M.~Levy, O.~Marre, C.~Monier, M.~Pananceau, and Y.~Fregnac.
\newblock Animation of natural scene by virtual eye-movements evokes high
  precision and low noise in v1 neurons.
\newblock {\em Front. Neural Circuits}, 7(206):1--29, 2013.

\bibitem{RuytervanSteveninck1997}
R.~de~Ruyter~van Steveninck, G.D. Lewen, R.~Strong, S.P.and~Koberle, and
  W.~Bialek.
\newblock Reproducibility and variability in neural spike trains.
\newblock {\em Science}, 275:1805--1808, 1997.

\bibitem{Hubel1959}
D.H. Hubel and T.N. Wiesel.
\newblock Receptive fields of single neurones in the cat's striate cortex.
\newblock {\em Journal of Physiology}, 148:574--91, 1959.

\bibitem{Jones1987}
J.P. Jones and L.A. Palmer.
\newblock The two-dimensional spatial structure of simple receptive fields in
  cat striate cortex.
\newblock {\em Journal of Neurophysiology}, 58(6):1187--1211, 1987.

\bibitem{Hasson2004}
U.~Hasson, Y.~Nir, I.~Levy, G.~Fuhrmann, and R.~Malach.
\newblock Intersubject synchronization of cortical activity during natural
  vision.
\newblock {\em Science}, 303:1634--1640, 2004.

\bibitem{Hasson2009}
R.~Hasson, U. andl~Malach and D.J. Heeger.
\newblock Reliability of cortical activity during natural stimulation.
\newblock {\em Trends in Cognitive Sciences}, 14(1):40--48, 2009.

\bibitem{Friston1998}
K.J. Friston, O.~Josephs, G.~Rees, and R.~Turner.
\newblock Nonlinear event-related responses in fmri.
\newblock {\em Magnetic resonance in medicine}, 39(1):41--52, 1998.

\bibitem{John2001}
E.R. John.
\newblock A field theory of consciousness.
\newblock {\em Consciousness and Cognition}, 10:184–213, 2001.

\bibitem{Pockett2000}
S.~Pockett.
\newblock {\em The Nature of Consciousness : A Hypothesis}.
\newblock Writers Club Press, 2000.

\bibitem{McFadden2002}
J.~McFadden.
\newblock Synchronous firing and its influence on the brain’s magnetic field.
\newblock {\em Journal of Consciousness Studies}, 9:23--50, 2002.

\bibitem{Juusola2007}
M.~Juusola, H.P. Robinson, and G.G. de~Polavieja.
\newblock Coding with spike shapes and graded potentials in cortical networks.
\newblock {\em Bioessays}, 29(2):178--187, 2007.

\bibitem{Schummers2008a}
J.~Schummers, H.~Yu, and M.~Sur.
\newblock Tuned responses of astrocytes and their influence on hemodynamic
  signals in the visual cortex.
\newblock {\em Science}, 320:1638--1643, 2008.

\bibitem{Preat2008}
T.~Preat and G.~Isabel.
\newblock Molecular and system analysis of olfactory memory in drosophila.
\newblock {\em In « Learning and Memory : A Comprehensive Reference ».
  Elsevier. Ed J.H. Byrne.}, 2008.

\bibitem{Soden2013}
M.~Soden, G.L. Jones, CA. Sanford, A.S. Chung, A.D. Guler, C.~Chavkin,
  R.~Lujan, and L.S. Zweifel.
\newblock Disruption of dopamine neuron activity pattern regulation through
  selective expression of a human kcnn3 mutation.
\newblock {\em Neuron}, 80(4):1010--1016, 2013.

\bibitem{Monod1965}
J.~Monod, J~Wyman, and J.P. Changeux.
\newblock On the nature of allosteric transition : a plausible model.
\newblock {\em J Mol Biol.}, 12:88--118, 1965.

\bibitem{Fuller2010}
D.~Fuller, W.~Chen, M.~Adler, A.~Groisman, H.~Levine, W.J. Rappel, and W.F.
  Loomis.
\newblock External and internal constraints on eukaryotic chemotaxis.
\newblock {\em Proc Natl Acad Sci U S A.}, 107(21):9656--9659, 2010.

\bibitem{Davies1987}
E.~Davies.
\newblock Action potentials as multifunctional signals in plants: a unifying
  hypothesis to explain apparently disparate wound responses.
\newblock {\em Plant, Cell and Environment}, 10(8):623--631, 1987.

\bibitem{Tinsley2016}
J.N. Tinsley, M.I. Molodtsov, R.~Prevedel, D.~Wartmann, J.~Espigule-Pons,
  M.~Lauwers, and A.~Vaziri.
\newblock Direct detection of a single photon by humans.
\newblock {\em Nat Commun}, 7(1217):1--9, 2016.

\bibitem{Holmes2015}
R.~Holmes, B.G. Christensen, R.~Wang, and P.~Kwiat.
\newblock Testing the limits of human vision with single photons.
\newblock {\em Frontiers in Optics, OSA Technical Digest}, 2015.

\bibitem{Bialek1985}
W.~Bialek and A.~Schweitzer.
\newblock Quantum noise and the threshold of hearing.
\newblock {\em Phys. Rev. Lett.}, 54(7):725--728, 1985.

\bibitem{Roy2009}
S.~Roy and R.~Llinas.
\newblock Relevance of quantum mechanics on some aspects of ion channel
  function.
\newblock {\em C R Biol.}, 332(6):517--522, 2009.

\bibitem{Bohr1958}
N.~Bohr.
\newblock {\em Atomic Physics and Human Knowledge}.
\newblock Chapman \& Hall, 1958.

\bibitem{Hameroff1996}
S.~Hameroff and R.~Penrose.
\newblock Conscious events as orchestrated space-time selections.
\newblock {\em J. Consc. Studies}, 3(1):36--53, 1996.

\bibitem{Penrose1989}
R.~Penrose.
\newblock {\em The Emperor's New Mind: Concerning Computers, Minds and The Laws
  of Physics}.
\newblock Oxford University Press, 1989.

\bibitem{Busemeyer2014}
J.R. Busemeyer and P.D. Bruza.
\newblock {\em Quantum Models of Cognition and Decision}.
\newblock Cambridge University Press, 2014.

\bibitem{Khrennikov2015}
A.~Khrennikov.
\newblock Quantum-like modeling of cognition.
\newblock {\em Front. Phys.}, 3(77), 2015.

\bibitem{Fusi2005}
S.~Fusi, P.~Drew, and L.F. Abbott.
\newblock Cascade models of synaptically stored memories.
\newblock {\em Neuron}, 45(4):599–611, 2005.

\bibitem{Shirkov1999}
D.~Shirkov.
\newblock Evolution of the bogoluibov renormalization group.
\newblock {\em arXiv:hep-th/9909024}, 1999.

\bibitem{Huang2013}
K.~Huang.
\newblock a critical history of renormalization.
\newblock {\em International Journal of Modern Physics A}, 28(29):1--27, 2013.

\bibitem{Kadanoff1966}
L.P. Kadanoff.
\newblock Scaling laws for ising models near tc.
\newblock {\em Physics}, 2(6):263--272, 1966.

\bibitem{Wilson1974b}
K~.G. Wilson and J.~Kogut.
\newblock The renormalization group and the epsilon expansion.
\newblock {\em Physics Reports - section C of Physics Letters}, 12(2):75--200,
  1974.

\bibitem{Dyson1949}
F.J. Dyson.
\newblock The radiation theories of tomonaga, schwinger, and feynman.
\newblock {\em Phys. Rev.}, 75(3):486--502, 1949.

\bibitem{Barlow1995}
H.~Barlow.
\newblock Banishing the homonculus.
\newblock {\em Perception as Bayesian inference, eds Knill and Richards,
  Cambridge University Press}, 1995.

\bibitem{Foldiack1993}
P.~Foldiack.
\newblock The ‘ideal homunculus’: Statistical inference from neural
  population responses.
\newblock {\em Computation and Neural Systems}, pages 55--60, 1993.

\bibitem{Hebb1949}
D.O. Hebb.
\newblock {\em The organization of behaviour}.
\newblock Wiley, New-York., 1949.

\bibitem{Malsburg1981}
C.~Von~der Malsburg.
\newblock The correlation theory of brain function.
\newblock {\em Internal Report 81-2, Dept. of Neurobiology, Max Planck
  Institute for Biophysical Chemistry, 3400 Gottingen, Germany}, 1981.

\bibitem{Abeles1982}
M.~Abeles.
\newblock {\em Local Cortical Circuits: An Electrophysiological study}.
\newblock Springer, Berlin, 1982.

\bibitem{Diesmann1999}
M.~Diesmann, M.O. Gewaltig, and A.~Aertsen.
\newblock Stable propagation of synchronous spiking in cortical neural
  networks.
\newblock {\em Nature}, 402:529--533, 1999.

\bibitem{Singer1995}
W.~Singer and C.M. Gray.
\newblock Visual feature integration and the temporal correlation hypothesis.
\newblock {\em Annu Rev Neurosci}, 18:555--586, 1995.

\bibitem{Ikegaya2004}
Y.~Ikegaya, Aaron., R.~Cossart, D.~Aronov, I.~Lampl, D.~Ferster, and R.~Yuste.
\newblock Synfire chains and cortical songs: temporal modules of cortical
  activity.
\newblock {\em Science}, 23(304):559--564, 2004.

\bibitem{Engel1999}
A.K. Engel, P.~Fries, P.~Koenig, M.~Brecht, and W.~Singer.
\newblock Temporal binding, binocular rivalry, and consciousness.
\newblock {\em Consciousness and Cognition}, 8(2):128–151, 1999.

\bibitem{Gold1999}
I.~Gold.
\newblock Does 40-hz oscillation play a role in visual consciousness?
\newblock {\em Consciousness and Cognition}, 8(2):186–195, 1999.

\bibitem{Golubitsky2006a}
M.~Golubitsky and I.~Stewart.
\newblock Nonlinear dynamics of networks: the groupoid formalism.
\newblock {\em Bull. Am. Math. Soc.}, 43(3):305–364, 2006.

\bibitem{Stilwell2011}
D.J. Stilwell, E.M. Bolt, and D.G. Robertson.
\newblock Synchronization of time varying networks under fast switching.
\newblock {\em Journal of Nonlinear Science}, 5:140, 2011.

\bibitem{Curto2008}
C.~Curto and V.~Itskov.
\newblock Cell groups reveal structure of stimulus space.
\newblock {\em PLOS comp. Biol.}, 4(10), 2008.

\bibitem{Singh2008}
G.~Singh, F.~Memoli, T.~Ishkhanov, G.~Sapiro, G.~Carlsson, and DL. Ringach.
\newblock Topological analysis of population activity in visual cortex.
\newblock {\em Journal of Vision}, 8(8):11:1--18, 2008.

\bibitem{Petri2014}
G.~Petri, P.~Expert, F.~Turkheimer, R.~Carhart-Harris, D.~Nutt, P.J. Hellyer,
  and F.~Vaccarino.
\newblock Homological scaffolds of brain functional networks.
\newblock {\em J R Soc Interface}, 6;11:101, 2014.

\bibitem{Kolmogorov1983}
A.~Kolmogorov.
\newblock Combinatorial foundations of information theory and the calculus of
  probabilities.
\newblock {\em Russ. Math. Surv.}, 38 29, 1983.

\bibitem{Jaynes2003}
E.T. Jaynes.
\newblock {\em Probability Theory: The Logic of Science}.
\newblock Cambridge University Press (posthum ed. 2003), 2003.

\bibitem{Griffiths2008}
T.~L. Griffiths, C.~Kemp, and J.~B. Tenenbaum.
\newblock Bayesian models of cognition.
\newblock {\em In Ron Sun (ed.), Cambridge Handbook of Computational Cognitive
  Modeling. Cambridge University Press.}, 2008.

\bibitem{Friston2012}
K.~Friston.
\newblock The history of the future of the bayesian brain.
\newblock {\em Neuroimage}, 62-248(2):1230–1233, 2012.

\bibitem{Kubo1966}
R.~Kubo.
\newblock The fluctuation-dissipation theorem.
\newblock {\em Reports on Progress in Physics}, 29(1):255--284, 1966.

\bibitem{Stevens1972}
C.F. Stevens.
\newblock Inferences about membrane properties from electrical noise
  measurements.
\newblock {\em Biophys J.}, 12(8):1028–1047, 1972.

\bibitem{Wiener1958}
N.~Wiener.
\newblock {\em Nonlinear problems in random theory.}
\newblock MIT press, John Wiley \& sons, 1958.

\bibitem{Palm1977}
G.~Palm and T.~Poggio.
\newblock The volterra representation and the wiener expansion: validity and
  pitfalls.
\newblock {\em SIAM J. Appl. Math.}, 33(2):195--216, 1977.

\bibitem{Palm1978}
G.~Palm and T.~Poggio.
\newblock Stochastic identification methods for nonlinear systems: an extension
  of wiener theory.
\newblock {\em SIAM J. Appl. Math.}, 34(3):524 –534, 1978.

\bibitem{Carandini2005}
M.~Carandini, J.B. Demb, V.~Mante, D.J. Tolhurst, Y.~Dan, B.A. Olshausen, J.L.
  Gallant, and N.C. Rust.
\newblock Do we know what the early visual system does?
\newblock {\em J Neurosci.}, 25(46):10577--97, 2005.

\bibitem{Hubel1962a}
D.H. Hubel and T.~Wiesel.
\newblock Receptive fields, binocular interaction and functional architecture
  in the cat's visual cortex.
\newblock {\em J. Physiol.}, 160(1):106–154, 1962.

\bibitem{DeAngelis1995}
G.C. DeAngelis, I.~Ohzawa, and R.D. Freeman.
\newblock Receptive-field dynamics in the central visual pathways.
\newblock {\em Trends Neurosci}, 8(10):451--458, 1995.

\bibitem{Fournier2014}
J.~Fournier, C.~Monier, M.~Levy, O.~Marre, Kisvarday Z.F., and Y.~Fregnac.
\newblock Hidden complexity of synaptic receptive fields in cat v1.
\newblock {\em J Neurosci}, 34(16):5515--5528, 2014.

\bibitem{Rieke1999}
F.~Rieke and R.~and‎ Bialek~W. Warland, D. and‎ de Ruyter van~Steveninck.
\newblock {\em Spikes: Exploring the Neural Code}.
\newblock A Bradford Book, Reprint edition (Computational Neuroscience), 1999.

\bibitem{Adrian1926}
E.D. Adrian and Y~Zotterman.
\newblock The impulses produced by sensory nerve endings: Part ii: The response
  of a single end organ.
\newblock {\em J Physiol}, 61:151–171, 1926.

\bibitem{Thorpe1996}
S.~Thorpe, D.~Fize, and C.~Marlot.
\newblock Speed of processing in the human visual system.
\newblock {\em Nature}, 382:520–522, 1996.

\bibitem{Gawne1996}
T.J. Gawne, T.W. Kjaer, and B.J. Richmond.
\newblock Latency: another potential code for feature binding in striate
  cortex.
\newblock {\em J Neurophysiol}, 76(2):1356--1360, 1996.

\bibitem{Strong1998}
S.P. Strong, R.R. de~Ruyter~van Steveninck, W.~Bialek, and R.~Koberle.
\newblock On the application of information theory to neural spike trains.
\newblock {\em Pac Symp Biocomput}, pages 621--32, 1998.

\bibitem{Bialek1991}
W.~Bialek, F.~Rieke, R.R. de~Ruyter~van Steveninck, and D.~Warland.
\newblock Reading a neural code.
\newblock {\em Science}, 252(5014):1854--1857, 1991.

\bibitem{Victor1996}
J.D. Victor and K.P. Purpura.
\newblock Nature and precision of temporal coding in visual cortex: a
  metric-space analysis.
\newblock {\em J Neurophysiol}, 76(2):1310--1326, 1996.

\bibitem{Mechler1998}
F.~Mechler, J.D. Victor, K.P. Purpura, and R.~Shapley.
\newblock Robust temporal coding of contrast by v1 neurons for transient but
  not for steady-state stimuli.
\newblock {\em J. Neurosci}, 18:6583--6598, 1998.

\bibitem{Debanne2013}
D.~Debanne, A.~Bialowas, and S.~Rama.
\newblock What are the mechanisms for analogue and digital signalling in the
  brain?
\newblock {\em Nature review neuroscience}, 14:63--69, 2013.

\bibitem{Zbili2016}
M.~Zbili, S.~Rama, and D.~Debanne.
\newblock Dynamic control of neurotransmitter release by presynaptic potential.
\newblock {\em Front Cell Neurosci}, 10(278), 2016.

\bibitem{RuytervanSteveninck1996}
R.R. de~Ruyter~van Steveninck and S.B. Laughlin.
\newblock The rate of information transfer at graded-potential synapses.
\newblock {\em Nature}, 379:642–645, 1996.

\bibitem{Rama2015}
S.~Rama, M.~Zbili, A.~Bialowas, L.~Fronzaroli-Molinieres, N.~Ankri, E.~Carlier,
  V.~Marra, and D.~Debanne.
\newblock Presynaptic hyperpolarization induces a fast analogue modulation of
  spike-evoked transmission mediated by axonal sodium channels.
\newblock {\em Nat Commun}, 6(10163), 2015.

\bibitem{Borst1999}
A.~Borst and F.E. Theunissen.
\newblock Information theory and neural coding.
\newblock {\em Nature Neurosci.}, 2:947--957, 1999.

\bibitem{Olshausen1996}
B.A. Olshausen and DJ. Field.
\newblock Emergence of simple-cell receptive field properties by learning a
  sparse code for natural images.
\newblock {\em Nature}, 381(6583):607--609, 1996.

\bibitem{Vinje2000}
W.E. Vinje and J.L. Gallant.
\newblock Sparse coding and decorrelation in primary visual cortex during
  natural vision.
\newblock {\em Science}, 287(5456):1273--6, 2000.

\bibitem{Vinje2002}
W.E. Vinje and J.L. Gallant.
\newblock Natural stimulation of the nonclassical receptive field increases
  information transmission efficiency in v1.
\newblock {\em J Neurosci}, 22(7):2904--2915, 2002.

\bibitem{Fregnac2005}
Y.~Fregnac, P.~Baudot, M.~Levy, and O.~Marre.
\newblock An intracellular view of time coding and sparseness of cortical
  representation in v1 neurons during virtual oculomotor exploration of natural
  scenes.
\newblock {\em Cosyne Computational and Systems Neuroscience. Proc.},
  https://drive.google.com/open?id=0B0QKxsVtOaTiOGhtSTg2T0t4cVE, 2005.

\bibitem{Butts2007}
D.A. Butts, C.~Weng, J.~Jin, C.I. Yeh, N.A. Lesica, J.M. Alonso, and G.B.
  Stanley.
\newblock Temporal precision in the neural code and the timescales of natural
  vision.
\newblock {\em Nature}, 6(449(7158)):92--5, 2007.

\bibitem{Haider2010}
B.~Haider, M.R. Krause, A.~Duque, Y.~Yu, J.~Touryan, J.A. Mazer, and D.A.
  McCormick.
\newblock Synaptic and network mechanisms of sparse and reliable visual
  cortical activity during nonclassical receptive field stimulation.
\newblock {\em Neuron}, 65(1):107--21, 2010.

\bibitem{Herikstad2011}
R.~Herikstad, J.~Baker, J.P. Lachaux, C.M. Gray, and S.C.Baker Yen.
\newblock Natural movies evoke spike trains with low spike time variability in
  cat primary visual cortex.
\newblock {\em J Neurosci.}, 31(44), 2011.

\bibitem{Martignon2000}
L.~Martignon, G.~Deco, K.~Laskey, M.~Diamond, W.~Freiwald, and E.~Vaadia.
\newblock Neural coding : Higher-order temporal patterns in the neurostatistics
  of cell assemblies.
\newblock {\em Neural Comput}, 12(11):2621--2653, 2000.

\bibitem{Amari2001a}
S.~Amari.
\newblock Information geometry on hierarchy of probability distributions.
\newblock {\em IEEE Transactions on Information Theory}, 47(5):1701--1711,
  2001.

\bibitem{Ma2006}
W.J. Ma, J.M. Beck, P.E. Latham, and A.~Pouget.
\newblock Bayesian inference with probabilistic population codes.
\newblock {\em Nat Neurosci}, 9(11):1432--1438, 2006.

\bibitem{Dehaene2005}
S.~Dehaene and J.P. Changeux.
\newblock Ongoing spontaneous activity controls access to consciousness: A
  neuronal model for inattentional blindness.
\newblock {\em PLoS Biol}, 3(5), 2005.

\bibitem{Wyart2009}
V.~Wyart and C.~Sergent.
\newblock The phase of ongoing eeg oscillations uncovers the fine temporal
  structure of conscious perception.
\newblock {\em J. Neurosci.}, 29(41):12839--12841, 2009.

\bibitem{Brown1828}
R.~Brown.
\newblock A brief account of microscopical observations made in the months of
  june, july and august 1827 on the particles contained in the pollen of plants
  and on the general existence of active molecules.
\newblock {\em Privately published, Science Museum London. Source book in
  physics, ed. Magie, W. F. (1965) Massachusetts: Harvard University Press.},
  1828.

\bibitem{Vilardi2009}
A.~Vilardi.
\newblock {\em The role of noise in brain dynamics processes}.
\newblock PhD Thesis. University of Trento. CIMeC, Centre for Mind/Brain
  Sciences, 2009.

\bibitem{Eldar2014}
A.~Eldar and M.B. Elowitz.
\newblock Functional roles for noise in genetic circuits.
\newblock {\em Nature}, 467(7312):167–173, 2014.

\bibitem{Bak1987}
P.~Bak, C.~Tang, and K.~Wiesenfeld.
\newblock Self-organized criticality: an explanation of 1/f noise.
\newblock {\em Physical Review Letters}, 59(4):381–384, 1987.

\bibitem{Shannon1948}
C.~E. Shannon.
\newblock A mathematical theory of communication.
\newblock {\em The Bell System Technical Journal}, 27:379--423, 1948.

\bibitem{Baji2005}
V.B. Baji and T.W. Tan.
\newblock {\em Information Processing and Living Systems}.
\newblock Imperial College Press, 2005.

\bibitem{Dayan1995}
P.~Dayan, G.~Hinton, R.M. Neal, and R.S. Zemel.
\newblock The helmholtz machine.
\newblock {\em Neural Computation}, 7:889--904, 1995.

\bibitem{Friston2006}
K.~Friston, J.~Kilner, and L.~Harrison.
\newblock A free energy principle for the brain.
\newblock {\em J Physiol Paris.}, 100(1–3):70–87, 2006.

\bibitem{Fechner1860}
G.T. Fechner.
\newblock {\em Elemente der Psychophysik (Elements of Psychophysics)}.
\newblock Breitkopf und Härtel, 1860.

\bibitem{Mengoli1670}
P.~Mengoli.
\newblock {\em Speculationi di musica}.
\newblock per l'herede del Benacci, 1670.

\bibitem{Planck1901}
M.~Planck.
\newblock On the law of distribution of energy in the normal spectrum.
\newblock {\em Annalen der Physik.}, 4:553--560 traduction:
  http://dbhs.wvusd.k12.ca.us/webdocs/Chem--History/Planck--1901/Planck--1901.html,
  1901.

\bibitem{Takahashi2006}
T.~Takahashi.
\newblock Time-estimation error following weber–fechner law may explain
  subadditive time-discounting.
\newblock {\em Medical Hypotheses}, 67(1372-1374), 2006.

\bibitem{Stevens1965}
J.C. Stevens and L.E Marks.
\newblock Cross-modality matching of brightness and loudness.
\newblock {\em Proceedings of the National Academy of Sciences.}, 54:407–411,
  1965.

\bibitem{Krueger1989}
L.E. Krueger.
\newblock Reconciling fechner and stevens: Toward a unified psychophysical law.
\newblock {\em Behavioral and Brain Sciences}, 12(2):251--267, 1989.

\bibitem{Laughlin1989}
SB. Laughlin.
\newblock The role of sensory adaptation in the retina.
\newblock {\em J Exp Biol.}, 146(39-62), 1989.

\bibitem{Kostala2016}
L.~Kostala and P.~Lansky.
\newblock Coding accuracy on the psychophysical scale.
\newblock {\em Scientific Report}, 6:23810, 2016.

\bibitem{Dzhafarov2012}
E.N. Dzhafarov.
\newblock Mathematical foundations of universal fechnerian scaling.
\newblock {\em Measurement with persons. Theory, methods and implementation
  areas. Psychology press}, 2012.

\bibitem{Aristotle350B.C.E}
Aristotle.
\newblock {\em Metaphysics}.
\newblock Translated by W. D. Ross
  http://classics.mit.edu/Aristotle/metaphysics.4.iv.html, 350 B.C.E.

\bibitem{Ross1951}
W.D. Ross.
\newblock {\em Plato's Theory of Ideas}.
\newblock Cambridge University Press, 1951.

\bibitem{Kohler1947}
W.~Kohler.
\newblock {\em Gestalt psychology: an introduction to new concepts in modern
  psychology}.
\newblock New York: Liveright, 1947.

\bibitem{Wertheimer1924}
M.~Wertheimer.
\newblock {\em Gestalt theory}.
\newblock In W. D. Ellis Ed., A source book of Gestalt psychology. London,
  England: Routledge \& Kegan Paul (1934), 1924.

\bibitem{Wertheimer1923}
M.~Wertheimer.
\newblock {\em Laws of organization in perceptual forms}.
\newblock In W. D. Ellis (Ed.), A source book of Gestalt psychology. London,
  England: Routledge \& Kegan Paul (1938)., 1923.

\bibitem{Rosenthal1999}
Y.~Rosenthal, V.;~Viseti.
\newblock Sens et temps de la gestalt.
\newblock {\em Intellectica}, 28:147--227., 1999.

\bibitem{Attneave1954}
F.~Attneave.
\newblock Some informational aspects of visual perception.
\newblock {\em Psychological Review.
  http://www.dc.uba.ar/materias/incc/teoricas/Attneave1954.pdf}, Vol 61 No.
  3:183--194, 1954.

\bibitem{Field1993}
D.J. Field, A.~Hayes, and R.F. Hess.
\newblock Contour integration by the human visual system: Evidence for a local
  “association field”.
\newblock {\em Vision Research}, 33(2):173--193, 1993.

\bibitem{Polat1993}
U.~Polat and D.~Sagi.
\newblock Lateral interactions between spatial channels: suppression and
  facilitation revealed by lateral masking experiments.
\newblock {\em Vision Research}, 33(7):993--999, 1993.

\bibitem{Georges2002}
S.~Georges, P.~Series, Y.~Fregnac, and J.~Lorenceau.
\newblock Orientation dependent modulation of apparent speed: psychophysical
  evidence.
\newblock {\em Vision Research}, 42(25):2757--2772, 2002.

\bibitem{Penfield1958}
W.~Penfield.
\newblock Some mechanisms of consciousness discovered during electrical
  stimulation of the brain.
\newblock {\em Proc. Natl Acad. Sci. USA}, 44:51–66, 1958.

\bibitem{Salzman1990}
C.D. Salzman, K.H. Britten, and W.T. Newsome.
\newblock Cortical microstimulation influences perceptual judgements of motion
  direction.
\newblock {\em Nature}, 346:174–177, 1990.

\bibitem{Parker1998}
A.J. Parker and W.T. Newsome.
\newblock Sense and the single neuron: probing the physiology of perception.
\newblock {\em Annu. Rev. Neurosci.}, 21:227–277, 1998.

\bibitem{Cicmil2015}
N.~Cicmil and K.~Krug.
\newblock Playing the electric light orchestra—how electrical stimulation of
  visual cortex elucidates the neural basis of perception.
\newblock {\em Philos Trans R Soc Lond B Biol Sci.}, 370(): 20140206(1677),
  2015.

\bibitem{Gilbert1989}
C.D. Gilbert and T.N. Wiesel.
\newblock Columnar specificity of intrinsic horizontal and corticocortical
  connections in cat visual cortex.
\newblock {\em J. Neurosci.}, 9:2432--2442, 1989.

\bibitem{Schmidt1997}
K.E. Schmidt, R.~Goebel, S.~Lowel, and W.~Singer.
\newblock The perceptual grouping criterion of colinearity is reflected by
  anisotropies of connections in the primary visual cortex.
\newblock {\em Eur J Neurosci.}, 9(1083-1089), 1997.

\bibitem{Yarbus1967}
A.L. Yarbus.
\newblock {\em Eye Movements and Vision}.
\newblock New York: Plenum Press, 1967.

\bibitem{Chavane2000}
F.~Chavane, C.~Monier, V.~Bringuier, P.~Baudot, L.~Borg-Graham, J.~Lorenceau,
  and Y.~Fregnac.
\newblock The visual cortical association field: a gestalt concept or a
  psychophysiological entity?
\newblock {\em Journal of Physiology (Paris)}, 94(5-6):333--342, 2000.

\bibitem{Gerard-Mercier2016}
F.~Gerard-Mercier, P.V. Carelli, M.~Pananceau, X.G. Troncoso, and Y.~Fregnac.
\newblock Synaptic correlates of low-level perception in v1.
\newblock {\em J Neurosci.}, 36(14):3925--3942, 2016.

\bibitem{Wehr2003}
M.~Wehr and A.M. Zador.
\newblock Balanced inhibition underlies tuning and sharpens spike timing in
  auditory cortex.
\newblock {\em Nature}, 426(6965):442--6, 2003.

\bibitem{Collins2007}
A.~Collins, M.~Stopfer, G.~Laurent, and Bazhenov M.
\newblock Adaptive regulation of sparseness by feedforward inhibition.
\newblock {\em Nature Neuroscience}, 10(9):1176--1184, 2007.

\bibitem{Olshausen2004}
B.~Olshausen and D.~Field.
\newblock What is the other 85
\newblock {\em Problems in system neuroscience. Oxford University Press}, 2004.

\bibitem{Barlow1961}
H.B. Barlow.
\newblock Possible principles underlying the transformation of sensory
  messages.
\newblock {\em In Sensory Communication, W.A. Rosenblith, ed. (Cambridge, MA:
  MIT Press)}, pages 217--234, 1961.

\bibitem{Thom1977}
R.~Thom.
\newblock {\em Stabilite struturelle et morphogenese}.
\newblock deuxieme edition, InterEdition, Paris, 1977.

\bibitem{Petitot1983}
J.~Petitot and R.~Thom.
\newblock {\em Semiotique et theorie des catastrophes}.
\newblock Presses Univ. Limoges, 1983.

\bibitem{Thom1990}
R.~Thom.
\newblock {\em Semio Physics: A Sketch}.
\newblock Addison Wesley, 1990.

\bibitem{Kendall1984}
D.~Kendall.
\newblock Shape manifolds, procrustean metrics and complex projective spaces.
\newblock {\em Bull. London Math. Soc.}, 18:81–121, 1984.

\bibitem{Dryden1998}
I.L. Dryden and K.V. Mardia.
\newblock {\em Statistical Shape Analysis}.
\newblock John Wiley \& Sons, 1998.

\bibitem{Michor2006}
P.W. Michor and D.~Mumford.
\newblock Riemannian geometries on spaces of plane curves.
\newblock {\em J. Eur. Math. Soc.}, 8:1--48, 2006.

\bibitem{Mumford2010}
D.~Mumford and A.~Desolneux.
\newblock {\em Pattern Theory, the Stochastic Analysis of Real World Signals}.
\newblock AKPeters/CRC Press,, 2010.

\bibitem{James1890}
W.~James.
\newblock {\em The Principles of Psychology}.
\newblock New York, Holt, 1890.

\bibitem{Llinas2002}
R.~Llinas.
\newblock {\em I of the Vortex. From Neurons to Self}.
\newblock The MIT Press - A Bradford Book, 2002.

\bibitem{Berthoz2000}
A.~Berthoz.
\newblock Physiologie de la perception et de l'action.
\newblock {\em Cours du collége de France}, 2000.

\bibitem{Berthoz2003}
A.~Berthoz.
\newblock {\em La Decision}.
\newblock O. Jacob, 2003.

\bibitem{Jeannerod2006}
M.~Jeannerod.
\newblock {\em Motor Cognition: What Actions Tell to the Self}.
\newblock Oxford University Press - Psychology Series 1st Edition, 2006.

\bibitem{Merleau-Ponty1945}
M.~Merleau-Ponty.
\newblock {\em Phenomenologie de la Perception}.
\newblock Gallimard, coll. « Tel », 1945.

\bibitem{Saraf-Sinik}
I.~Saraf-Sinik, E.~Assa, and E.~Ahissar.
\newblock Motion makes sense: An adaptive motor-sensory strategy underlies the
  perception of object location in rats.
\newblock {\em Journal of Neuroscience}, 35(23):8777--8789, 2015.

\bibitem{Pellionisz1979}
A.~Pellionisz and R.~Llinas.
\newblock Brain modeling by tensor network theory and computer simulation. the
  cerebellum: distributed processor for predictive coordination.
\newblock {\em Neuroscience}, 4(3):323--348, 1979.

\bibitem{ORegan2001}
J.K. O'Regan and A.~Noe.
\newblock A sensorimotor account of vision and visual consciousness.
\newblock {\em Behav Brain Sci}, 24:939--973, 2001.

\bibitem{Piaget1964}
J.~Piaget.
\newblock {\em Six etudes de psychologie. Problemes de psychologie genetique.}
\newblock
  http://www.fondationjeanpiaget.ch/fjp/site/presentation/index.php?PRESMODE=1\&DOCID=1042,
  1964.

\bibitem{Phillips1981}
J.~L Phillips.
\newblock {\em Piaget s Theory A Primer .}
\newblock Freeman, 1981.

\bibitem{Lomo1966}
T.~Lomo.
\newblock Frequency potentiation of excitatory synaptic activity in the dentate
  area of the hippocampal formation.
\newblock {\em Acta Physiologica Scandinavica}, 68(277):128, 1966.

\bibitem{Bliss1973}
T.~Bliss and T.~Lomo.
\newblock Long-lasting potentiation of synaptic transmission in the dentate
  area of the anaesthetized rabbit following stimulation of the perforant path.
\newblock {\em J. Physiol.}, 232(2):331--356, 1973.

\bibitem{Chaitin2006}
G.~Chaitin.
\newblock Epistemology as information theory: From leibniz to omega.
\newblock {\em Collapse}, Volume I:27--51, 2006.

\bibitem{Maguire2016}
P.~Maguire, P.~Moser, and R.Moser Maguire.
\newblock Understanding consciousness as data compression.
\newblock {\em Journal of Cognitive Science}, 17(1):63--94, 2016.

\bibitem{Linsker1986}
R.~Linsker.
\newblock From basic network principles to neural architecture.
\newblock {\em Proc. Nat. Acad. Sci. USA}, 83:7508--7512, 1986.

\bibitem{Linsker1988}
R.~Linsker.
\newblock Self-organization in a perceptual network.
\newblock {\em Computer}, 21(3):105--117, 1988.

\bibitem{Srinivasan1981}
M.V. Srinivasan, S.~Laughlin, and A.~Dubs.
\newblock Predictive coding, a fresh view of infhibition in the retina.
\newblock {\em Proc Royal Soc London}, 216(1205):427--459, 1981.

\bibitem{Nadal1999}
N.~Nadal, J.-P. ;~Parga.
\newblock Sensory coding: information maximization and redundancy reduction.
\newblock {\em Neural information processing, G. Burdet, P. Combe and O. Parodi
  Eds. World Scientific Series in Mathematical Biology and Medecine}, Vol. 7:p.
  164--171, 1999.

\bibitem{Nadal1994}
N.~Nadal, J.-P. ;~Parga.
\newblock Nonlinear neurons in the low noise limit: a factorial code maximizes
  information transfer.
\newblock {\em Network Computation in Neural Systems}, 5:565--581, 1994.

\bibitem{Bell1995}
T.J. Bell, A.J. ;~Sejnowski.
\newblock An information maximisation approach to blind separation and blind
  deconvolution.
\newblock {\em Neural Computation}, 7, 6:1129--1159, 1995.

\bibitem{Laughlin1981}
S.~Laughlin.
\newblock A simple coding procedure enhances the neuron's information capacity.
\newblock {\em Z. Naturforsch}, 36(c):910--912, 1981.

\bibitem{Ackley1985}
D.H. Ackley, G.E. Hinton, and T.~J. Sejnowski.
\newblock A learning algorithm for boltzmann machines.
\newblock {\em Cognitive Science}, 9(1):147--169, 1985.

\bibitem{Hopfield1982}
J.J. Hopfield.
\newblock Neural networks and physical systems with emergent collective
  computational abilities.
\newblock {\em PNAS}, 79:2554--2558, 1982.

\bibitem{Marr1982}
D.~Marr.
\newblock {\em Vision}.
\newblock W.H. Freeman and Co., 1982.

\bibitem{Hinton2012}
G.~Hinton and R.~Salakhutdinov.
\newblock A better way to pretrain deep boltzmann machines.
\newblock {\em Advances in Neural.}, 3:1–9, 2012.

\bibitem{Smolensky1986}
P.~Smolensky.
\newblock Chapter 6: Information processing in dynamical systems: Foundations
  of harmony theory" (pdf).
\newblock {\em In Rumelhart, David E.; McLelland, James L. Parallel Distributed
  Processing: Explorations in the Microstructure of Cognition, Volume 1:
  Foundations. MIT Press.}, page 194–281, 1986.

\bibitem{Mehta2014}
P.~Mehta and D.J. Schwab.
\newblock An exact mapping between the variational renormalization group and
  deep learning.
\newblock {\em arXiv: 1410.3831 .}, 2014.

\bibitem{Rahimi2017}
A.~Rahimi and B.~Recht.
\newblock Back when we were kids.
\newblock In {\em NIPS 2017 Test-of-Time Award presentation}, 2017.

\bibitem{Baudot2006a}
P.~Baudot, O.~Marre, M.~Levy, and Y~Fregnac.
\newblock Nature is the code: reliable and efficient dissipation in v1.
\newblock {\em Published in Thesis "Natural Computation: much ado about
  nothing"},
  https://drive.google.com/open?id=1hV8WLLT3rgMW7QmDpEBy1yyQhRT1sZNW, 2006.

\bibitem{Fregnac2015}
Y.~Fregnac and B.~Bathellier.
\newblock Cortical correlates of low-level perception: From neural circuits to
  percepts.
\newblock {\em Neuron}, 88(7), 2015.

\bibitem{Debanne1998}
D.~Debanne, D.E. Shulz, and Y.~Fregnac.
\newblock Activity-dependent regulation of 'on' and 'off' responses in cat
  visual cortical receptive fields.
\newblock {\em Journal of Physiology}, 508(2):523--548, 1998.

\bibitem{Monier2003}
C.~Monier, F.~Chavane, P.~Baudot, L.~Borg-Graham, and Y.~Fregnac.
\newblock Orientation and direction selectivity of synaptic inputs in visual
  cortical neurons: a diversity of combinations produces spike tuning.
\newblock {\em Neuron}, 37(4):663--680, 2003.

\bibitem{Markov2013}
N.T. Markov, J.~Vezoli, and P.~et~al. Chameau.
\newblock Anatomy of hierarchy: Feedforward and feedback pathways in macaque
  visual cortex.
\newblock {\em The Journal of Comparative Neurology}, 522(1):225--259, 2013.

\bibitem{Roxin2005}
A.~Roxin, N.~Brunel, and D.~Hansel.
\newblock Role of delays in shaping spatiotemporal dynamics of neuronal
  activity in large networks.
\newblock {\em Phys Rev Lett.}, 94(23):238103, 2005.

\bibitem{Debanne1994}
D.~Debanne, B.H. Gahwiler, and S.M. Thompson.
\newblock Asynchronous pre- and postsynaptic activity induces associative
  long-term depression in area ca1 of the rat hippocampus in vitro.
\newblock {\em Proc Natl Acad Sci USA}, 91(3):1148--1152, 1994.

\bibitem{Abbott2000}
L.F. Abbott and S.B. Nelson.
\newblock Synaptic plasticity: Taming the beast.
\newblock {\em Nat Neurosci}, 3:1178--1183, 2000.

\bibitem{Graupner2010}
M.~Graupner and N.~Brunel.
\newblock Mechanisms of induction and maintenance of spike-timing dependent
  plasticity in biophysical synapse models.
\newblock {\em Front Comput Neurosci.}, 4(136), 2010.

\bibitem{Graupner2012}
M.~Graupner and N.~Brunel.
\newblock Calcium-based plasticity model explains sensitivity of synaptic
  changes to spike pattern, rate, and dendritic location.
\newblock {\em Proc Natl Acad Sci USA}, 109(10):3991–3996, 2012.

\bibitem{Galvan2010}
A.~Galvan.
\newblock Neural plasticity of development and learning.
\newblock {\em Hum Brain Mapp}, 31(6):879–890, 2010.

\bibitem{Baudot2018}
P.~Baudot, M.~Tapia, and J.~Goaillard.
\newblock Topological information data analysis: Poincare-shannon machine and
  statistical physic of finite heterogeneous systems.
\newblock {\em Preprints 2018040157}, 2018.

\bibitem{Jacob1961}
F.~Jacob and J.~Monod.
\newblock Genetic regulatory mechanisms in the synthesis of proteins.
\newblock {\em J. Mol. Biol.}, vol. 3:318--356, 1961.

\bibitem{Cowan1965}
J.D. Cowan.
\newblock The problem of organismic reliability.
\newblock {\em Progress in Brain Research}, 17:9--63, 1965.

\bibitem{Danchin2003}
A.~Danchin.
\newblock {\em The delfic boat: What genomes tell us}.
\newblock Harvard University Press, 2003.

\bibitem{Marder2006}
E.~Marder and J.M. Goaillard.
\newblock Variability, compensation and homeostasis in neuron and network
  function.
\newblock {\em Nat Rev Neurosci.}, 7:563--74, 2006.

\bibitem{Swensen2003}
A.M. Swensen and B.P. Bean.
\newblock Ionic mechanisms of burst firing in dissociated purkinje neurons.
\newblock {\em J. Neurosci.}, 23:9650–9663, 2003.

\bibitem{Bucher2005}
D.~Bucher, A.A. Prinz, and E.J. Marder.
\newblock Animal-to-animal variability in motor pattern production in adults
  and during growth.
\newblock {\em Neurosci.}, 25:1611–1619, 2005.

\bibitem{Turrigiano2008}
G.G. Turrigiano.
\newblock The self-tuning neuron: synaptic scaling of excitatory synapses.
\newblock {\em Cell}, 135:422–435, 2008.

\bibitem{Watt2000}
A.J. Watt, M.C. van Rossum, K.M. MacLeod, S.B. Nelson, and G.G. Turrigiano.
\newblock Activity coregulates quantal ampa and nmda currents at neocortical
  synapses.
\newblock {\em Neuron}, 26:659–670., 2000.

\bibitem{Watt2010}
A.J. Watt and N.S. Desai.
\newblock Homeostatic plasticity and stdp: Keeping a neuron's cool in a
  fluctuating world.
\newblock {\em Front Synaptic Neurosci.}, 2(5), 2010.

\bibitem{Woods2013}
H.A. Woods and J.K. Wilson.
\newblock An information hypothesis for the evolution of homeostasis.
\newblock {\em Trends in Ecology and Evolution}, 28(5):283--289, 2013.

\bibitem{Taylor2006a}
A.L. Taylor, T.J. Hickey, A.A. Prinz, and E.~Marder.
\newblock Structure and visualization of high-dimensional conductance spaces.
\newblock {\em J Neurophysiol.}, 96(2):891--905, 2006.

\bibitem{Tapia2018}
M.~Tapia, P.~Baudot, C.~Formizano-Treziny, M.~Dufour, S.~Temporal, M.~Lasserre,
  B.~Marqueze-Pouey, J.~Gabert, K.~Kobayashi, and Goaillard J.M.
\newblock Neurotransmitter identity and electrophysiological phenotype are
  genetically coupled in midbrain dopaminergic neurons.
\newblock {\em Submitted to Nat. Comm}, 2018.

\bibitem{Martin2008}
E.~Martin.
\newblock A dictionary of biology (6th ed.).
\newblock {\em Oxford: Oxford University Press.}, page 315–316, 2008.

\bibitem{Suzuki2002}
M.~Suzuki, S.;~Grabowecky.
\newblock Evidence for perceptual trapping and adaptation in mulstistable
  binocular rivalry.
\newblock {\em Neuron}, 36:143--157, 2002.

\bibitem{Proute2008}
A.~Proute.
\newblock Le raisonnement par l absurde.
\newblock {\em La negation de l antiquite grecque a nos jours. Institut Henri
  Poincare. http://www.univ-paris-diderot.fr/philomathique/Proute21-05-08.pdf},
  2008.

\bibitem{Bauer2017}
A.~Bauer.
\newblock Five stages of accepting constructive mathematics.
\newblock {\em Bulletin of the American Mathematical Society}, 54(3):481--498,
  2017.

\bibitem{Necker1832}
L.A. Necker.
\newblock Observations on some remarkable optical phaenomena seen in
  switzerland; and on an optical phaenomenon which occurs on viewing a figure
  of a crystal or geometrical solid.
\newblock {\em London and Edinburgh Philosophical Magazine and Journal of
  Science.}, 1(5):329–337, 1832.

\bibitem{Hill1887–1962}
W.E Hill.
\newblock My wife and my mother-in-law. they are both in this picture – find
  them.
\newblock {\em Prints \& Photographs Online Catalog. Library of Congress.},
  1887–1962.

\bibitem{Rubin1915}
E.~Rubin.
\newblock Synsoplevede figurer.
\newblock {\em PhD thesis of University of Copenhagen}, 1915.

\bibitem{Ramachandran1985}
V.S. Ramachandran and S.M. Anstis.
\newblock Perceptual organization in multistable apparent motion.
\newblock {\em Perception}, 14(2):135--143, 1985.

\bibitem{Carter2008}
O.~Carter, T.~Konkle, Q.~Wang, V.~Hayward, and C.~Moore.
\newblock Tactile rivalry demonstrated with an ambiguous apparent-motion
  quartet.
\newblock {\em Current Biology}, 18:1050--1054, 2008.

\bibitem{Fisher1967}
G.H. Fisher.
\newblock Preparation of ambiguous stimulus materials.
\newblock {\em Perception \& Psychophysics}, 2:421--422., 1967.

\bibitem{Tononi2016}
G.~Tononi, M.~Boly, M.~Massimini, and C.~Koch.
\newblock Integrated information theory: From consciousness to its physical
  substrate.
\newblock {\em Nature Reviews Neuroscience}, 17(7), 2016.

\bibitem{Oizumi2014}
M.~Oizumi, L.~Albantakis, and G.~Tononi.
\newblock From the phenomenology to the mechanisms of consciousness: Integrated
  information theory 3.0.
\newblock {\em PLOS comp. Biol.}, 10(5), 2014.

\bibitem{Cantor1895}
G.~Cantor.
\newblock Beiträge zur begründung der transfiniten mengenlehre.
\newblock {\em Mathematische Annalen}, 46(4):481--512, 1895.

\bibitem{Whithead1910}
A.N. Whithead and B.~Russel.
\newblock {\em Principia Mathematica}.
\newblock Cambridge University Press 2nd ed. 1963, 1910.

\bibitem{Whithead1929}
A.~Whithead.
\newblock {\em Process and Reality. An Essay in Cosmology.}
\newblock Gifford Lectures Delivered in the University of Edinburgh During the
  Session 1927–1928. Macmillan, New York, Cambridge University Press,, 1929.

\bibitem{Hilbert1930}
D.~Hilbert.
\newblock Naturerkennen und logik (logic and the knowledge of nature).
\newblock In {\em The Conference on Epistemology of the Exact Sciences}, 1930.

\bibitem{Godel1931}
K.~Godel.
\newblock {\em On Formally Undecidable Propositions of Principia Mathematica
  and Related Systems.}
\newblock Dover, 1962, 1931.

\bibitem{Turing1937}
A.M. Turing.
\newblock On computable numbers, with an application to the
  entscheidungsproblem.
\newblock {\em Proc. Lond. Math. Soc. Ser.}, 42:230--265, 1937.

\bibitem{Villani2013}
C.~Villani.
\newblock {\em Théorème vivant}.
\newblock Le livre de poche, 2013.

\bibitem{MacLane1998}
S.~Mac~Lane.
\newblock {\em Categories for the Working Mathematician.}
\newblock Graduate Texts in Mathematics. Springer 2nd ed., 1998.

\bibitem{Lawvere2005}
F.W. Lawvere.
\newblock Taking categories seriously.
\newblock {\em Reprints in Theory and Applications of Categories. Reprint of
  "Revista Colombiana de Matematicas" XX (1986) 147-178}, No. 8,:pp. 1--24.,
  2005.

\bibitem{Baez2014}
J.~C. Baez and T.~Fritz.
\newblock A bayesian characterization of relative entropy.
\newblock {\em Theory and Applications of Categories,}, Vol. 29, No. 16:p.
  422--456, 2014.

\bibitem{Baez2011}
J.C. Baez, T.~Fritz, and T.~Leinster.
\newblock A characterization of entropy in terms of information loss.
\newblock {\em Entropy}, 13:1945--1957, 2011.

\bibitem{Vigneaux2017}
J.P. Vigneaux.
\newblock The structure of information: from probability to homology.
\newblock {\em arXiv:1709.07807}, 2017.

\bibitem{Proute2013}
A.~Proute.
\newblock {\em Introduction a la Logique Categorique}.
\newblock Prepublications, Paris 7 Logique www.logique.jussieu.fr/~alp/, 2013.

\bibitem{Riemann1854b}
B.~Riemann.
\newblock Ueber die darstellbarkeit einer funktion durch eine trigonometrische
  reihe.
\newblock {\em Traduction publiée dans le Bulletin des Sciences mathém. et
  astron. tome V ; juillet 1873}, 1854.

\bibitem{Lebesgue1901}
H.~Lebesgue.
\newblock Sur une generalisation de l integrale definie.
\newblock {\em Comptes rendus des seances de l Academie des sciences},
  132:1025--1027, 1901.

\bibitem{Borel1898}
E.~Borel.
\newblock {\em Lecons sur la theorie des fonctions}.
\newblock Paris, Gauthier-Villars et fils.
  https://archive.org/details/leconstheoriefon00borerich, 1898.

\bibitem{Ciesielski1997}
K.~Ciesielski.
\newblock {\em Set Theory for the Working Mathematician}.
\newblock Cambridge University Press, 1997.

\bibitem{Howard1998}
P.~Howard and J.E. Rubin.
\newblock {\em Consequences of the axiom of choice}.
\newblock Mathematical Surveys and Monographs, 1998.

\bibitem{Bell2015}
J.L. Bell.
\newblock The axiom of choice.
\newblock {\em The Stanford Encyclopedia of Philosophy (Summer 2015 Edition)},
  2015.

\bibitem{Ageron2002}
P.~Ageron.
\newblock L'autre axiom du choix.
\newblock {\em Revue d’histoire des mathematiques}, 8:113--140, 2002.

\bibitem{Wagon1986}
S.~Wagon.
\newblock {\em The Banach-Tarski Paradox}.
\newblock Cambridge Univ. Press., 1986.

\bibitem{Dewdney1989a}
A.K. Dewdney.
\newblock A matter fabricator provides matter for thought.
\newblock {\em Scientific American}, 60(4):116--119, 1989.

\bibitem{Diaconescu1975}
R.~Diaconescu.
\newblock Axiom of choice and complementation.
\newblock {\em Proc. Amer. Math. Soc.}, 51:176--178, 1975.

\bibitem{Dehn1901}
M.~Dehn.
\newblock Ueber den rauminhalt.
\newblock {\em Mathematische Annalen}, 55(3):465--478, 1901.

\bibitem{Dupont1982}
J.~Dupont and C.H. Sah.
\newblock Scissors congruences, ii.
\newblock {\em Journal of Pure and Applied Algebra}, 25:159--195, 1982.

\bibitem{Dupont2001}
J.L. Dupont.
\newblock {\em Scissors Congruences, Group Homology and Characteristic
  Classes}.
\newblock Nankai Tracts in Mathematics, V. 1., 2001.

\bibitem{Hales2005}
T.C. Hales.
\newblock What is motivic measure?
\newblock {\em BULLETIN (New Series) OF THE AMERICAN MATHEMATICAL SOCIETY},
  Vol. 42, Nb 2:P. 119�135, 2005.

\bibitem{Tao2012}
T.~Tao.
\newblock A cheap version of nonstandard analysis.
\newblock {\em What's new. Tao's blog.
  https://terrytao.wordpress.com/2012/04/02/a-cheap-version-of-nonstandard-analysis/},
  2012.

\bibitem{Solovay1970}
Robert~M. Solovay.
\newblock A model of set-theory in which every set of reals is lebesgue
  measurable.
\newblock {\em Annals of Mathematics. Second Series}, 92:1--56, 1970.

\bibitem{Kolmogorov1933a}
A.~N. Kolmogorov.
\newblock {\em Grundbegriffe der Wahrscheinlichkeitsrechnung.(English
  translation (1950): Foundations of the theory of probability.)}.
\newblock Springer, Berlin (Chelsea, New York)., 1933.

\bibitem{Boole1854}
G.~Boole.
\newblock {\em An Investigation Of The Laws Of Thought On Which Are Founded The
  Mathematical Theories Of Logic And Probabilities.}
\newblock McMillan and Co., 1854.

\bibitem{Nahin2012}
P.~J. Nahin.
\newblock {\em The Logician and the Engineer: How George Boole and Claude
  Shannon Created the Information Age}.
\newblock Princeton University Press, 2012.

\bibitem{Hume1738}
D.~Hume.
\newblock {\em A Treatise of Human Nature}.
\newblock Oxford: Clarendon Press, 1738.

\bibitem{Gromov2015}
M.~Gromov.
\newblock Symmetry, probabiliy, entropy: Synopsis of the lecture at maxent
  2014.
\newblock {\em Entropy}, 17:1273--1277, 2015.

\bibitem{Gromov2015a}
M.~Gromov.
\newblock Memorandum ergo. ,.
\newblock {\em Internal report IHES}, 2015.

\bibitem{Gromov2013}
M.~Gromov.
\newblock In a search for a structure, part 1: On entropy.
\newblock {\em unpublished manuscript
  http://www.ihes.fr/~gromov/PDF/structre-serch-entropy-july5-2012.pdf}, 2013.

\bibitem{Gromov2018}
M.~Gromov.
\newblock Alternative probabilities.
\newblock {\em Bernoully Lecture, March 27, 2018 EPFL}, 2018.

\bibitem{Hawking}
S.~Hawking.
\newblock Does god play dice?
\newblock {\em Online personnal website
  http://www.hawking.org.uk/does-god-play-dice.html}, 2016.

\bibitem{Natarajan2008}
V.~Natarajan.
\newblock What einstein meant when he said “god does not play dice ...".
\newblock {\em Resonance}, pages 655--661, 2008.

\bibitem{Bourbaki1968}
N.~Bourbaki.
\newblock {\em Theory of Sets - Elements of Mathematic.}
\newblock Addison Wesley publishing company. Hermann, 1968.

\bibitem{Maxwell1855}
J.C. Maxwell.
\newblock Experiments on colour as perceived by the eye, with remarks on
  colour-blindness.
\newblock {\em Transactions of the royal society of Edinburgh}, XXI(II), 1855.

\bibitem{Cencov1982}
N.N. Cencov.
\newblock {\em Statistical Decision Rules and Optimal Inference.}
\newblock Translations of Mathematical Monographs. Amer Mathematical Society,
  1982.

\bibitem{Morton2013}
J.~Morton.
\newblock Relations among conditional probabilities.
\newblock {\em Journal of Symbolic Computation}, 50:478--492, 2013.

\bibitem{Cox1961}
R.T. Cox.
\newblock Algebra of probable inference.
\newblock {\em Baltimore, Maryland: The Johns Hopkins Press.}, 1961.

\bibitem{Dupre2009}
M.~Dupre and Tipler F.
\newblock New axioms for rigorous bayesian probability.
\newblock {\em Bayesian Analysis}, 4(3):599--606, 2009.

\bibitem{Zadeh1965}
L.A. Zadeh.
\newblock Fuzzy sets.
\newblock {\em Information and Control}, 8(3):338--353, 1965.

\bibitem{Hajek2013}
P.~Hajek, L.~Godo, and F.~Esteva.
\newblock Fuzzy logic and probability.
\newblock {\em arXiv:1302.4953}, 2013.

\bibitem{Wierman2010}
M.J. Wierman.
\newblock {\em An Introduction to the Mathematics of Uncertainty}.
\newblock Creighton University., 2010.

\bibitem{Knuth2005}
K.~Knuth.
\newblock Lattice duality: The origin of probability and entropy.
\newblock {\em Neurocomputing}, 67:245--274, 2005.

\bibitem{Knuth2009}
K.~Knuth.
\newblock Measuring on lattices.
\newblock {\em arXiv:0909.3684 [math.GM]}, 2009.

\bibitem{Cartier2001}
P.~Cartier.
\newblock A mad day's work: from grothendieck to connes and kontsevich the
  evolution of concepts of space and symmetry.
\newblock {\em Bull. Amer. Math. Soc.}, 38(4):389--408, 2001.

\bibitem{Lawvere2014}
F.W. Lawvere.
\newblock Comments on the development of topos theory.
\newblock {\em Reprints in Theory and Applications of Categories}, 24:1–22,
  2014.

\bibitem{Artin1964}
M.~Artin, A.~Grothendieck, and J.L. Verdier.
\newblock {\em Theorie des topos et cohomologie etale des schemas - (SGA 4) Vol
  I,II,III}.
\newblock Seminaire de Geometrie Algebrique du Bois Marie. Berlin; New York,
  Springer-Verlag, coll. e Lecture notes in mathematics ,? 1972, 1964.

\bibitem{Grothendieck1985}
A.~Grothendieck.
\newblock {\em Realcoltes et Semailles,Reeflexions et temoignage sur un passe
  de mathématicien}.
\newblock Unpublished
  http://lipn.univ-paris13.fr/~duchamp/Books\&more/Grothendieck/RS/pdf/RetS.pdf,
  1985.

\bibitem{Andre2007a}
Y.~Andre.
\newblock Espace i. topos.
\newblock {\em Chap 1.}, 2007.

\bibitem{Reyes1977}
G.E. Reyes.
\newblock Sheaves and concepts: a model theoretic interpretation of
  grothendieck topoi.
\newblock {\em Cahier de topologie et g�om�trie differentielle
  categorique.}, 18, N.2:105--137, 1977.

\bibitem{Lawvere1972}
F.W Lawvere.
\newblock {\em Toposes, Algebraic Geometry and Logic,}.
\newblock Lecture Notes in Math., Springer-Verlag, 1972.

\bibitem{Barr1985}
M.~Barr and C.~Wells.
\newblock Toposes, triples and theories.
\newblock {\em Reprints in Theory and Applications of Categories (2005)
  Originally published by Springer-Verlag, NewYork}, 12:1–288, 1985.

\bibitem{Doering2008}
A.~Doering and C.~Isham.
\newblock A topos foundation for theories of physics: I. formal languages for
  physics.
\newblock {\em J.Math.Phys.}, 49, 2008.

\bibitem{Doering2012}
A.~Doering and C.J. Isham.
\newblock Classical and quantum probabilities as truth values.
\newblock {\em Journal of Mathematical Physics}, Vol: 53, 2012.

\bibitem{Baudot2015}
P.~Baudot and D.~Bennequin.
\newblock Topological forms of information.
\newblock {\em AIP Conf. Proc.}, 1641:213--221, 2015.

\bibitem{Simpson2012}
A.~Simpson.
\newblock Measure, randomness and sublocales.
\newblock {\em Annals of Pure and Applied Logic}, 163(11):1642--1659, 2012.

\bibitem{TapiaPacheco2017}
M.~Tapia, P.~Baudot, M.~Dufour, C.~Formizano-Treziny, S.~Temporal, M.~Lasserre,
  K.~Kobayashi, and Goaillard J.M.
\newblock Information topology of gene expression profile in dopaminergic
  neurons.
\newblock {\em BioArXiv168740}, 2017.

\bibitem{Stanley2011}
R.~Stanley.
\newblock {\em Enumerative Combinatorics. 2nd edition}.
\newblock Cambridge Studies in Advanced Mathematics, 2011.

\bibitem{Andrews}
G.E. Andrews.
\newblock Partitions.
\newblock {\em
  https://www.math.psu.edu/vstein/alg/antheory/preprint/andrews/chapter.pdf},
  chp 8., 1998.

\bibitem{Andrews1998}
G.~Andrews.
\newblock {\em The Theory of Partitions}.
\newblock Cambridge University Press, Cambridge, 1998.

\bibitem{Macdonald1995}
I.~G. Macdonald.
\newblock {\em Symmetric Functions and Hall Polynomials}.
\newblock Oxford Mathematical Monographs [2 ed.] Oxford University Press, 1995.

\bibitem{Godel1932}
K.~Godel.
\newblock On the intuitionistic propositional calculus.
\newblock {\em In Solomon Feferman et al., editor, Kurt G ̈odel: Collected
  Works. Oxford University Press, New York, 1986.}, I:222–5, 1932.

\bibitem{Gottwald2004}
S.~Gottwald.
\newblock Many-valued logic.
\newblock {\em in Stanford Encyclopedia of Philosophy 2008 ed.}, 2004.

\bibitem{Sorensen2006}
M.H.B. Sorensen and P.~Urzyczyn.
\newblock {\em Lectures on the Curry-Howard Isomorphism}.
\newblock Studies in Logic and the Foundations of Mathematics. Elsevier., 2006.

\bibitem{Hu1962}
Kuo~Ting Hu.
\newblock On the amount of information.
\newblock {\em Theory Probab. Appl.}, 7(4):439--447, 1962.

\bibitem{McGill1954}
W.J. McGill.
\newblock Multivariate information transmission.
\newblock {\em Psychometrika}, 19:p. 97--116, 1954.

\bibitem{Watanabe1960}
S.~Watanabe.
\newblock Information theoretical analysis of multivariate correlation.
\newblock {\em IBM Journal of Research and Development}, 4:66--81, 1960.

\bibitem{Studeny1999}
M.~Studeny and J.~Vejnarova.
\newblock The multiinformation function as a tool for measuring stochastic
  dependence.
\newblock {\em in M I Jordan, ed., Learning in Graphical Models, MIT Press,
  Cambridge}, pages 261--296, 1999.

\bibitem{Matsuda2001}
H.~Matsuda.
\newblock Information theoretic characterization of frustrated systems.
\newblock {\em Physica A: Statistical Mechanics and its Applications.}, 294
  (1-2):180--190, 2001.

\bibitem{Weibel1995}
C.~Weibel.
\newblock {\em An introduction to homological algebra}.
\newblock Cambridge University Press, 1995.

\bibitem{Yeung2003}
R.~Yeung.
\newblock On entropy, information inequalities, and groups.
\newblock {\em Communications, Information and Network Security}, Volume 712 of
  the series The Springer International Series in Engineering and Computer
  Science:333--359, 2003.

\bibitem{Yeung2007}
R.W. Yeung.
\newblock {\em Information Theory and Network Coding.}
\newblock Springer, 2007.

\bibitem{Yeung1997}
R.W. Yeung.
\newblock A framework for linear information inequalities.
\newblock {\em IEEE Transactions on Information Theory (New York)},
  43(6):1924–1934, 1997.

\bibitem{Csiszar2011}
J.~Csisz�r, I.;~K�rner.
\newblock {\em Information Theory: Coding Theorems for Discrete Memoryless
  Systems [Second ed.]}.
\newblock Cambridge University Press, 2011.

\bibitem{Shannon1953}
C.E. Shannon.
\newblock A lattice theory of information.
\newblock {\em Trans. IRE Prof. Group Inform. Theory}, 1:105--107, 1953.

\bibitem{Han1975}
T.~S. Han.
\newblock Linear dependence structure of the entropy space.
\newblock {\em Information and Control.}, vol. 29:p. 337--368, 1975.

\bibitem{Karatsuba1992}
A.A Karatsuba and S.~M. Voronin.
\newblock {\em The Riemann Zeta-Function}.
\newblock De Gruyter Expositions in Mathematics. Walter de Gruyter, 1992.

\bibitem{Voronin1975}
S.M. Voronin.
\newblock Theorem on the 'universality' of the riemann zeta-function.
\newblock {\em Izv. Akad. Nauk SSSR, Ser. Matem. (1975) (in Russian). English
  translation in: Math. USSR Izvestija 9 (1975)}, 39:443--453, 1975.

\bibitem{Zvontin1970a}
A.K. Zvontin and L.A. Levin.
\newblock The complexity of finite objects and the development of the concepts
  of information and randomness by means of the theory of algorithms.
\newblock {\em Russ. Math. Surv.}, 256:83--124, 1970.

\bibitem{Hardy1979}
G.H. Hardy and E.M. Wright.
\newblock {\em Introduction to the theory of numbers}.
\newblock Oxford University Press,. New York, 1979.

\bibitem{Khrennikov2004}
A.Y. Khrennikov and M.~Nilson.
\newblock {\em P-adic Deterministic and Random Dynamics}.
\newblock Springer, 2004.

\bibitem{Goedel1931}
K.~G�del.
\newblock �ber formal unentscheidbare s�tze der principia mathematica und
  verwandter systeme i.
\newblock {\em Monatsheft f�r Math. und Physik}, 38:p. 173--198, 1931.

\bibitem{Nagel1959}
J.R. Nagel, E.;~Newman.
\newblock {\em G�del's Proof}.
\newblock New York University Press, 1959.

\bibitem{Ornstein1970}
D.S. Ornstein.
\newblock Bernoulli shifts with the same entropy are isomorphic,.
\newblock {\em Advances in Math.}, 4.:337--352, 1970.

\bibitem{Milnor2011}
J.~Milnor.
\newblock Differential topology forty-six years later.
\newblock {\em Notices A.M.S.}, 58:804--809, 2011.

\bibitem{Weibel1999}
C.~Weibel.
\newblock History of homological algbera.
\newblock {\em The History of Topology ed. I.M. James, Elsevier,}, pages
  797--836, 1999.

\bibitem{Dieudonne1989}
J.A. Dieudonne.
\newblock {\em A History of Algebraic and Differential Topology 1900-1960}.
\newblock Modern Birkhauser Classics, 1989.

\bibitem{Leibniz1679}
G.W. Leibniz.
\newblock Characteristica geometrica.
\newblock {\em Echeverría, J. and Parmentier, M. eds. (1995)}, pages 148--9,
  1679.

\bibitem{Uchii2015}
S.~Uchii.
\newblock Monadology, information, and physics part 1: Metaphysics and
  dynamics.
\newblock {\em PhilSci-Archive 11523}, 2014.

\bibitem{Uchii2014}
S.~Uchii.
\newblock Monadology, information, and physics part 2: Space and time.
\newblock {\em PhilSci-Archive 11647}, 2014.

\bibitem{Uchii2014a}
S.~Uchii.
\newblock Monadology, information, and physics part 3 (revised): Inertia and
  gravity.
\newblock {\em PhilSci-Archive 11125}, 2014.

\bibitem{Euler1759}
L.~Euler.
\newblock Solutio problematis ad geometriam situs pertinentis.
\newblock {\em Mémoires de l'Académie des sciences de Berlin}, 8:128--140,
  1759.

\bibitem{Riemann1857}
B.~Riemann.
\newblock Theorie der abel'schen functionen.
\newblock {\em Journal fur die reine und angewandte Mathematik}, 54:101--155,
  1857.

\bibitem{Poincare1895a}
H.~Poincare.
\newblock {\em Analysis Situs.}, volume~2.
\newblock Journal de l'École Polytechnique, 1895.

\bibitem{Bass2000}
H.~Bass, H.~Cartan, P.~Freyd, A.~Heller, and S.~Mac~Lane.
\newblock Samuel eilenberg 1913-1998 - a bibliographical memoir.
\newblock {\em Bibliographical memoir - The national academy press Washington},
  79, 2000.

\bibitem{Grothendieck1997}
A.~Grothendieck.
\newblock {\em Long March through Galois theory}.
\newblock Cambridge University Press, 1997.

\bibitem{Deleuze2000}
G.~Deleuze.
\newblock {\em La difference et la repetition.}
\newblock Presse Universite de France. Epimethee, 2000.

\bibitem{Andre2008}
Y.~Andre.
\newblock Ambiguity theory, old and new.
\newblock {\em arXiv:0805.2568}, 2008.

\bibitem{Andre2007b}
Y.~Andre.
\newblock {\em Symétries I. Idées galoisiennes.}
\newblock Ircam online courses, 2007.

\bibitem{Bennequin1994}
D.~Bennequin.
\newblock Questions de physique galoisienne.
\newblock {\em Passion des Formes Dynamique qualitative, Sémiophysique et
  Intelligibilité A René Thom, Michèle Porte coordonateur Paris, ENS
  Editions Fontenay-St Cloud}, pages 311--410, 1994.

\bibitem{Cook2009}
M.~Cook.
\newblock {\em Mathematicians: An Outer View of the Inner World}.
\newblock Princeton University Press, 2009.

\bibitem{Berger2009}
M.~Berger.
\newblock {\em Geometry I.}
\newblock Springer, Universitext, Corrected edition, 2009.

\bibitem{Chevalley1948}
Cl. Chevalley and S.~Eilenberg.
\newblock Cohomology theory of lie groups and lie algebras.
\newblock {\em Transactions of the American Mathematical Society},
  63(1):85--124, 1948.

\bibitem{Alexandroff1932a}
P.~Alexandroff.
\newblock {\em Elementary concepts of Topology}.
\newblock Dover Publication Inc., 1932.

\bibitem{Hatcher2002}
A.~Hatcher.
\newblock {\em Algebraic Topology}.
\newblock Cambridge University Press, 2002.

\bibitem{Ghrist2014}
R.~Ghrist.
\newblock {\em Elementary Applied Topology ed. 1.0}.
\newblock Createspace., 2014.

\bibitem{Erdoes1959}
P.~Erd�s, P.s and A.~R�nyi.
\newblock On random graphs. i.
\newblock {\em Publicationes Mathematicae}, 6:290--297, 1959.

\bibitem{Watts1998}
D.~J. Watts and S.~H. Strogatz.
\newblock Collective dynamics of 'small-world' networks.
\newblock {\em Nature}, 393 (6684):440--442, 1998.

\bibitem{Dorogovtsev2008}
S.N. Dorogovtsev, A.V. Goltsev, and J.F.F. Mendes.
\newblock Critical phenomena in complex networks.
\newblock {\em Rev. Mod. Phys.}, 80, 4:1275, 2008.

\bibitem{Newman2010}
M.~Newman.
\newblock {\em Networks: an introduction.}
\newblock Oxford University Press, 2010.

\bibitem{Banchoff1970}
T.F. Banchoff.
\newblock Critical points and curvature for embedded polyhedral surfaces.
\newblock {\em The American Mathematical Monthly}, Vol. 77, No. 5:475--485,
  1970.

\bibitem{Milnor1964}
J.~Milnor.
\newblock {\em Morse theory}.
\newblock Princeton University Press, 1964.

\bibitem{Forman2002}
R.~Forman.
\newblock A user's guide to discrete morse theory.
\newblock {\em S�minaire Lotharingien de Combinatoire.
  http://www.emis.de/journals/SLC/wpapers/s48forman.pdf}, 48:48c, 35 p., 2002.

\bibitem{Postnikov1951}
A.~Postnikov.
\newblock Determination of the homology groups of a space by means of the
  homotopy invariants.
\newblock {\em Doklady Akademii Nauk SSSR}, 76:359–362, 1951.

\bibitem{Milnor1954}
J.~Milnor.
\newblock Link groups.
\newblock {\em Annals of Mathematics}, 59 (2):177--195, 1954.

\bibitem{Carlsson2009}
G.~Carlsson.
\newblock Topology and data.
\newblock {\em Bull. Amer. Math. Soc.}, 46:p.255--308, 2009.

\bibitem{Port2018}
A.~Port, I.~Gheorghita, D.~Guth, J.M. Clark, C.~Liang, S.~Dasu, and
  M.~Marcolli.
\newblock Persistent topology of syntax.
\newblock {\em Mathematics in Computer Science}, 1(12):33--50, 2018.

\bibitem{Zeidler2011}
E.~Zeidler.
\newblock {\em Quantum Field Theory III: Gauge Theory. A Bridge between
  Mathematicians and Physicists}.
\newblock Springer-Verlag Berlin Heidelberg, 2011.

\bibitem{Wheeler1982}
J.A. Wheeler.
\newblock Physics and austerity, law without law.
\newblock {\em Anhui Science and Technology Publications, Anhui, China}, 1982.

\bibitem{Misner1973}
C.W Misner, K.S. Thorne, and J.A. Wheeler.
\newblock {\em Gravitation}.
\newblock San Francisco, W.H. Freeman, 1973.

\bibitem{Rovelli2008}
C.~Rovelli.
\newblock Notes for a brief history of quantum gravity.
\newblock {\em arXiv:gr-qc/0006061v3}, 2008.

\bibitem{Wheeler1990}
J.A. Wheeler.
\newblock Information, physics, quantum: The search for links.
\newblock {\em in Complexity, Entropy and the Physics of Information ed.,
  Wojciech H. Zurek}, 1990.

\bibitem{Cathelineau1988}
J.L. Cathelineau.
\newblock Sur l'homologie de sl2 a coefficients dans l'action adjointe.
\newblock {\em Math. Scand.}, 63:51--86, 1988.

\bibitem{Kontsevitch1995}
M.~Kontsevitch.
\newblock The 11/2 logarithm.
\newblock {\em Unpublished note. Reproduced in Elbaz-Vincent \& Gangl, 2002 On
  poly(ana)logs I. Compositio Mathematica}, 1995.

\bibitem{Elbaz-Vincent2002}
Ph. Elbaz-Vincent and H.~Gangl.
\newblock On poly(ana)logs i.
\newblock {\em Compositio Mathematica}, 130(2):161--214, 2002.

\bibitem{Connes2009}
A.~Connes and C.~Consani.
\newblock Characteristic 1, entropy and the absolute point.
\newblock {\em preprint arXiv:0911.3537v1.}, 2009.

\bibitem{Marcolli2011}
M.~Marcolli and R.~Thorngren.
\newblock Thermodynamic semirings.
\newblock {\em arXiv 10.4171/JNCG/159}, Vol. abs/1108.2874, 2011.

\bibitem{Fresse2004}
B.~Fresse.
\newblock Koszul duality of operads and homology of partitionn posets.
\newblock {\em Contemp. Math. Amer. Math. Soc.}, 346:pp. 115--215, 2004.

\bibitem{Schoeller2018}
F.~Schoeller, L.~Perlovsky, and D.~Arseniev.
\newblock Physics of mind: experimental confirmations of theoretical
  predictions.
\newblock {\em Physics of Life Reviews}, 2018.
  https://doi.org/10.1016/j.plrev.2017.11.021 [this issue].

\bibitem{Costa2002}
M.~Costa, A.L. Goldberger, and C.K. Peng.
\newblock Multiscale entropy to distinguish physiologic and synthetic rr time
  series.
\newblock {\em Computers in Cardiology}, 29:137--140, 2002.

\bibitem{Costa2005}
M.~Costa, A.L. Goldberger, and C.K. Peng.
\newblock Multiscale entropy analysis of biological signals.
\newblock {\em Phys Rev E}, 71 ::021906, 2005.

\bibitem{Hochschild1945}
G.~Hochschild.
\newblock On the cohomology groups of an associative algebra.
\newblock {\em Annals of Mathematics. Second Series,}, 46:58–67, 1945.

\bibitem{Tate1991}
J.~Tate.
\newblock Galois cohomology.
\newblock {\em online course}, 1991.

\bibitem{Cartan1956}
H.~Cartan and S.~Eilenberg.
\newblock {\em Homological Algebra}.
\newblock The Princeton University Press, Princeton, 1956.

\bibitem{MacLane1975}
S.~Mac~Lane.
\newblock {\em Homology.}
\newblock Classic in Mathematics, Springer, Reprint of the 1975 edition, 1975.

\bibitem{Kendall1964}
D.G. Kendall.
\newblock Functional equations in information theory.
\newblock {\em Z. Wahrscheinlichkeitstheorie}, 2:p. 225--229, 1964.

\bibitem{Lee1964}
P.M. Lee.
\newblock On the axioms of information theory.
\newblock {\em The Annals of Mathematical Statistics}, Vol. 35, No. 1:pp.
  415--418, 1964.

\bibitem{Gerstenhaber1987}
M.~Gerstenhaber and S.D. Schack.
\newblock A hodge-type decomposition for commutative algebra cohomology.
\newblock {\em Journal of Pure and Applied Algebra}, 48(1-2):229--247, 1987.

\bibitem{Kassel2004}
C.~Kassel.
\newblock Homology and cohomology of associative algebras- a concise
  introduction to cyclic homology.
\newblock {\em Advanced Course on non-commutative geometry}, 2004.

\bibitem{Hilbert1924}
D.~Hilbert.
\newblock Sur l'infini. hilbert's lectures on the infinite.
\newblock {\em Traduit par Andre Weil Paris (1926). edited in David Hilbert's
  Lectures on the Foundations of Arithmetic and Logic 1917-1933. Springer},
  1924.

\bibitem{Pethel2014}
S.D. Pethel and D.W. Hahs.
\newblock Exact test of independence using mutual information.
\newblock {\em Entropy}, 16:2839--2849, 2014.

\bibitem{Adami1999}
C.~Adami and N.J. Cerf.
\newblock Prolegomena to a non-equilibrium quantum statistical mechanics.
\newblock {\em Chaos, Solitons \& Fractals}, 10(10):1637--1650, 1999.

\bibitem{Kapranov2011}
M.~Kapranov.
\newblock Thermodynamics and the moment map.
\newblock {\em arXiv:1108.3472}, 2011.

\bibitem{Cover1994}
T.M. Cover.
\newblock Which processes satisfy the second law?
\newblock {\em in: Physical Origins of Time Asymmetry, , eds. J. J. Halliwell,
  J. Perez-Mercader and W. H. Zurek}, pages 98--107, 1994.

\bibitem{Baez2013}
J.~Baez and B.~Fong.
\newblock A noether theorem for markov processes.
\newblock {\em Journal of Mathematical Physics}, 54:013301, 2013.

\bibitem{Barnett2009}
L.~Barnett, A.K. Barrett, and Seth A.K.
\newblock Granger causality and transfer entropy are equivalent for gaussian
  variables.
\newblock {\em Phys. Rev. Lett.}, 103:238701, 2009.

\bibitem{Schreiber2000}
T.~Schreiber.
\newblock Measuring information transfer.
\newblock {\em Physical Review Letters}, 85(2):461–464, 2000.

\bibitem{Dirac1929}
P.~Dirac.
\newblock Discussion of the infinite distribution of electrons in the theory of
  the positron.
\newblock {\em Proc. Camb. Phil. Soc.}, 25:62, 1929.

\bibitem{Feynman1985}
R.~Feynman.
\newblock {\em QED. The Strange Theory of Light and Matter.}
\newblock Princeton University Press, 1985.

\bibitem{Vannimenus1977}
J.~Vannimenus and G.~Toulouse.
\newblock Theory of the frustration effect. ii. ising spins on a square
  lattice.
\newblock {\em Journal of Physics C: Solid State Physics}, 10(18), 1977.

\bibitem{Mezard2009a}
M.~Mezard and A.~Montanari.
\newblock {\em Information, Physics, and Computation}.
\newblock Oxford University Press, 2009.

\bibitem{Griffiths1984}
R.B. Griffiths.
\newblock Consistent histories and the interpretation of quantum mechanics.
\newblock {\em J. Stat. Phys.}, 35:219, 1984.

\bibitem{Omnes1988}
R.~Omnes.
\newblock Logical reformulation of quantum mechanics i. foundations.
\newblock {\em Journal of Statistical Physics}, 53:893--932, 1988.

\bibitem{Gell-Mann1990}
M.~Gell-Mann and J.B. Hartle.
\newblock Quantum mechanics in the light of quantum cosmology.
\newblock {\em W. H. Zurek (ed.), Complexity, entropy and the physics of
  information. Redwood City, Calif.: Addison-Wesley}, pages 425--458, 1990.

\bibitem{Rudrauf2017}
D.~Rudrauf, D.~Bennequin, I.~Granic, G.~Landini, K.~Friston, and K.~Williford.
\newblock A mathematical model of embodied consciousness.
\newblock {\em J Theor Biol.}, 428:106--131, 2017.

\bibitem{DAgostino2014a}
G.~D'Agostino and A.~Scala.
\newblock {\em Networks of Networks: The Last Frontier of Complexity}.
\newblock Springer - Understanding Complex Systems, 2014.

\bibitem{Reshef2011}
D.N. Reshef, Y.A. Reshef, H.K. Finucane, S.R. Grossman, G.~McVean, P.J.
  Turnbaugh, E.S. Lander, M.~Mitzenmacher, and P.C. Sabeti.
\newblock Detecting novel associations in large data sets.
\newblock {\em Science}, 334:1518, 2011.

\bibitem{Brenner2000}
N.~Brenner, S.~Strong, R.~Koberle, and W.~Bialek.
\newblock Synergy in a neural code.
\newblock {\em Neural Computation.}, 12:1531--1552, 2000.

\bibitem{Schroedinger1944}
E.~Schrodinger.
\newblock {\em What is Life?}
\newblock Based on lectures delivered under the auspices of the Dublin
  Institute for Advanced Studies at Trinity College, Dublin, in February 1943,
  1944.

\bibitem{Williams2010}
P.~Williams and R.~Beer.
\newblock Nonnegative decomposition of multivariate information.
\newblock {\em arXiv:1004.2515v1}, 2010.

\bibitem{Olbrich2015}
E.~Olbrich, N.~Bertschinger, and J.~Rauh.
\newblock Information decomposition and synergy.
\newblock {\em entropy}, 17(5):3501--3517, 2015.

\bibitem{Bertschinger2014}
N.~Bertschinger, J.~Rauh, E.~Olbrich, J.~Jost, and N.~Ay.
\newblock Quantifying unique information.
\newblock {\em Entropy}, 16:2161–2183, 2014.

\bibitem{Griffith2014}
V.~Griffith and C.~Koch.
\newblock Quantifying synergistic mutual information.
\newblock {\em In Guided Self-Organization: Inception; Prokopenko, M., Ed.;
  Springer: Berlin/Heidelberg, Germany}, pages 159--190, 2014.

\bibitem{Wibral2017}
M.~Wibral, C.~Finn, P.~Wollstadt, J.T Lizier, and V.~Priesemann.
\newblock Quantifying information modification in developing neural networks
  via partial information decomposition.
\newblock {\em Entropy}, 19(9):494, 2017.

\bibitem{Kay2017}
J.W. Kay, R.A.A. Ince, B.~Dering, and W.~Phillips.
\newblock Partial and entropic information decompositions of a neuronal
  modulatory interaction.
\newblock {\em Entropy}, 19(11):560, 2017.

\bibitem{Seth2006}
A.K. Seth, E.~Izhikevich, G.N. Reeke, and G.M. Edelman.
\newblock Theories and measures of consciousness: An extended framework.
\newblock {\em PNAS}, 103(28):10799--804, 2006.

\bibitem{Atlan1979}
H.~Atlan.
\newblock {\em Entre le cristal et la fumee. Essai sur l'organisation du
  vivant.}
\newblock Seuil, 1979.

\bibitem{Gibson1979}
J.J. Gibson.
\newblock {\em The ecological approach to visual perception}.
\newblock Boston: Houghton Mifflin, 1979.

\bibitem{Cassirer1938}
E.~Cassirer.
\newblock The concept of group and the theory of perception.
\newblock {\em A. Gurwitsch, trans. Philosophy and Phenomenological Research
  (1944)}, 5:1--35, 1938.

\bibitem{Piaget1970}
J.~Piaget.
\newblock {\em Structuralism}.
\newblock New York: Basic Books, 1970.

\bibitem{Stratton1896}
G.M. Stratton.
\newblock Some preliminary experiments on vision without inversion of the
  retinal image.
\newblock {\em Psychological Review}, 3(6):611--617, 1896.

\bibitem{Stratton1897}
G.M. Stratton.
\newblock Upright vision and the retinal image.
\newblock {\em Psychological Review}, 4(2):182--187, 1897.

\bibitem{Kohler1962}
I.~Kohler.
\newblock Experiments with goggles.
\newblock {\em Scientific American}, 206(5):62--72, 1962.

\bibitem{Gibson1933}
J.J. Gibson.
\newblock Adaptation, after-effect and contrast in the perception of curved
  lines.
\newblock {\em Experimental Psychology}, 16(1):1--31, 1933.

\bibitem{Sharma2000}
J.~Sharma, A.~Angelucci, and M.~Sur.
\newblock Induction of visual orientation modules in auditory cortex.
\newblock {\em Nature}, 404:841--847, 2000.

\bibitem{Sur2001}
M.~Sur and C.~Leamey.
\newblock Development and plasticity of cortical areas and networks.
\newblock {\em Nat. Rev. Neurosci.}, 2:251–262, 2001.

\bibitem{Sur2005}
M.~Sur and J.L.R. Rubenstein.
\newblock Patterning and plasticity of the cerebral cortex.
\newblock {\em Science}, 310(5749):805--807, 2005.

\bibitem{Roe1990}
A.W. Roe, S.L. Pallas, J.O. Hahm, and M.~Sur.
\newblock A map of visual space induced in primary auditory cortex.
\newblock {\em Science}, 250:818–820, 1990.

\bibitem{Boring1952}
E.~Boring.
\newblock Visual perception as invariance.
\newblock {\em Psychological Review}, 59(2):141--148, 1952.

\bibitem{Cutting1983}
J.E. Cutting.
\newblock Observations - four assumptions about invariance in perception.
\newblock {\em Journal of Experimental Psychology}, 9(2):310--317, 1983.

\bibitem{Todd2001}
J.T. Todd, A.H. Oomes, J.J. Koenderink, and A.M. Kappers.
\newblock On the affine structure of perceptual space.
\newblock {\em Psychol Sci.}, 12(3):191--196, 2001.

\bibitem{Koenderink2002}
J.~Koenderink, A.J. Doorn, and A.~Kappers.
\newblock Pappus in optical space.
\newblock {\em Perception \& Psychophysics}, 64(3):380--391, 2002.

\bibitem{Battro1976}
A.M. Battro, S.P. Netto, and R.J.A. Rozestraten.
\newblock Riemannian geometries of variable curvature in visual space: Visual
  alleys, horopters, and triangles in big open fields.
\newblock {\em Perception}, 5:9--23, 1976.

\bibitem{Indow1991}
T.~Indow.
\newblock A critical review of luneburg’s model with regard to global
  structure of visual space.
\newblock {\em Psychological Review}, 98:430–453, 1991.

\bibitem{Koenderink2000}
J.~Koenderink, A.~Doorn, and J.~Lappin.
\newblock Direct measurement of the curvature of visual space.
\newblock {\em Perception}, 29:69–80, 2000.

\bibitem{Suppes1977}
P.~Suppes.
\newblock Is visual space euclidean?
\newblock {\em Synthese}, 35:397–421, 1977.

\bibitem{Ermentrout1979}
G.B. Ermentrout and J.D. Cowan.
\newblock A mathematical theory of visual hallucinations.
\newblock {\em Biol. Cybern.}, 34:137--150, 1979.

\bibitem{Bressloff2001}
PC~Bressloff, J.D. Cowan, M.~Golubitsky, P.J. Thomas, and M.C. Wiener.
\newblock Geometric visual hallucinations, euclidean symmetry, and the
  functional architecture of striate cortex.
\newblock {\em Phil. Trans. Royal Soc. London}, 356:299–330, 2001.

\bibitem{Golubitsky2006b}
M.~Golubitsky.
\newblock Symmetry and neuroscience.
\newblock {\em Bulletion of the AMS, Amercian Mathematical Society}, January
  2006.

\bibitem{Thom2002}
R.~Thom.
\newblock Logos et theorie des catastrophes. expose de rene thom au colloque
  international de cerisy 1982.
\newblock {\em Annales de la Fondation Louis de Broglie}, 27 (4):575--595,
  2002.

\bibitem{Riemann1854}
B.~Riemann.
\newblock On the hypotheses which lie at the bases of geometry. translated by
  william kingdon clifford.
\newblock {\em Nature}, VIII N�183-184:14--17,36, 37, 1854.

\bibitem{Klein1872}
F.~Klein.
\newblock Vergleichende bertrachtungen uber neuere geometrische forschungen
  erlangen.
\newblock {\em Reprinted in Mathematische Annalen 1893}, 1:460--97., 1872.

\bibitem{Birkhoff1988}
G.~Birkhoff and M.K. Bennett.
\newblock Felix klein and his "erlanger programm".
\newblock {\em In History and Philosophy of Modern Mathematics - U of Minnesota
  Press,}, pages 145--176, 1988.

\bibitem{Smale1967}
S.~Smale.
\newblock Differentiable dynamical systems.
\newblock {\em Bulletin of the American Mathematics Society}, 73:747--817,
  1967.

\bibitem{Mumford1994}
D.~Mumford, J.~Fogarty, and F.~Kirwan.
\newblock {\em Geometric invariant theory}.
\newblock Results in Mathematics and Related Areas (3rd ed.), Berlin, New York:
  Springer-Verlag, 1994.

\bibitem{Izhikevich2007}
E.M. Izhikevich.
\newblock {\em Dynamical Systems in Neuroscience: The Geometry of Excitability
  and Bursting.}
\newblock The MIT press, 2007.

\bibitem{Sompolinsky1988}
H.~Sompolinsky, A.~Crisanti, and H.J. Sommers.
\newblock Chaos in random neural networks.
\newblock {\em Physical Review Letters}, 61:259, 1988.

\bibitem{Bergelson2010}
V.~Bergelson, T.~Tao, and T.~Ziegler.
\newblock An inverse theorem for the uniformity seminorms associated with the
  action of fω.
\newblock {\em Geom. Funct. Anal.}, 6:1539--1596, 2010.

\bibitem{Tao2008a}
T.~Tao.
\newblock Cohomology for dynamical systems.
\newblock {\em Online Blog
  https://terrytao.wordpress.com/2008/12/21/cohomology-for-dynamical-systems/},
  2008.

\bibitem{Ornstein1971}
D.S. Ornstein.
\newblock Some new results in the kolmogorov-sinai theory of entropy and
  ergodic theory.
\newblock {\em BULLETIN OF THE AMERICAN MATHEMATICAL SOCIETY}, Volume 77,
  Number 6,:878--888, 1971.

\bibitem{Katok2007}
A.~Katok.
\newblock fifty years of entropy in dynamics: 1958-2007.
\newblock {\em Journal of modern dynamics}, 1(4):545–596, 2007.

\bibitem{Gaboriau2016}
D.~Gaboriau.
\newblock Entropie sofique d’après lewis bowen, david kerr et hanfeng li.
\newblock {\em Séminaire Bourbaki du 16 janvier 2016. 2016. <hal-01347306>},
  2016.

\bibitem{Ornstein1987}
D.~Ornstein and B.~Weiss.
\newblock Entropy and isomorphism theorems for actions of amenable groups.
\newblock {\em J. Analyse Math.}, 48:1–141, 1987.

\bibitem{Neumann1929}
J.~von Neumann.
\newblock Zur allgemeinen theorie des maßes.
\newblock {\em Fund. Math.}, 13(1):73–111, 1929.

\bibitem{Ornstein2004}
D.~Ornstein.
\newblock Kolmogorov, random processes, and newtonian dynamics.
\newblock {\em Russian Math. Surveys}, 59(2):121–126, 2004.

\bibitem{Pesin1977}
Y.~Pesin.
\newblock Characteristic lyapunov exponents and smooth ergodic theory.
\newblock {\em Russ. Math. Surveys}, 32:55--114, 1977.

\bibitem{Gaspard1993}
P.~Gaspard and X.J. Wang.
\newblock Noise, chaos, and (t,e)-entropy per unit time.
\newblock {\em Phys. Rep.}, 235(6):321--373, 1993.

\bibitem{Destexhe1992}
A.~Destexhe.
\newblock {\em Nonlinear Dynamics of the Rhythmical Activity of the Brain}.
\newblock Doctoral Dissertation (Université Libre de Bruxelles, Brussels),
  1992.

\bibitem{El-Boustani2010}
S.~El-Boustani and A.~Destexhe.
\newblock Brain dynamics at multiple scales: Can one reconcile the apparent
  low-dimensional chaos of macroscopic variables with the seemingly stochastic
  behavior of single neurons?
\newblock {\em International Journal of Bifurcation and Chaos},
  20(6):1687--1702, 2010.

\bibitem{Stern2017}
P.~Stern.
\newblock Neuroscience: In search of new concepts.
\newblock {\em Science}, 358(6362):464--465, 2017.

\bibitem{Bertot2015}
Yves Bertot and P.~Casteran.
\newblock {\em Le Coq’ Art (V8)}.
\newblock Online http://www-sop.inria.fr/members/Yves.Bertot/coqartF.pdf, 2015.

\bibitem{Silver2017}
D.~Silver, Schrittwieser J., Simonyan K., and al.
\newblock Mastering the game of go without human knowledge.
\newblock {\em Nature}, 550(7676):354--359, 2017.

\bibitem{Mnih2015}
V.~Mnih, K.~Kavukcuoglu, and D.~Silver.
\newblock Human-level control through deep reinforcement learning.
\newblock {\em Nature}, 518(7540):529--33, 2015.

\bibitem{Kohn1999}
W.~Kohn.
\newblock Electonic structure of matter - wave functions and density
  functional.
\newblock {\em Nobel Lecture, January 28, 1999}, 1999.

\bibitem{Aloupis2012}
G.~Aloupis, E.~Demaine, A.~Guo, and G.~Viglietta.
\newblock Classic nintendo games are (computationally) hard.
\newblock {\em arXiv:1203.1895v3}, 2012.

\bibitem{Nakagaki2000}
T.~Nakagaki, H.~Yamada, and A.~Toth.
\newblock Intelligence: Maze-solving by an amoeboid organism.
\newblock {\em Nature}, 407:470, 2000.

\bibitem{Wigner1960}
E.P. Wigner.
\newblock The unreasonable effectiveness of mathematics in the natural
  sciences.
\newblock {\em Communications on Pure and Applied Mathematics.}, 13:1--14,
  1960.

\bibitem{Blazewicz2012}
Jacek Blazewicz and Marta Kasprzak.
\newblock Complexity issues in computational biology.
\newblock {\em Fundamenta Informaticae}, 118(4):385--401, 2012.

\bibitem{Atick1992}
J.J. Atick.
\newblock Could information theory provide an ecological theory of sensory
  processing.
\newblock {\em Network: Computation in Neural Systems.}, 3:213--251, 1992.

\bibitem{Dieudonne1987a}
J.~Dieudonne.
\newblock {\em Pour l’honneur de l’esprit humain}.
\newblock Les mathematiques aujourd’hui - Hachette, 1987.

\bibitem{Fritz2018}
T.~Fritz and P.~Perrone.
\newblock Bimonoidal structure of probability monads.
\newblock {\em arXiv:1804.03527v1}, 2018.

\end{thebibliography}

\end{document}